\newcommand{\mass}[1]{\ensuremath{#1\, \mathrm{M}_\odot}}

\newcommand{\Teff}{\ensuremath{T{_\mathrm{\!\!eff}}}}

\newcommand{\chem}[2]{\ensuremath{^{#2}\kern-0.8pt\mathrm{#1}}}
\newcommand{\reac}[6]{\ensuremath{\,^{#2}\kern-0.8pt\mathrm{#1}\,({#3}\,,{#4})\,{}^{#6}\kern-0.8pt\mathrm{#5}\,}}
\newcommand{\raHour}[3]{\ensuremath{#1^{\mathrm{h}}#2^{\mathrm{m}}#3^{\mathrm{s}}}}
\newcommand{\decDeg}[3]{\ensuremath{#1^{\mathrm{d}}#2^{\mathrm{m}}#3^{\mathrm{s}}}}

\documentclass[structabstract]{aa}  
\usepackage{graphicx}
\usepackage{enumerate}
\usepackage{txfonts}
\usepackage{natbib}

\usepackage{ulem}
\usepackage{color}

\def\toReferee#1{#1}

\begin{document}
   \title{Stellar variability in open clusters}

   \subtitle{I. A new class of variable stars in NGC~3766}

   \author{N. Mowlavi\inst{1}
          \and
          F. Barblan\inst{1}
          \and
          S. Saesen\inst{1}
          \and
          L. Eyer\inst{1}
          }

   \institute{$^1$Astronomy Department, Geneva Observatory, chemin des Maillettes, 1290 Versoix, Switzerland
                \email{Nami.Mowlavi@unige.ch}
             }

   \date{Accepted 16/04/2013}


  \abstract
   {}
   {We analyze the population of periodic variable stars in the open cluster NGC~3766 based on a 7-year multiband monitoring campaign conducted on the 1.2~m Swiss Euler telescope at La Silla, Chili.
   }
   {The data reduction, light curve cleaning, and period search procedures, combined with the long observation time line, allowed us to detect variability amplitudes down to the mmag level.
   The variability properties were complemented with the positions in the color-magnitude and color-color diagrams to classify periodic variable stars into distinct variability types.
   }
   {We find a large population (36 stars) of new variable stars between the red edge of slowly pulsating B (SPB) stars and the blue edge of $\delta$~Sct stars, a region in the Hertzsprung-Russell (HR) diagram where no pulsation is predicted to occur based on standard stellar models.
\toReferee{The bulk of their periods ranges from 0.1 to 0.7~d}, with amplitudes between 1 and 4~mmag for the majority of them.
About 20\% of stars in that region of the HR diagram are found to be variable, but the number of members of this new group is expected to be higher, with amplitudes below our mmag detection limit.

The properties of this new group of variable stars are summarized and arguments set forth in favor of a pulsation origin of the variability, with $g$-modes sustained by stellar rotation.
Potential members of this new class of low-amplitude periodic (most probably pulsating) A \toReferee{and late-}B variables in the literature are discussed.

We additionally identify 16 eclipsing binary, 13 SPB, 14 $\delta$~Sct, and 12 $\gamma$~Dor candidates, as well as 72 fainter periodic variables.
   All are new discoveries.
   }
   {We encourage searching for this new class of variables in other young open clusters, especially in those hosting a rich population of Be stars.}


   \maketitle

\section{Introduction}
\label{Sect:introduction}

Stars are known to pulsate under certain conditions that translate into instability strips in the Hertzsprung-Russell (HR) or color-magnitude (CM) diagram \citep[see reviews by, e.g.,][]{GautschySaio95,GautschySaio96,EyerMowlavi08}.
On the main sequence (MS), $\delta$~Sct stars (A- and early F-type stars pulsating in $p$-modes) are found in the classical instability strip triggered by the $\kappa$~mechanism acting on H and He.
At higher luminosities, $\beta$~Cep ($p$-mode) and slowly pulsating B (SPB; $g$-mode) stars are found with pulsations triggered by the $\kappa$~mechanism acting on the iron-group elements, while $\gamma$~Dor stars (early F-type; $g$-mode) lie at the low-luminosity edge of the classical instability strip, with pulsations triggered by the `convective blocking' mechanism \citep{Pesnell87,GuzikKayeBradley00}.

The detection and analysis of the pulsation frequencies of pulsating stars lead to unique insights into their internal structure and the physics in play.
Asteroseismology of $\beta$~Cep stars, for example, has opened the doors in the past decade to study their interior rotation and convective core \citep[][and references therein]{AertsThoulDaszynska_etal03,DupretThoulScuflaire_etal04,AusseloosScuflaireThoul_etal04,LovekinGoupil10,BriquetNeinerAerts_etal12}.
SPB stars also offer high promise for asteroseimology, with their $g$-modes probing the deep stellar interior \citep{DeCat07}.
For those stars, however, determining the degrees and orders of their modes is complex because, as pointed out by \cite{AertsDeCatKuschnig06}, a) the predicted rich eigenspectra lead to non-unique solutions and b) stellar rotation splits the frequencies and causes multiplets of adjacent radial orders to overlap.
Space-based observations by the MOST satellite have shown the advantage of obtaining continuous light curves from space \citep[e.g.][]{BalonaPigulskiCat_etal11,GruberSaioKuschnig12}, and the data accumulated by satellites like CoRoT and Kepler may bring important breakthroughs in this field in the near future \citep{McNamaraJackiewiczMcKeever12}.
The same is true for $\gamma$~Dor stars \citep[see][for a review of $\gamma$~Dor stars]{UytterhoevenMoyalGrigahcene11,BalonaGuzikUytterhoeven_etal11,Pollard09}, as well as for $\delta$~Sct stars \citep[e.g.][]{FoxMachado_PerezHernandez_Suarez_etala06}.
CoRoT has observed hundreds of frequencies in a $\delta$~Sct star \citep{PorettiMichelGarrido09}, and yet, mode selection and stellar properties derivation remain difficult.  
Much effort continues to be devoted to analyzing the light curves of all those pulsating stars, considering the good prospects for asteroseismology.
Obviously, increasing the number of known pulsators like them is essential.

Another aspect of stellar variability concerns the question why some stars pulsate and others do not.
As a matter of fact, not all stars in an instability strip pulsate.
\cite{BriquetHubrigDeCat_etal07}, for example, address this question for pulsating B stars, while \cite{BalonaDziembowski11} note that the identification of constant stars, observed by the Kepler satellite, in the $\delta$~Sct instability strip is not due to photometric detection threshold.
\cite{UshomirskyBildsten98} provide an interesting view of this issue in relation with SPB stars, arguing that stellar rotation, combined with the viewing angle relative to the rotation axis, may be at the origin of not detecting pulsations in stars that are nevertheless pulsating.
Conversely, pulsations are observed when not expected.
A historical example is given by the unexpected discovery of rapidly oscillating Ap (roAp) stars \citep{Kurtz82}, the pulsation of which has been attributed to strong magnetic fields \citep[][for a review]{Kochukhov07}.
The potential existence of pulsators in regions outside instability strips must also be addressed.
Such a gap exists on the MS in the region between $\delta$~Sct and SPB stars, as shown in Fig.~3 of \citet{Pamyatnykh99}, in Fig.~22 of \cite{Christensen-Dalsgaard04}, or in Fig.~4 of \cite{Pollard09}, but we show in this paper evidence for the presence of periodic variables in this region of the HR diagram.

Stellar clusters are ideal environments to study stellar variability because some basic properties and the evolutionary status of individual star members can be derived from the properties of the cluster.
It, however, requires extensive monitoring on an as-long-as-possible time base line.
This requirement may explain why not many clusters have been studied for their variability content so far, compared to the number of known and characterized clusters.
In order to fill this shortcoming, an observation campaign of twenty-seven Galactic open clusters was triggered by the Geneva group in 2002.
It ended in 2009, resulting in a rich database for a wide range of cluster ages and metallicities \citep[][Saesen et al., in preparation]{GrecoMowlaviEyer_etal09}.
Preliminary explorations of the data have been published in
\cite{CherixCarrierBurki_etal06} for NGC~1901,
\cite{CarrierSaesenCherix09} for NGC~5617, and 
\cite{GrecoMowlaviEyer_etal09,GrecoMowlaviEyer_etal10} for IC~4651.

\begin{table}
\centering
\caption{NGC~3766}
\begin{tabular}{l l l}
\hline
Parameter            & Value                                      & Ref.\\
\hline
($\alpha_{2000}$, $\delta_{2000}$) & (\raHour{11}{36}{14}, \decDeg{61}{36}{30}) & \\
$(l, b)$             & $(294.117\deg, -0.03\deg)$                 & \\
Angular size         & 9.24' ($V<17$~mag)                         & 1\\
Minimum mass content & \mass{793}                                 & 1\\
Distance             & 1.9 to 2.3~kpc                             & 2\\
Distance modulus $(V-M_V)_0$ & $11.6 \pm 0.2$                     & 2, 3\\
Age                  & 14.5 to 25, 31~Myr                         & 2, 3\\
Reddening $E(B-V)$   & $0.22 \pm 0.03$~mag                        & 2, 3\\
\hline
\end{tabular}
{\textbf{References.}
(1) \cite{MoitinhoAlfaroYun_etal97}; (2) \citet{McSwainHuangGies_etal08}; (3) \cite{AidelmanCidaleZorec_etal12}.
}
\label{Tab:ngc3766}
\end{table}

In this paper, we present an extensive analysis of the periodic variables in NGC~3766.
It is the first paper in a series analyzing in more details the variability content of the clusters in our survey.
The characteristics of NGC~3766 are summarized in Table~\ref{Tab:ngc3766}.
Among the first authors to have published photometric data for this cluster, we can cite \cite{Ahmed62} for $UBV$ photometry, \cite{Yilmaz76} for RGU photometry, \cite{Shobbrook85,Shobbrook87} for $uvby \beta$ photometry, \cite{MoitinhoAlfaroYun_etal97} for Charge Coupled Device (CCD) $UBV$ photometry and \cite{PiattiClariaBica98} for CCD $VI$ photometry.
Thanks to the relative closeness of the cluster, lying at $\sim$2~kpc from the Sun, the upper main sequence in the HR diagram stands rather clearly out from field stars, making the identification of cluster members easier.
With an age of about 20~Myr, the cluster is very rich in B stars, and a good fraction of them have been identified to be Be stars \citep{McSwainHuangGies_etal08, McSwain08, AidelmanCidaleZorec_etal12}.
\toReferee{
Strangely, almost no information about the metallicity of the cluster exists in the literature.
Authors studying this cluster usually assume solar metallicity \citep[e.g.,][]{MoitinhoAlfaroYun_etal97, McSwainHuangGies_etal08}.
The only metallicity determination available in the literature to date, to our knowledge, is [Fe/H]=-0.47 proposed by \cite{Tadross03} and repeated by \cite{PaunzenHeiterNetopil_etal10}, based on the mean ultraviolet excess $\delta(U-B)$ of bright dwarfs from the photometry of \cite{MoitinhoAlfaroYun_etal97}.
The reliability of the estimation cannot be easily established, though (membership of the dwarfs, uncertainty of the metallicity estimate), and we therefore do not report this value in Table~\ref{Tab:ngc3766}.
}
No blue straggler is known in this cluster, the only candidate reported in the literature \citep{Mermilliod82} having since then been re-classified as a Be star \citep{McSwainHuangGies09}.
\toReferee{Finally, it must be mentioned that} no in depth study of periodic variable stars (which we simply call \textit{periodic variables} in the rest of this paper) has been published so far for this cluster.

We describe our data reduction and time series extraction procedures in Sect.~\ref{Sect:observations}, where we also present the overall photometric properties of the cluster.
Section~\ref{Sect:variabilityAnalysis} describes the variability analysis procedure, the results of which are presented in Sect.~\ref{Sect:EBs} for eclipsing binaries, and in Sect.~\ref{Sect:periodicVariables} for other periodic variables.
Various properties of the periodic variables are discussed in Sect.~\ref{Sect:discussion_periodics}.
The nature of the new class of MS periodic variables in the variability `gap' in between the regions of $\delta$~Sct and SPB stars in the HR diagram is addressed in Sect.~\ref{Sect:discussion_group2}.
Conclusions are finally drawn in Sect.~\ref{Sect:conclusions}.

A list of the periodic variables and of their properties is given in Appendices~\ref{Appendix:EBs} and \ref{Appendix:periodicVariables}, together with their light curves.

\section{Observations and data reduction}
\label{Sect:observations}

\begin{figure}
  \centering
  \includegraphics[width=1.0\columnwidth]{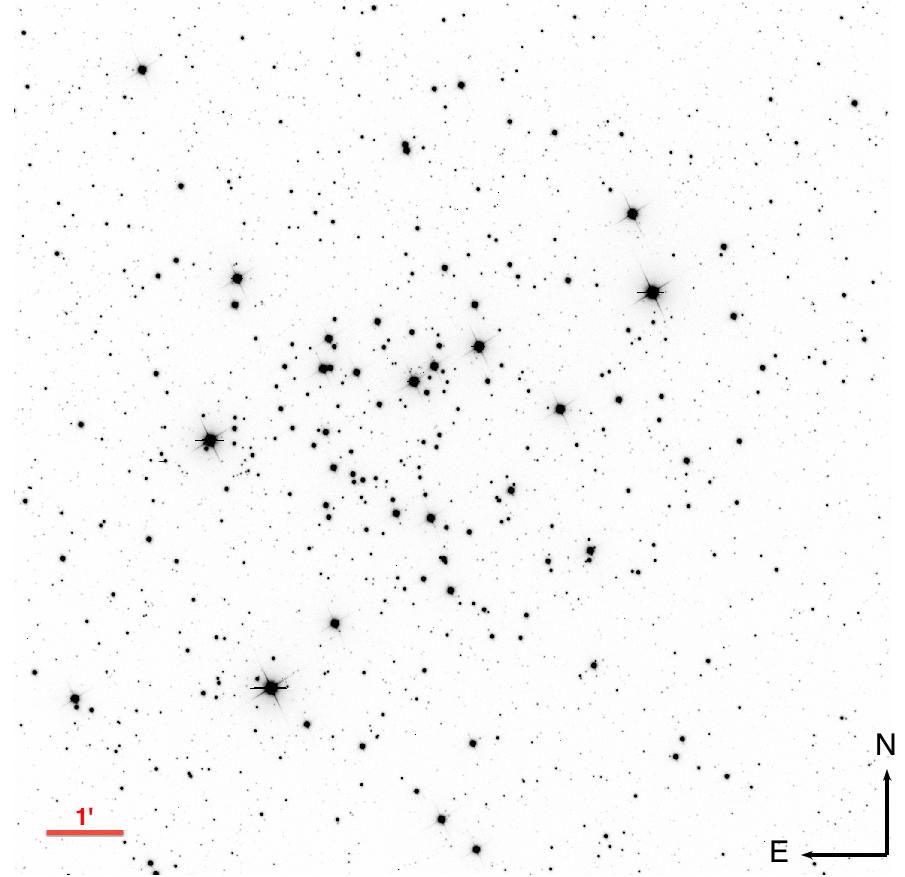}
  \caption{Reference image of NGC~3766 used in the data reduction process.
  }
\label{Fig:referenceImage}
\end{figure}

\begin{figure}
  \centering
  \includegraphics[width=1.0\columnwidth]{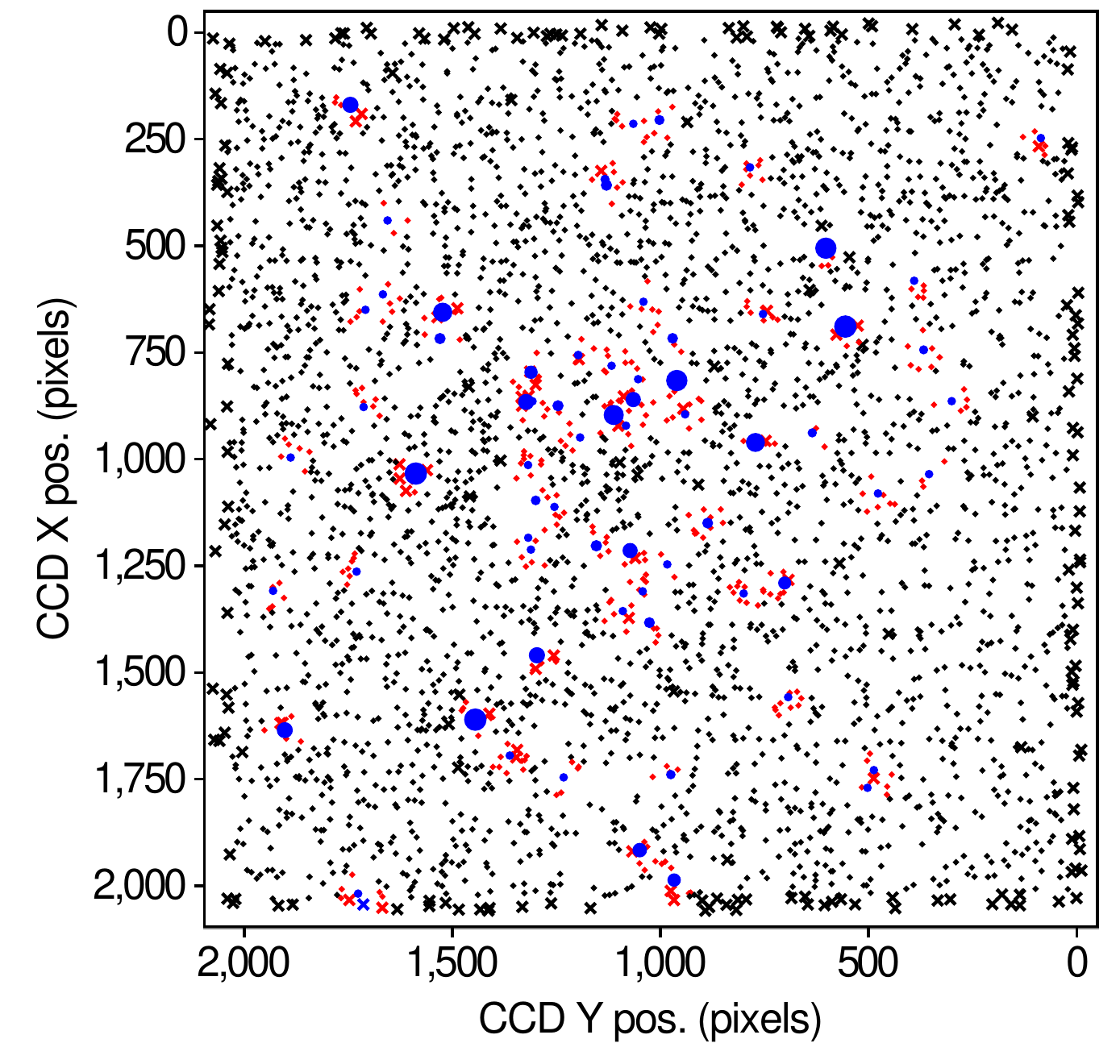}
  \caption{Location on the CCD of all stars with more than 100 good points in their light curve.
           The \toReferee{seventy} brightest stars are plotted in blue, with the size of the marker proportional to the brightness of the star.
           Stars that lie closer than 50 pixels on the CCD from one of those bright stars are plotted in red.
           Crosses identify stars that have more than 20\% bad points in their light curves (see Sect.~\ref{Sect:lcSelection}).
  }
\label{Fig:ccdAllStars}
\end{figure}

Observations of NGC~3766 have been performed from 2002 to 2009 on the 1.2~m Swiss Euler telescope at La Silla, Chile.
CCD images have been taken in the Geneva $V$, $B$ and $U$ photometric bands, centered on $\alpha$~=~\raHour{11}{36}{14} and $\delta$~=~-\decDeg{61}{36}{30} with a field of view of $11.5' \times 11.5'$ (Fig.~\ref{Fig:referenceImage}).
Those bands are not strictly identical to the standard Geneva system, which were defined in the era of photomultipliers.
While the filter+CCD responses have been adjusted to reproduce as closely as possible the standard Geneva system, they are not identical.
A calibration would thus be necessary before using the photometry to characterize stars.
We however do not perform such a (delicate) calibration in this study, and solely focus on the variability analysis of the time series, where differential photometric quantities are sufficient.
In order to keep in mind the uncalibrated nature of the time series, we hereafter note by $V'$, $B'$ and $U'$ the reduced CCD measurements in the Geneva $V$, $B$ and $U$ filters, respectively.

The program was part of a wider observational campaign of 27 open clusters, 12 in the Southern and 15 in the Northern hemispheres \citep[][Saesen et al. in preparation]{GrecoMowlaviEyer_etal09}.
Several observation runs per year, each 10 to 15 nights long, were scheduled in La Silla for the Southern open clusters, during which all observable clusters were monitored at least once every night.
In addition, two clusters were chosen during each observation run for a denser follow-up each night.
For NGC~3766, the denser follow-ups occurred at HJD-2,450,000~=~1844 -- 1853, 2538 -- 2553 and 2923 -- 2936\footnote{
All epochs in this paper are given in days relative to HJD$_0$~=~2,452,000.
}.
In total, 2545 images were taken in $V'$, 430 in $B'$ and 376 in $U'$, with integration times chosen to minimize the number of saturated bright stars while performing an as-deep-as-possible survey of the cluster
\toReferee{(typically 25~sec in $V'$, 30~sec in $B'$ and 300~sec in $U'$, though a small fraction of the observations were made with shorter and longer exposure times as well).
}
As a result, stars brighter than $V'\simeq 11$~mag were saturated.

All images were calibrated, stars therein identified and their light curves obtained with the procedure described in Sect.~\ref{Sect:dataReduction}.
Stars with potentially unreliable photometric measurements were flagged and disregarded, and the light curves of the remaining stars cleaned from bad points (Sect.~\ref{Sect:lcCleaning}).
Good light curves in $V'$, $B'$ and $U'$ are finally selected from the set of cleaned light curves based on the number of their good and bad points (Sect.~\ref{Sect:lcSelection}).
They form the basis for the variability analysis of the paper.
A summary of our data in the color-magnitude and color-color diagrams is given in Sect.~\ref{Sect:summaryDiagrams}.

\subsection{Data reduction}
\label{Sect:dataReduction}

The data reduction process is presented in \cite{SaesenCarrierPigulski_etal10}, and we summarize here only the main steps.
First, the raw science images were calibrated in a standard way, by removing the bias level using the overscan data, correcting for the shutter effects, and flat fielding using master flat fields.
Then, \textsc{daophot~II} \citep{Stetson87} and \textsc{allstar} \citep{StetsonHarris88} were used to extract the raw light curves.
We started with the construction of an extensive master star list, containing 3547~stars, which was converted to each frame taking a small shift and rotation into account.
The magnitudes of the stars in each frame were calculated with an iterative procedure using a combined aperture and point spread function (PSF) photometry.
Multi-differential photometric magnitudes and error estimates were derived, using about 40~reference stars, to reduce the effects of atmospheric extinction and other non-stellar noise sources.
The light curves were finally de-trended with the Sys-Rem algorithm \citep{TamuzMazehZucker05} to remove residual trends.

\subsection{Star selection and light curve cleaning}
\label{Sect:lcCleaning}

\begin{figure}
  \centering
  \includegraphics[width=1.0\columnwidth]{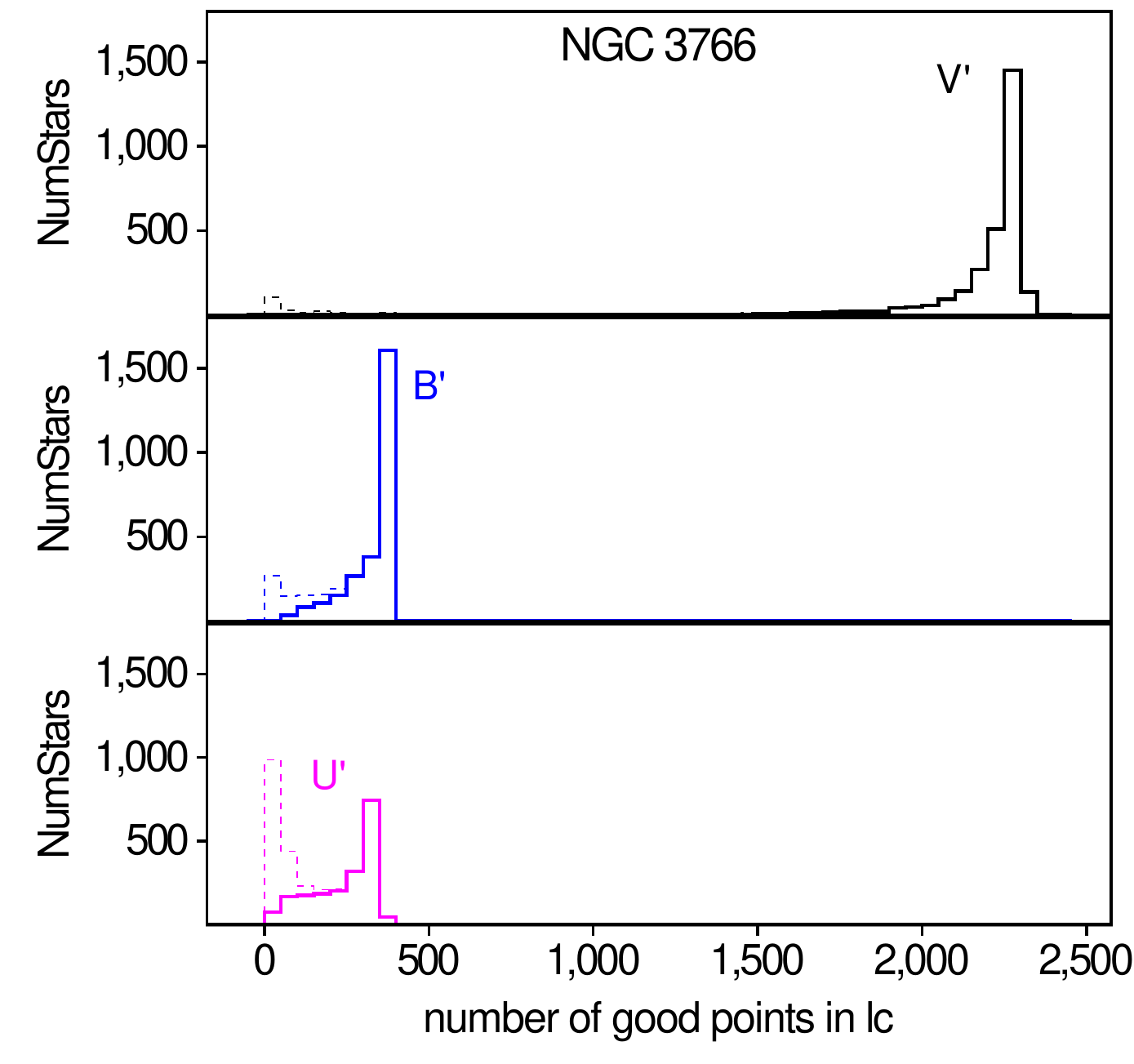}
  \caption{Histogram of the number of good points in the $V'$ (top panel), $B'$ (middle panel) and $U'$ (bottom panel) time series.
   Dashed lines give the histograms considering all stars (see Sect.~\ref{Sect:lcCleaning}).
   Continuous lines give the histograms of selected light curves (see Sect.~\ref{Sect:lcSelection}).
  }
\label{Fig:histoNumGoodPoints}
\end{figure}

Three procedures are used to disregard potentially problematic stars and wrong flux estimates.
The first procedure tackles stars polluted by neighboring bright stars,
the second deals with stars whose flux distributions are badly determined on individual images, and the third cleans the light curves from their outliers.

\paragraph{Stars contaminated by bright stars:}

The analysis of individual light curves reveals a pollution of stars located close to bright stars (we consider here the `bright' stars to be the --somehow arbitrarily defined-- seventy brightest stars, the other ones being `faint' stars).
This is especially true if the bright star is saturated.
We then disregard all faint stars that lie closer than 50 pixels to one of the bright stars.
Their distribution on the CCD is shown in Fig.~\ref{Fig:ccdAllStars}.

\paragraph{Stars offset from expected position in selected images:}

The flux of a star measured in a given image can be unreliable if the position of the star on the CCD is not accurate enough, e.g. due to a wrong convergence in the iterative PSF fitting.
We therefore disregard a flux measure if the position offset of a certain star in a certain image relative to the reference image deviates more than $5 \sigma$ sigma from the position offset distribution of all stars in that image.
The procedure is iterated five times in each image, removing each time from the list of all stars the ones with large position offsets.
We also remove measurements of stars in a given image that fall closer than eleven pixels to one of the borders of the CCD.

\paragraph{Light curve cleaning:}

In addition to the two cleaning steps described above that involve the position of the stars in the images, we remove all points in a given light curve that have a magnitude error larger than 0.5~mag or that are at more than 3 sigmas above the mean error.
We also disregard measurements if the magnitude deviates more than 5 sigmas from the mean magnitude.
This last step, classically called sigma clipping on the magnitudes, is iterated twice.
We however keep the points if at least four of them would be disregarded in a row, in order to avoid removing good points in eclipsing binaries.

The light curve cleaning is applied to each $V'$, $B'$ and $U'$ time series.
The histograms of the resulting number of good points for the three photometric bands are shown in dashed lines in Fig.~\ref{Fig:histoNumGoodPoints}.

\subsection{Light curve selection}
\label{Sect:lcSelection}

\begin{figure}
  \centering
  \includegraphics[width=1.0\columnwidth]{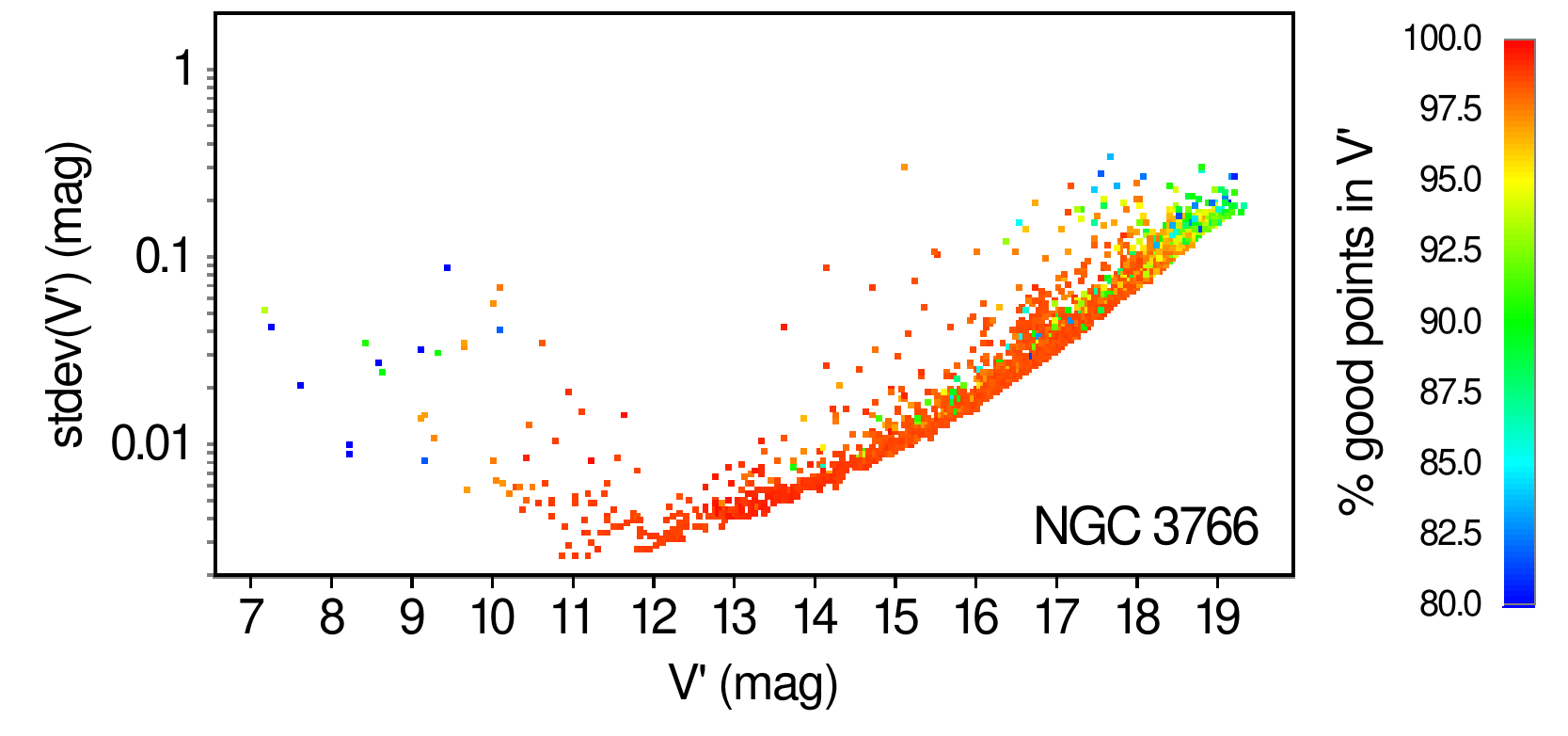}
  \includegraphics[width=1.0\columnwidth]{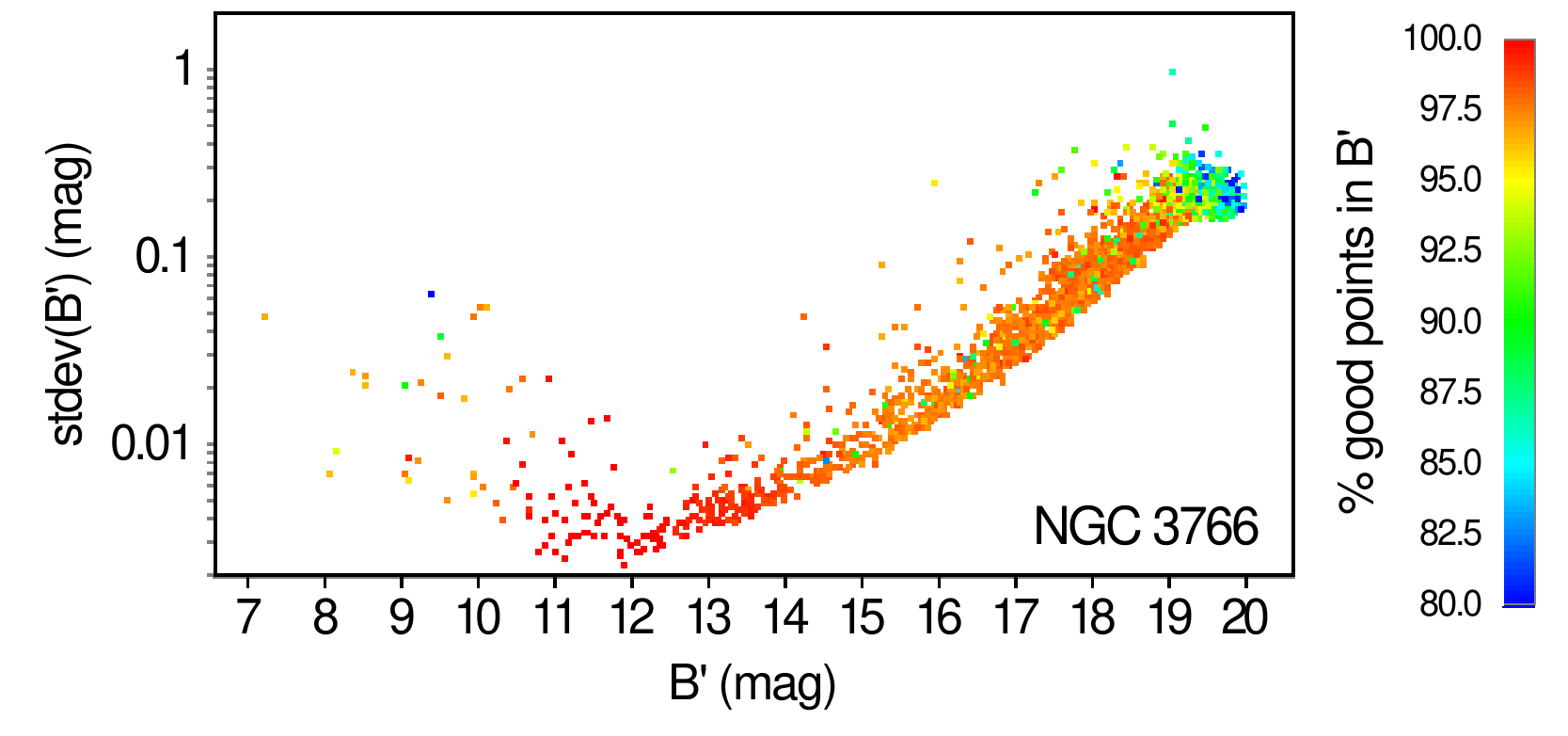}
  \includegraphics[width=1.0\columnwidth]{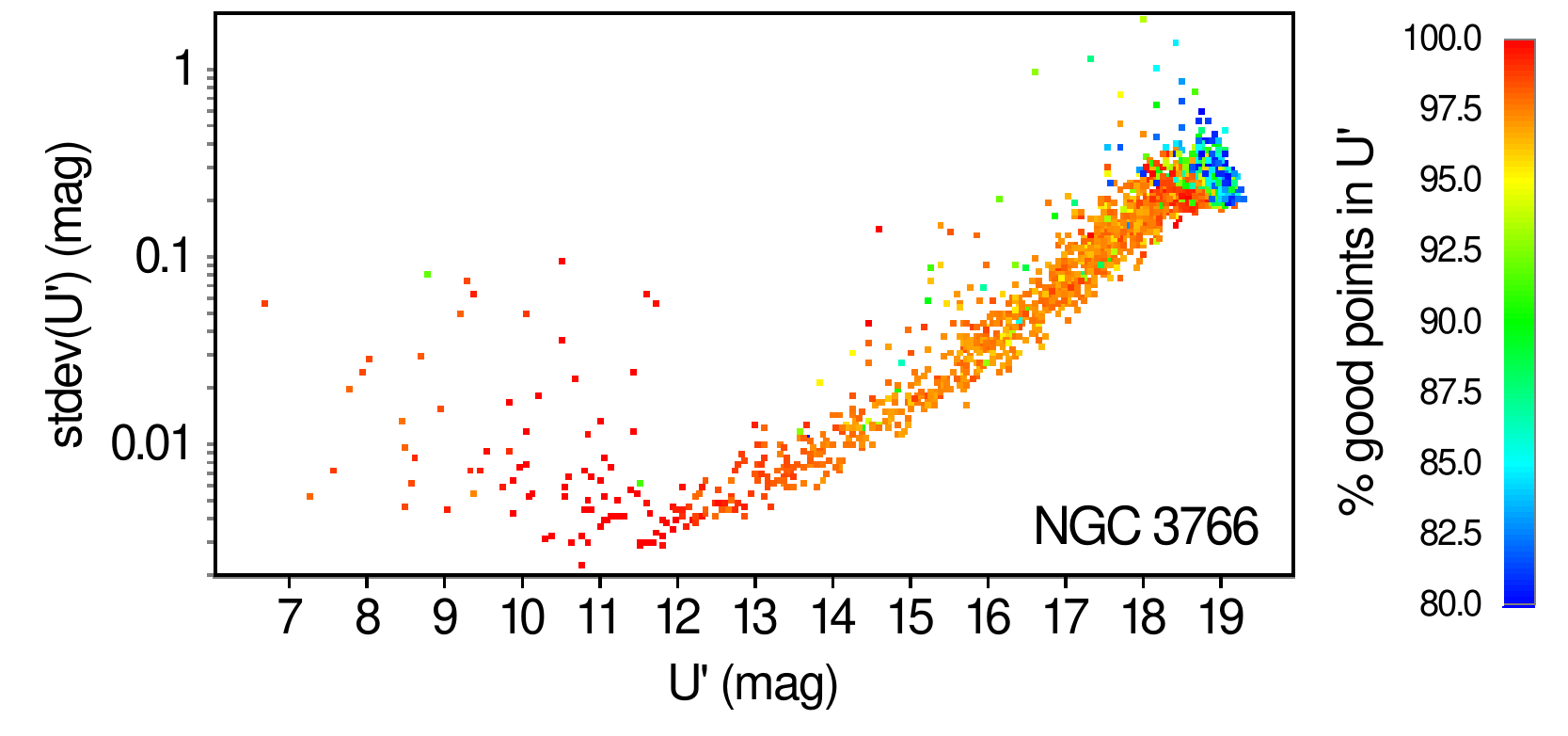}
  \caption{Standard deviations of the time series of good points as a function of magnitudes in the $V'$ (top), $B'$ (middle) and $U'$ (bottom) bands.
  The percentage of good points in the original light curves is shown in color according to the color scales displayed on the right.
  }
\label{Fig:sigmas}
\end{figure}

The light curve cleaning procedure described in Sect.~\ref{Sect:lcCleaning} removes a certain number of bad points in each $V'$, $B'$ and $U'$ time series.
The number of points disregarded in this way is actually a relevant indication of the quality of the time series of a given star.
The more points disregarded, the more problematic the flux computation may be in the images, whether it be due to a position issue on the CCD (closeness to a bright source for example), CCD issues (bad pixels for example), or data reduction difficulties (in a crowded region for example).
We therefore disregard all time series that contain more than 20\% of bad points if the star is fainter than 10.5~mag.
Their distribution on the CCD is shown in Fig.~\ref{Fig:ccdAllStars}.
We do not apply this filter to brighter stars because they may have a larger number of bad points due to saturation while the remaining good points may still be relevant.
The histograms of the number of good points for all good time series are shown in solid lines in Fig.~\ref{Fig:histoNumGoodPoints}.
The tails of the histograms at the side of small numbers of points are seen to be significantly reduced, especially in $U'$.
In total, there are 2858 good times series in $V'$, 2637 in $B'$ and 1907 in $U'$.
The $U'$ time series are expected to provide the least reliable measurements.
This however does not affect our variability study, which mainly relies on the $V'$ time series.

\label{Sect:lcPrecisions}

\begin{figure}
  \centering
  \includegraphics[width=1.0\columnwidth]{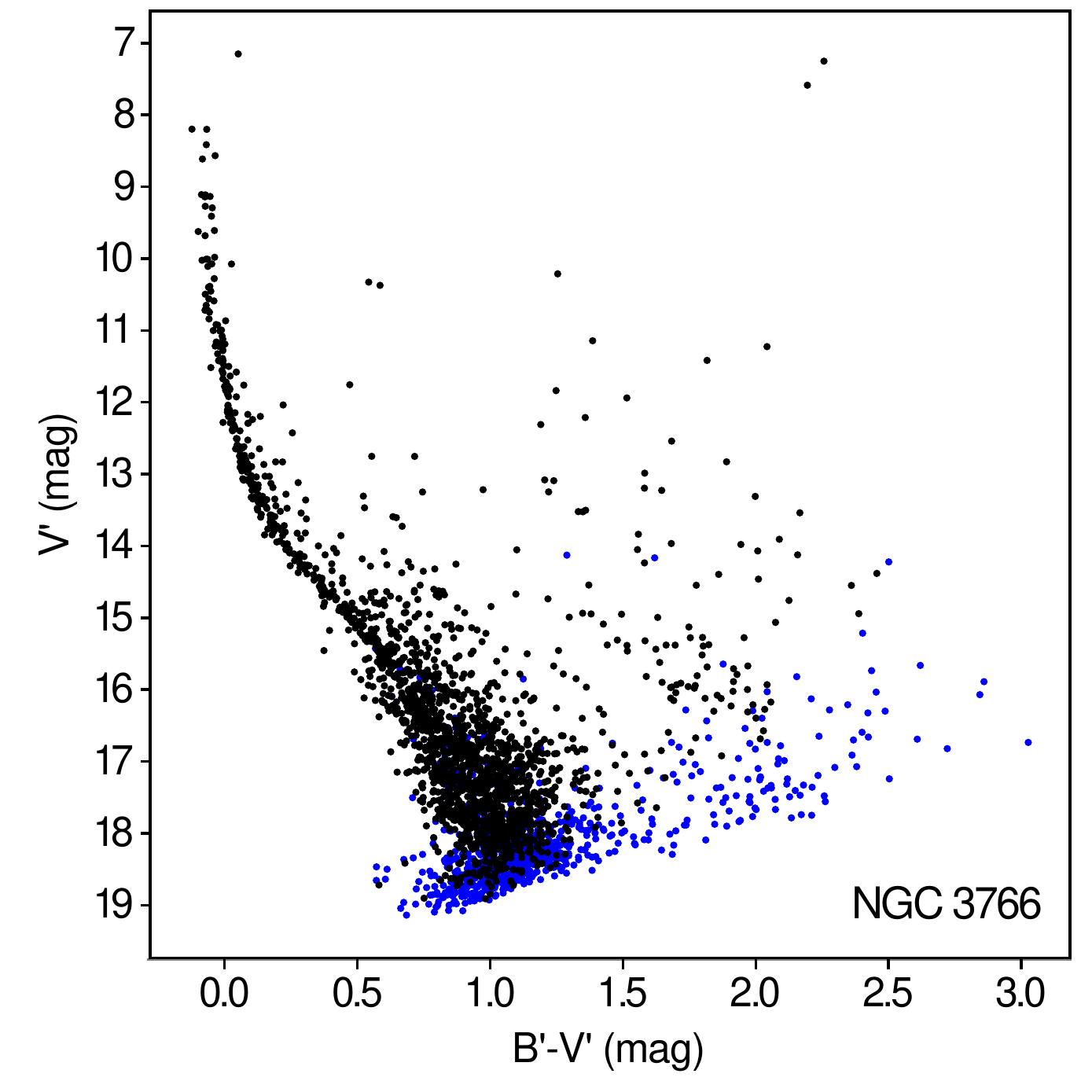}
  \caption{Color-magnitude diagram ($B'-V'$, $V'$) of stars with good light curves in both $V'$ and $B'$.
  Black points have good $U'$ time series as well, blue points don't.
  }
\label{Fig:cmGoodLcs}
\end{figure}

The photometric precisions reached in $V'$, $B'$ and $U'$ can be estimated from the standard deviations $\sigma$ of the time series as a function of magnitude.
These are shown in Fig.~\ref{Fig:sigmas} for the three photometric bands.
The lower envelopes in those diagrams give the best precisions obtained from our data at any given magnitude.
We call them the \textsl{envelopes of constant stars} (ECSs), as they identify the location expected for constant stars in those diagrams.

A precision better than 3~mmag is reached around $V'=11-12$~mag, as seen in Fig.~\ref{Fig:sigmas} (we note that the precision reached in \textit{amplitude} detection of periodic variables reaches 1~mmag, see Sect.~\ref{Sect:periodicVariables}).
For brighter stars, the standard deviation increases with increasing brightness, due to both photometric saturation and intrinsic variability.
Stars fainter than $V'=12$~mag have worse precisions, as expected.
The precisions are about 10, 30 and 60~mmag at $V'=15.2$, 17 and 18~mag, respectively.
They are about similar in $B'$, though the dispersion in $\sigma$ of the ECS at a given magnitude is larger in $B'$ than in $V'$.
The $U'$ time series have the least precise photometry, though still good for bright stars, with a precision of $\sim$3~mmag at $U'=11-12$~mag.
It is less good for faint stars, reaching a precision of 100~mmag at $U'=18$~mag.
We can thus expect to have reliable $B'-V'$ colors, but much less reliable $U'-B'$ colors for fainter stars.
We have also to keep in mind that there are much fewer good time series in $U'$ than in $V'$ and $B'$.

\subsection{Summary diagrams}
\label{Sect:summaryDiagrams}

The color-magnitude diagram ($B'-V'$, $V'$) is shown in Fig.~\ref{Fig:cmGoodLcs}.
Stars which have good light curves in all three bands are shown in black, those with only good $V'$ and $B'$ light curves in blue.
The upper MS is rather clear and well detached from background field stars due to the relative closeness of the cluster to us.
\toReferee{Based on the density of stars on the red side of the MS in the diagram, we may expect} field stars to \toReferee{start contaminating} the MS \toReferee{mainly at magnitudes} above $V'\simeq 15$~mag.
\toReferee{
Of course, only radial velocities, proper motions and good parallaxes will be able to distinguish members from field stars.
The future Gaia mission of the European Space Agency, to be launched end of 2013, will be a key contributor in this respect.
}

\begin{figure}
  \centering
  \includegraphics[width=1.0\columnwidth]{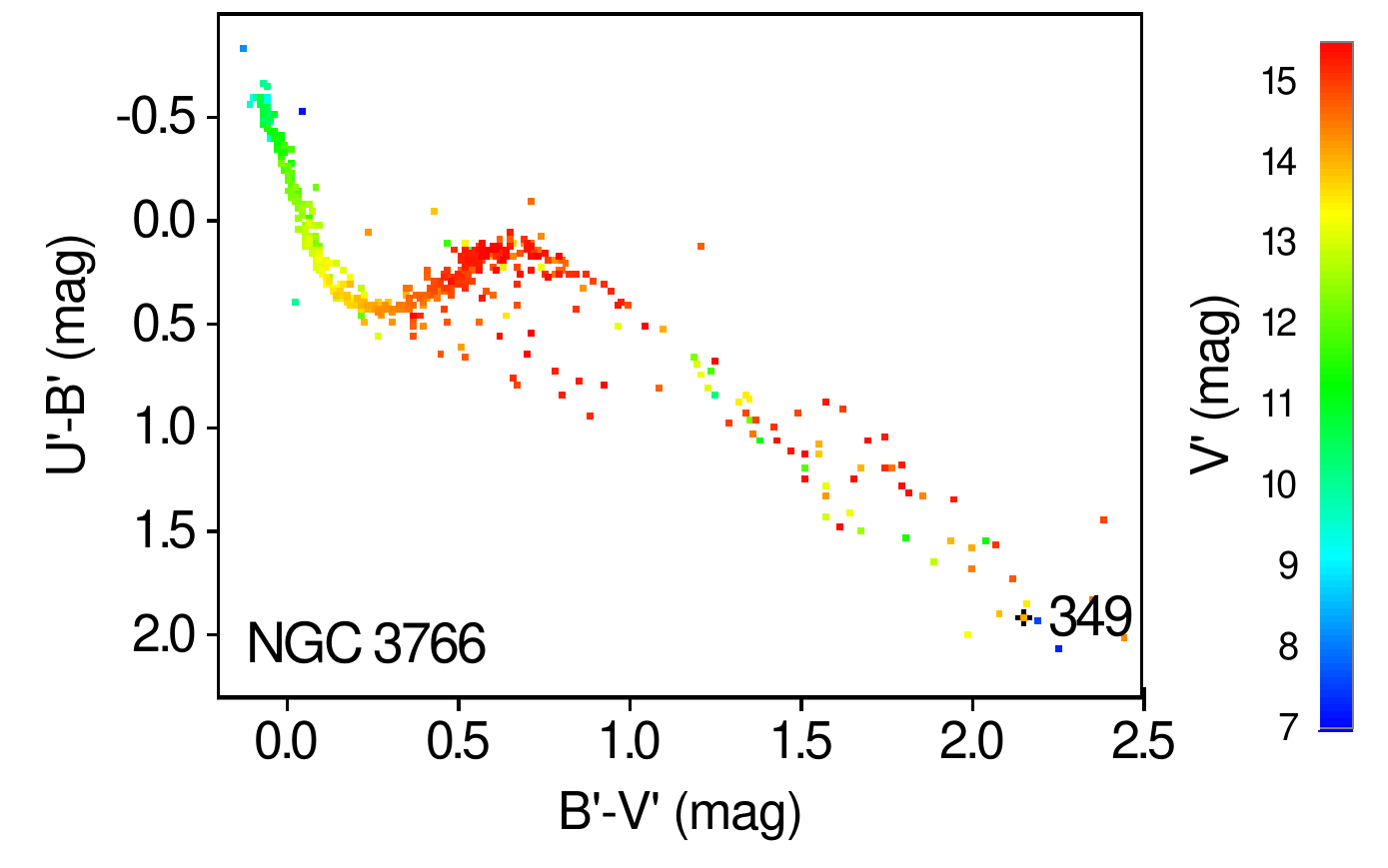}
  \caption{Color-color diagram ($B'-V'$, $U'-B'$) of stars brighter than $V'=15.5$~mag and with good light curves in $V'$, $B'$ and $U'$. 
  The $V'$ magnitude is color coded according to the color scale on the right.
  }
\label{Fig:ccGoodLcs1}
\end{figure}

The color-color diagram ($B'-V'$, $U'-B'$) constructed with all stars brighter than $V'=15.5$~mag is shown in Fig.~\ref{Fig:ccGoodLcs1}.
The S shape (rotated 45$^\mathrm{o}$ anti-clockwise) characteristic of that diagram is well visible.
Stars located below the S shape at 0.3~mag $\lesssim$ $B'-V'$ $\lesssim$ 1~mag and 0.5~mag $\lesssim$ $U'-B'$ $\lesssim$ 1~mag are expected to be field stars behind the cluster that are more reddened by interstellar extinction (reddening moves a point to the lower-right direction in the diagram).
This statement is supported by the thinness of the sequence of B stars in the diagram, which indicates no important differential reddening for cluster members.

\section{Periodic variability analysis}
\label{Sect:variabilityAnalysis}

Variability detection and characterization is based on $V'$ time series only, which are much better sampled than the $B'$ and $U'$ ones and with better precisions.
The identification of periodic variable candidates is performed in two ways. 
The method is adapted and simplified from the one described in Section~6 of \cite{SaesenCarrierPigulski_etal10}, where it was used to find variable stars in data from a multisite campaign on a cluster.

In a first step, we calculate the generalized Lomb-Scargle diagrams \citep{ZechmeisterKurster09}, weighted with the inverse square of each measurement error, for each of the stars from 0~to 50~d$^{-1}$.
We pick stars that have at least one significant peak, i.e. with a signal-to-noise ratio (S/N) above 4.5, and for which the selected frequency does not occur in several stars (at least 3), pointing to unwanted instrumental effects.
We calculate the S/N as the ratio of the amplitude of a frequency to the noise level, where the noise is computed as the mean amplitude around this frequency.
The interval over which we evaluate the average changes according to the frequency value.
We use an interval of 1~d$^{-1}$ for $f \in$ [0--3]~d$^{-1}$, of 1.9~d$^{-1}$ for $f \in$ [3--6]~d$^{-1}$, of 3.9~d$^{-1}$ for $f \in$ [6--11]~d$^{-1}$ and of 5~d$^{-1}$ for $f \in$ [11--50]~d$^{-1}$.
This is done in order to account for the increasing noise at lower frequencies.
All of the retained stars are then checked manually for variability to remove remaining spurious detections (see Sect.~\ref{Sect:periodicVariables}).
The light curves of all other stars are visually examined to identify missing eclipsing binaries and other non-periodic variables.

In a second step, the stars where one genuine frequency peak was detected are submitted to an automated frequency analysis to deduce all significant frequencies till 50 d$^{-1}$ present in the light curves.
The procedure uses the same frequency search parameters as described above and is based on standard prewhitening.
When no significant peaks are present any more in the periodogram of the residuals, the frequency search is stopped.
All extracted frequencies are however not necessarily relevant or independent of each other and therefore a manual check for each star is performed to assess the validity of each frequency and to identify the non-independent frequencies, e.g., harmonic or alias frequencies.
In this way, we obtain a final frequency list for each of the periodic variables.

We exclude from the variability analysis all stars brighter than $V'=10$~mag because their light curves are affected by pixel saturation effects.
This includes all potential $\beta$~Cep stars of the cluster.
The study of those stars would require a specific light curve analysis procedure that we may perform in a future study.
The distributions of the detected periodic variables in the ($B'-V'$, $V'$), ($B'-V'$, $U'-B'$) and ($V'$, $\sigma(V')$) diagrams are shown in Figs~\ref{Fig:cmClass}, \ref{Fig:ccClass} and \ref{Fig:sigmaClass}, respectively.
The eclipsing binaries among them are presented in Sect.~\ref{Sect:EBs} and the other periodic variables analyzed in Sect.~\ref{Sect:periodicVariables}.

\section{Eclipsing binaries}
\label{Sect:EBs}

\begin{figure}
  \centering
  \includegraphics[width=1.0\columnwidth]{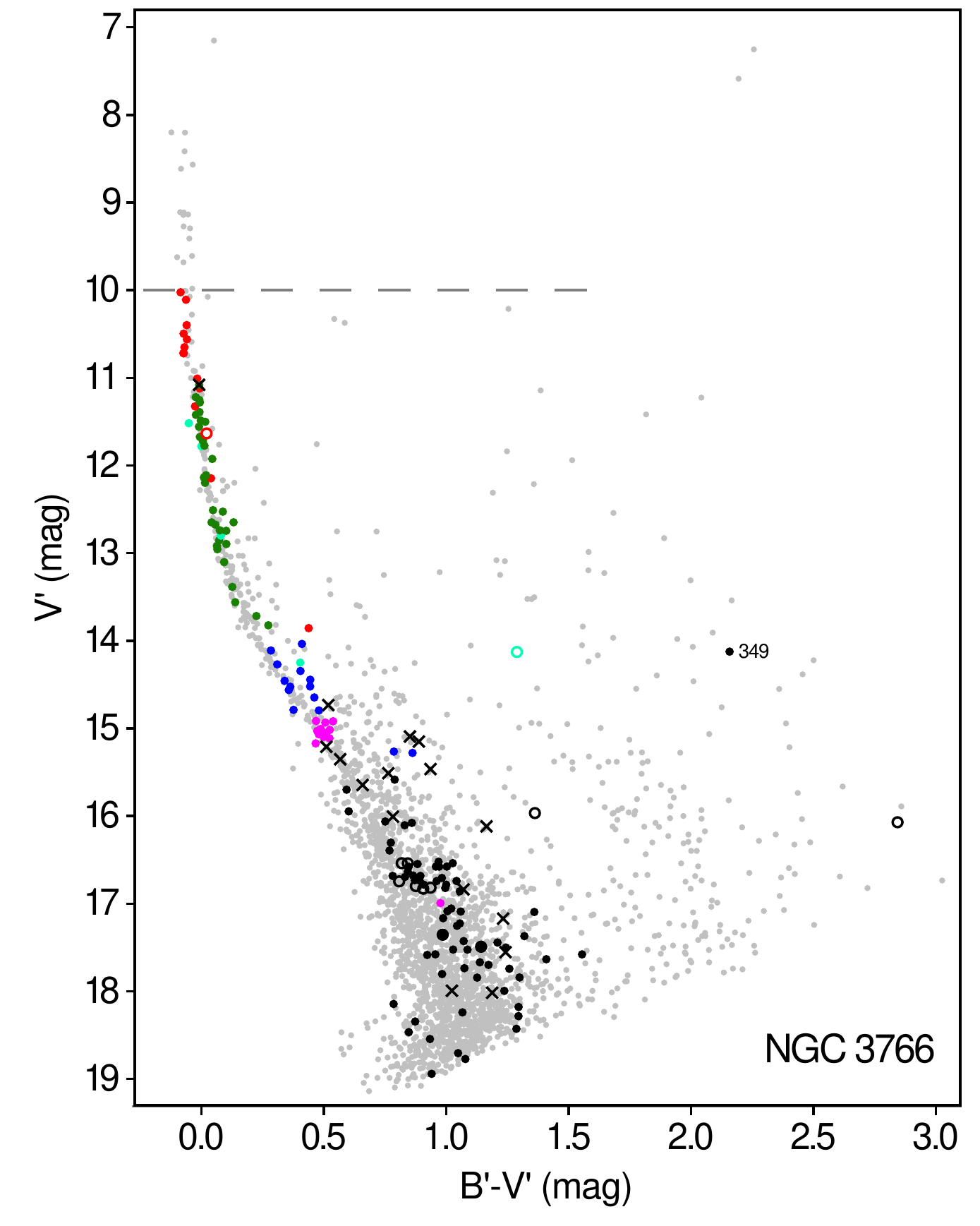}
  \caption{Distribution in the color-magnitude diagram of the periodic variables studied in this paper.
  Filled colored and black circles represent stars with secured periods, while open circles locate periodic variables whose periods need confirmation.
  Their colors identify the different groups of variable stars identified in Sect.~\ref{Sect:periodicVariables}.
  Black crosses identify eclipsing binaries.
  Light-gray points locate all other stars with good light curves in $V'$ and $B'$.
  The horizontal dashed line indicates the $V'$ magnitude above which periodic variables have been searched for.
  The location of red giant star 349 is also indicated.
  }
\label{Fig:cmClass}
\end{figure}

\begin{figure}
  \centering
  \includegraphics[width=1.0\columnwidth]{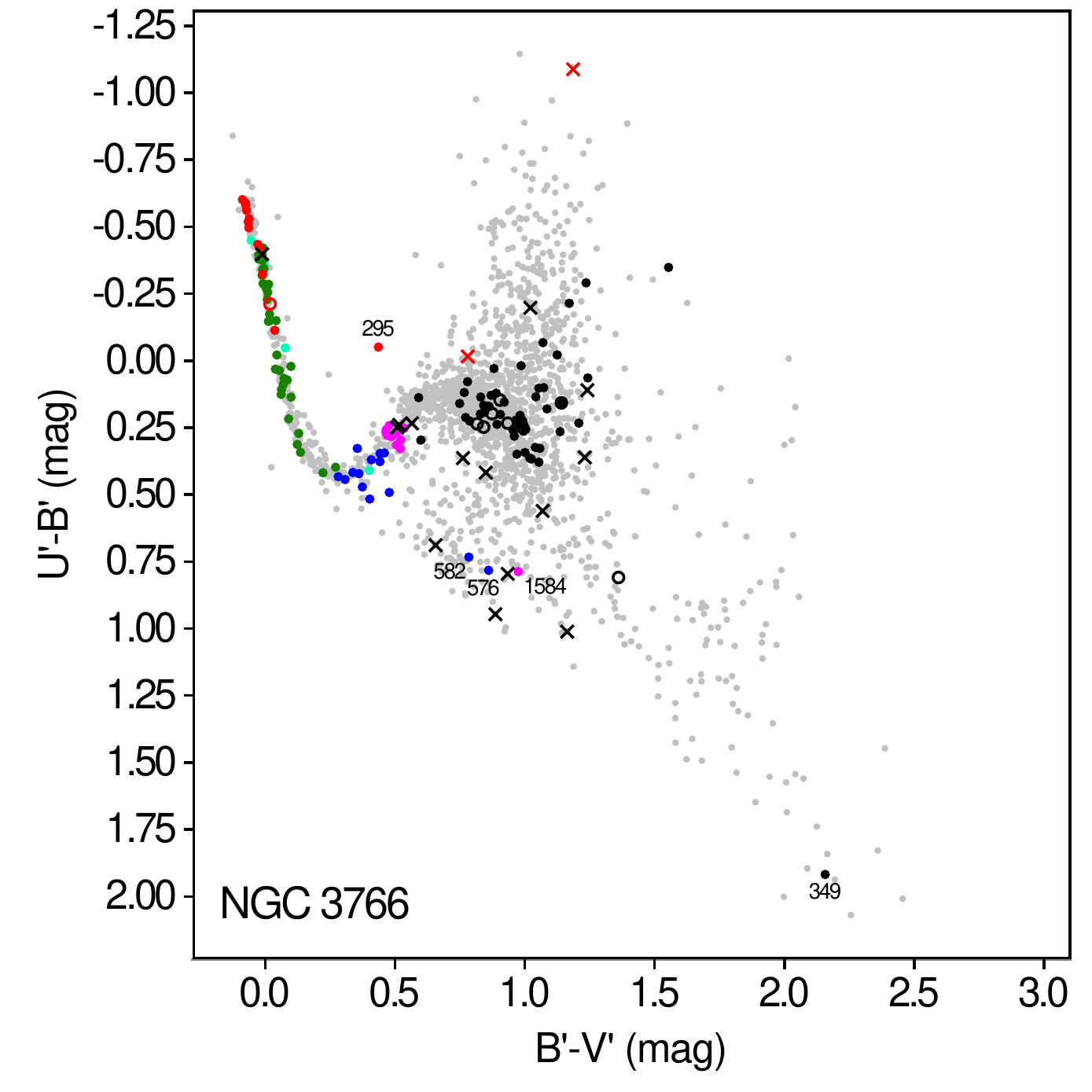}
  \caption{Color-color diagram, with symbols as in Fig.~\ref{Fig:cmClass}.
  Only stars with good light curves in $V'$, $B'$ and $U'$ are plotted, except for binary stars which are plotted in red if their $U'$ light curve is not good.
  Individual periodic variables whose variability classification are debated in Sect.~\ref{Sect:periodicVariables} are labeled with their star id next to the marker.
  }
\label{Fig:ccClass}
\end{figure}

Eclipsing binaries are of great interest in the fields of stellar formation, stellar evolution and distance determinations, to cite only a few of them.
For stellar formation studies, the analysis of populations of binary systems versus those of single stars provides information on the initial conditions in star forming regions.
For stellar evolution, precise photometric and radial velocity curves of binary stars enable derivation of the masses, radii, surface gravities and densities of both components with a high precision, and confront them with stellar evolution model predictions.
Analysis of the spectra of both components would further allow the determination of their effective temperatures, whereby their luminosities can be inferred.
The distance to the system can then be derived \cite[e.g.][]{Southworth12}.
The study of eclipsing binaries in clusters is particularly interesting in this respect since it benefits from the determination of the cluster properties such as the age, chemical composition and distance, \textit{if} their membership is ensured.
Conversely, they can also contribute to characterize the host clusters \citep[e.g.][]{GrundahlClausenHardis_etal08}.
Despite the efforts invested in this area \cite[e.g.][and references therein]{ThompsonKaluznyPych_etal01,ThompsonKaluznyRucinski10}, not many cluster eclipsing binaries have so far been analyzed with high precision, probably due to both stringent observational and data analysis requirements, such as well sampled photometric and radial velocity curves, high resolution spectra, and the need to disentangle both components in the blended spectra.
We refer for example to \cite{Yildiz11} and \cite{BrogaardVandenBergBruntt_etal12} for some relatively well characterized eclipsing binaries in open clusters in the past two years.
While it is obviously out of the scope of this article to perform such detailed analysis for the eclipsing binaries found in our data, it is worth listing them, which we do here.

\begin{figure}
  \centering
  \includegraphics[width=1.0\columnwidth]{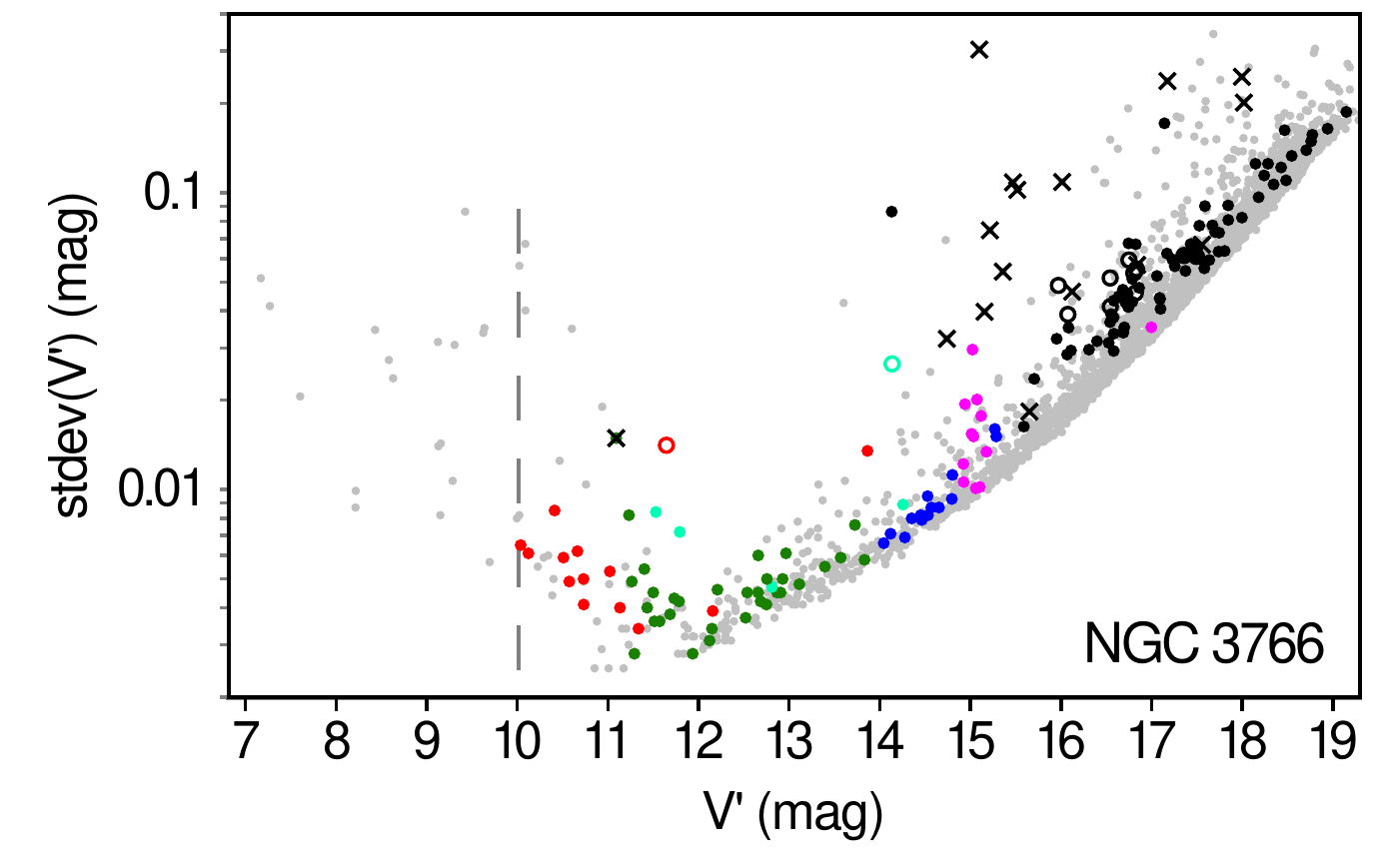}
  \caption{$(V', \sigma(V'))$ diagram, with symbols as in Fig.~\ref{Fig:cmClass}.
  The vertical dashed line indicates the $V'$ magnitude above which periodic variables have been searched for.
  }
\label{Fig:sigmaClass}
\end{figure}

Sixteen eclipsing binaries are identified in our FOV of NGC~3766, with one of them being uncertain (eclipsing binary 693 in our identification numbering scheme).
They are marked by crosses in Figs~\ref{Fig:cmClass} to \ref{Fig:sigmaClass}.
All of them are new discoveries.
From the morphology of their folded light curves, nine are classified as Algol-type (EA), four as $\beta$~Lyrae-type (EB) and three as W Ursae Majoris-type (EW) binary candidates.
Their characteristics are summarized in Table~\ref{Tab:EBs} of Appendix~\ref{Appendix:EBs}, and their folded light curves displayed in Fig.~\ref{Fig:foldedLcsEBs}.
The question of their membership to NGC~3766 is important if one wants to study them in combination with the cluster properties.
From their position in the color-magnitude (Fig.~\ref{Fig:cmClass}) and color-color (Fig.~\ref{Fig:ccClass}) diagrams, five of them may be cluster members, three of which are of type EA, one of type EB and one of type EW.

\section{Periodic variable stars}
\label{Sect:periodicVariables}

\begin{figure}
  \centering
  \includegraphics[width=1.0\columnwidth]{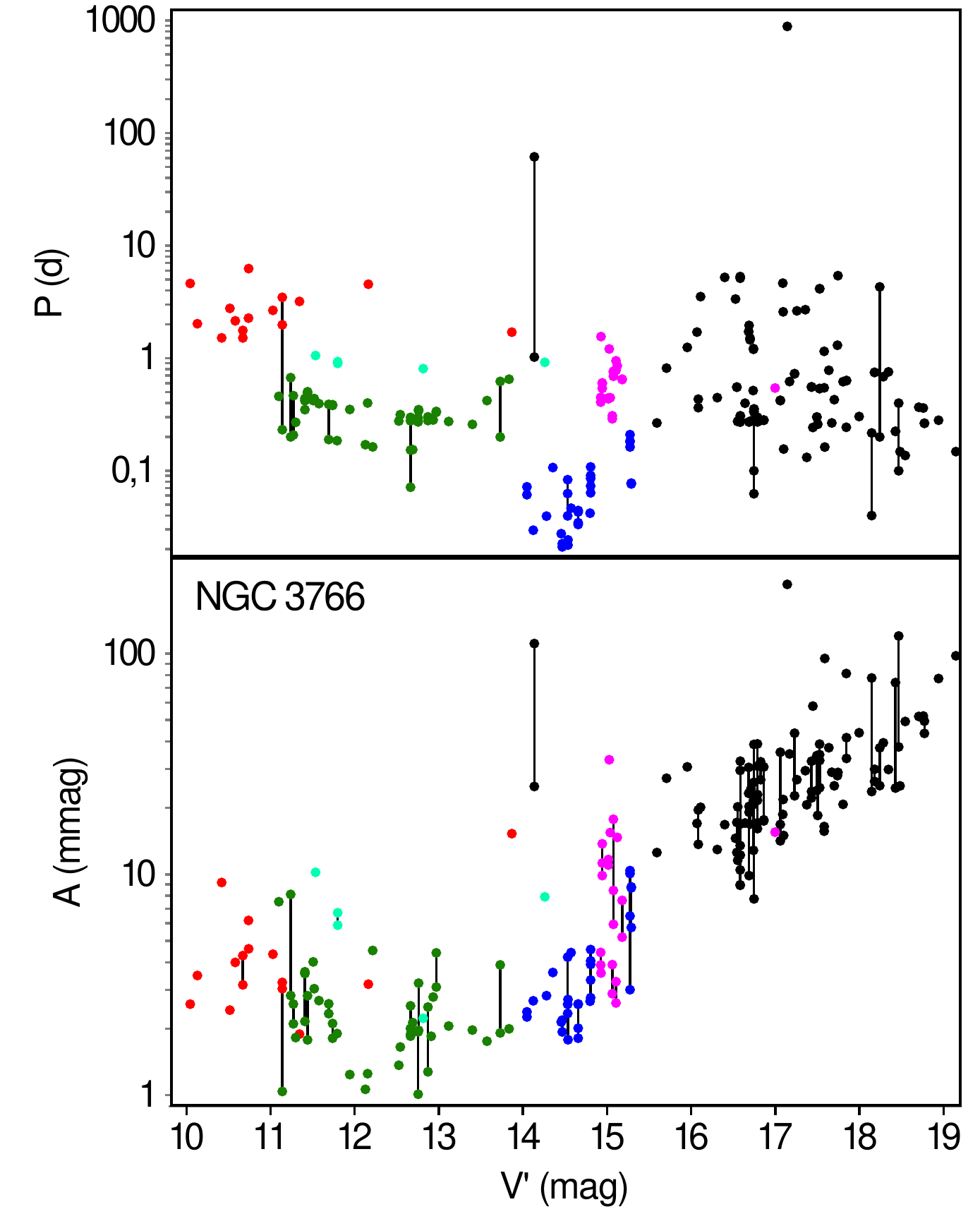}
  \caption{Periods (top panel) and associated pulsation amplitudes (bottom panel) of the periodic variables (other than purely eclipsing binaries) as a function of their $V'$ magnitude.
  Multiple periods of the same object are connected with solid lines.
  The color of each marker indicates the group to which the star belongs: red for group~1 (SPB), \toReferee{green and} cyan for group~2 (new class\toReferee{, see text}), blue for group~3 ($\delta$~Sct), magenta for group~4 ($\gamma$~Dor) and black for either red giants or periodic variables fainter than $V'=15.5$~mag, unless clearly belonging to one of groups~1 to 4 (see text).
  }
\label{Fig:VClass}
\end{figure}

\begin{figure}
  \centering
  \includegraphics[width=1.0\columnwidth]{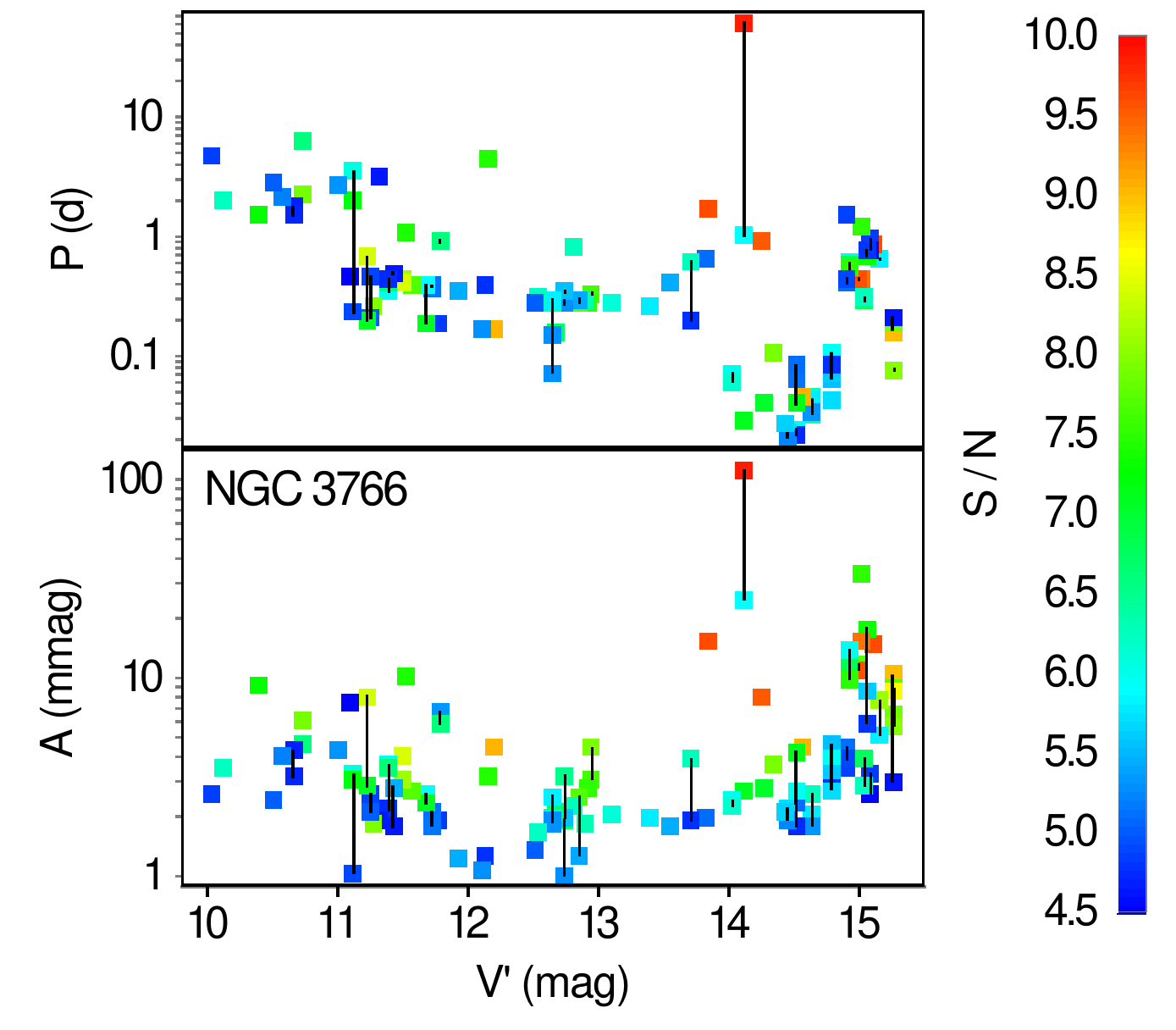}
  \caption{Periods (top panel) and associated pulsation amplitudes (bottom panel) of the periodic variables brighter than $V'=15.5$~mag, as a function of their $V'$ magnitude.
  Multiple periods of the same object are connected with solid lines (note that the period separations of multiperiodic variables are often much smaller than the periods themselves, resulting in the vertical solid lines being reduced to a black point inside the markers in the top panel).
  The color of each point indicates the S/N ratio of the peak in the periodogram corresponding to the given frequency, according to the color scale displayed on the right of the figure.
  When overlapping, the marker of the highest S/N ratio appears in the plot. 
  }
\label{Fig:VPeriodicsBright}
\end{figure}

There are 159 stars (excluding purely eclipsing binaries) for which at least one significant frequency is found in the periodogram with the procedure described in Sect.~\ref{Sect:variabilityAnalysis}.
A visual check of the frequencies and periodograms of those stars led us to consider the values of the frequencies as reliable for 147 of them. 
The remaining 12 stars have frequencies that are either \toReferee{too} close to each other for a given star \toReferee{(they differ by less than the frequency resolution $\sim$$1/T$, with $T$ being the duration of the observation campaign)}, or separated by frequency intervals reminiscent of aliases.
Their periods need confirmation, possibly based on additional observations to be performed, and are excluded from the rest of this study.
All those 12 stars, except one, are fainter than $V'=16$~mag.
They are identified by open circles in Figs~\ref{Fig:cmClass} to \ref{Fig:sigmaClass}.

Among the periodic variables with presumably secured frequencies, 85 stars are detected with one significant (i.e. with S/N$>$4.5) frequency, 45 stars with two, 13 with three, 3 with four and 1 with five independent frequencies.
The values of the periods are shown in Fig.~\ref{Fig:VClass} as a function of $V'$.
The majority of the periods are longer than 0.1~d, except for a group of stars at $V'$ between 14.0 and 14.9~mag, which have periods that can be as low as 1/2~h.
These stars with low periods are $\delta$~Sct candidates, and offer a nice way to differentiate the different groups of periodic variables.
Two groups, that we note 1 and 2, are identified at the brighter side of the $\delta$~Sct candidates, which is itself labeled group~3.
Two other groups, 4 and 5, are identified at their fainter side.
They are color-coded in Fig.~\ref{Fig:VClass} and defined according to their $V'$ magnitude and period $P$ in the following way:
\begin{itemize}
\vskip 1mm
\item \textbf{group~1:} $V'<12.5$~mag and $P>1.1$~d (red points in Fig.~\ref{Fig:VClass}).
They are two magnitudes or more brighter than the most luminous $\delta$~Sct candidate.
One fainter star at $V'=13.86$~mag is also added to this group (see Sect.~\ref{Sect:group1});
\vskip 1mm
\item \textbf{group~2:} $11<V'<14$~mag and $0.1\mathrm{~d}<P<1.1\mathrm{~d}$ (\toReferee{green and} cyan points).
They fill the magnitude range between the $\delta$~Sct and group~1 stars;
\vskip 1mm
\item  \textbf{group~3:} $14<V'<14.9$~mag and $P<0.1$~d (blue points).
It is the group of $\delta$~Sct candidates.
Two fainter stars at $V'\simeq 15.3$~mag are also added to this group (see Sect.~\ref{Sect:group3});
\vskip 1mm
\item \textbf{group~4:} $14.9<V'<15.5$~mag and $P>0.1$~d (magenta points).
One fainter star at $V' \simeq 17$~mag is also added to this group (see Sect.~\ref{Sect:group4});
\vskip 1mm
\item \textbf{group 5:} all other periodic variables (black points), i.e. those fainter than $V'=15.5$~mag and the red giants that were excluded from the other groups.
\end{itemize}

The distributions of those groups in the color-magnitude and color-color diagrams are shown in Figs~\ref{Fig:cmClass} and \ref{Fig:ccClass}, respectively.
Thanks to the closeness of NGC~3766 to us, all periodic variables brighter than $V'=15$~mag have a high probability to belong to the cluster.
This covers groups 1 to 4.
Their periods and amplitudes as a function of $V'$ are shown in more detail in Fig.~\ref{Fig:VPeriodicsBright}, with the S/N ratio of each detected frequency color coded.

The properties of the periodic variables are listed in Table~\ref{Tab:periodicVariables} of Appendix~\ref{Appendix:periodicVariables}.
In the next sections, we analyze each group in more details, starting with the most luminous one.

\subsection{Group 1 of periodic variables: SPB candidates}
\label{Sect:group1}

\begin{figure}
  \centering
  \includegraphics[width=0.8\columnwidth]{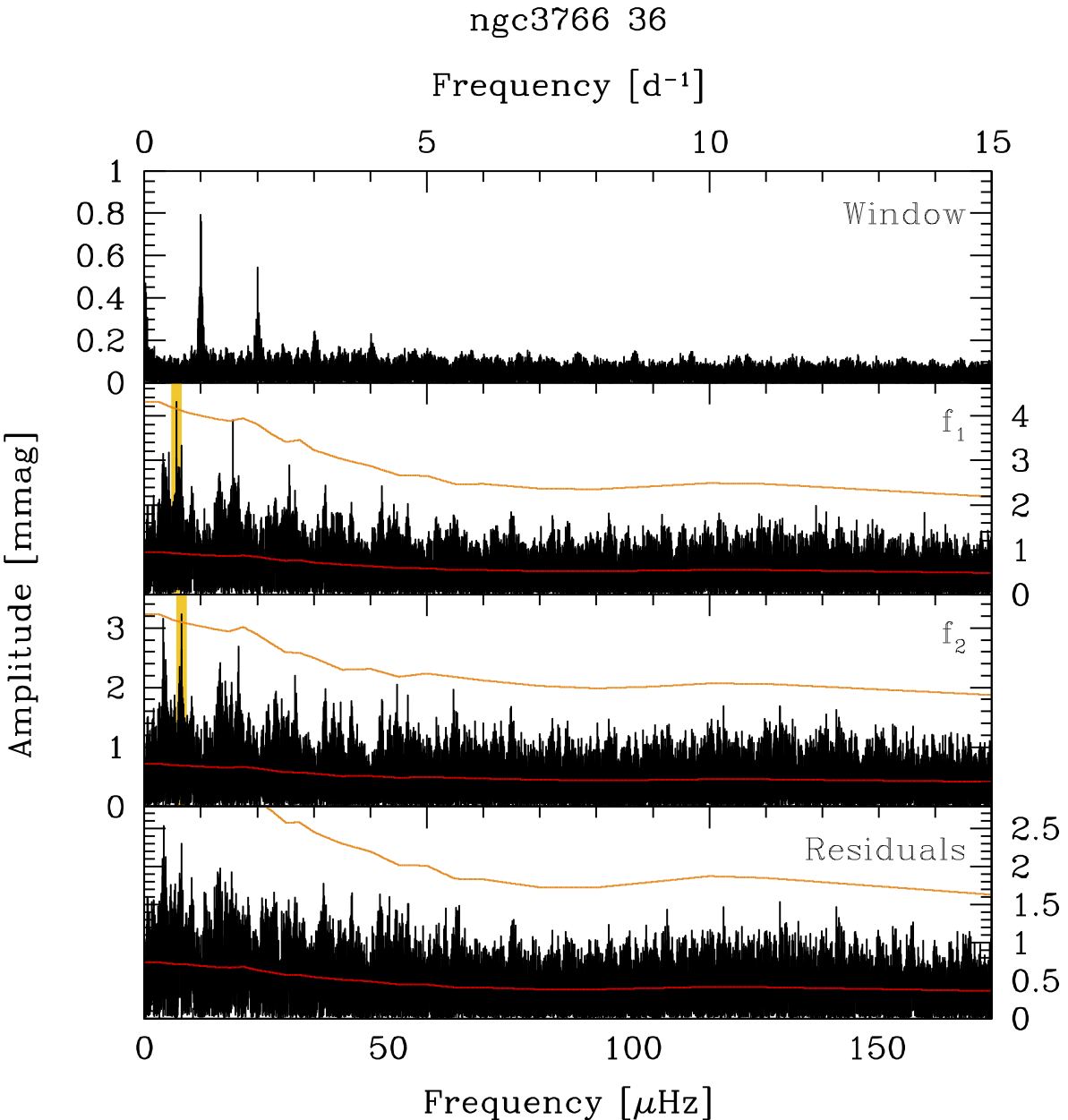}
  \caption{Successive periodograms of the multiperiodic variable 36 of group 1.
  The top panel gives the spectral window.
  The second panel gives the periodogram of the original light curve.
  The next panels give the periodograms from the subsequent pre-whitening steps.
  In each periodogram, the red line gives the background noise at 1~$\sigma$ and the orange line the limit at 4.5~$\sigma$.
  The yellow vertical bands in the second and third panels locate the dominant frequency in the respective periodogram.
  }
\label{Fig:star36Group1_periodogram}
\end{figure}

Thirteen stars belong to group~1.
Their folded light curves are shown in Figs~\ref{foldedLcsGroup1_monoperiodic} and \ref{Fig:foldedLcsGroup1_multiperiodic} of Appendix~\ref{Appendix:periodicVariables}.
All are monoperiodic, to our mmag detection limit, except two.
The two exceptions are stars 36, which has two periods, and 51, with three periods.
The successive periodograms of the biperiodic variable 36 ($P=1.7631$ and 1.5127~d) are shown in Fig.~\ref{Fig:star36Group1_periodogram} as an illustration.

All periodic variables in our field of view that are brighter than $V'=11$~mag belong to this group. Periodic variables with $V'$ between 11 and 11.3~mag, on the other hand, contain a mixture of group~1 and 2 stars.
The two groups are nevertheless distinct in the $(V',P)$ diagram (Fig.~\ref{Fig:VClass}).
Interestingly, star 51 is a hybrid star belonging to both groups~1 and 2, with two periods in group~1 ($P=3.4692$ and 1.9777~d) and one period in group~2 ($P=0.23111$~d).
We also note that star 107, while having $V'=12.15$~mag, well in the range of magnitudes expected for group 2, \toReferee{has a period of $P=4.536$~d reminiscent of group~1 stars.
We therefore assign it to this group, but keep in mind that it could, instead, be a spotted or binary star.}

Star~295 is also added to this group.
Its magnitude of $V'=13.86$~mag makes it an outlier in this respect.
Inspection of its position in the color-color diagram (Fig.~\ref{Fig:ccClass}) reveals that it is a highly reddened star located behind NGC~3766 and not belonging to the cluster.
De-reddening would bring it back, in that diagram, among the population of group~1 stars on the sequence of B stars.
Its period of 1.70109~d is compatible with this conclusion.

\toReferee{
The magnitudes of group~1 stars fall in the range expected for SPB stars.
From Fig.~3 of \cite{Pamyatnykh99}, SPB stars are predicted to have luminosities $\log(L/L_\odot) \gtrsim 1.8$.
Observed SPBs reported in this figure even start at higher luminosities.
$\delta$~Sct stars, on the other hand, are expected to have $\log(L/L_\odot) \simeq 1$ in NGC 3766 (in which they must be located at beginning of their MS phase given the age of the cluster).
SPB stars are thus expected to be at least 2~mag ($\Delta \log L = 0.8$) brighter than $\delta$~Sct stars.
The magnitudes of our group~1 stars in Fig.~\ref{Fig:VClass} agree with this statement.
}

\toReferee{The periods of group~1 stars also fall in the range expected for SPB stars.} 
They pulsate in $g$-modes ($\kappa$ mechanism on iron-group elements) with periods that are generally between 0.5 and 5~d and amplitudes of few mmags \toReferee{\citep{DeCat02}}.
This is consistent with our group~1 stars.

\subsection{Group 2 of periodic variables}
\label{Sect:group2}

\begin{figure}
  \centering
  \includegraphics[width=0.8\columnwidth]{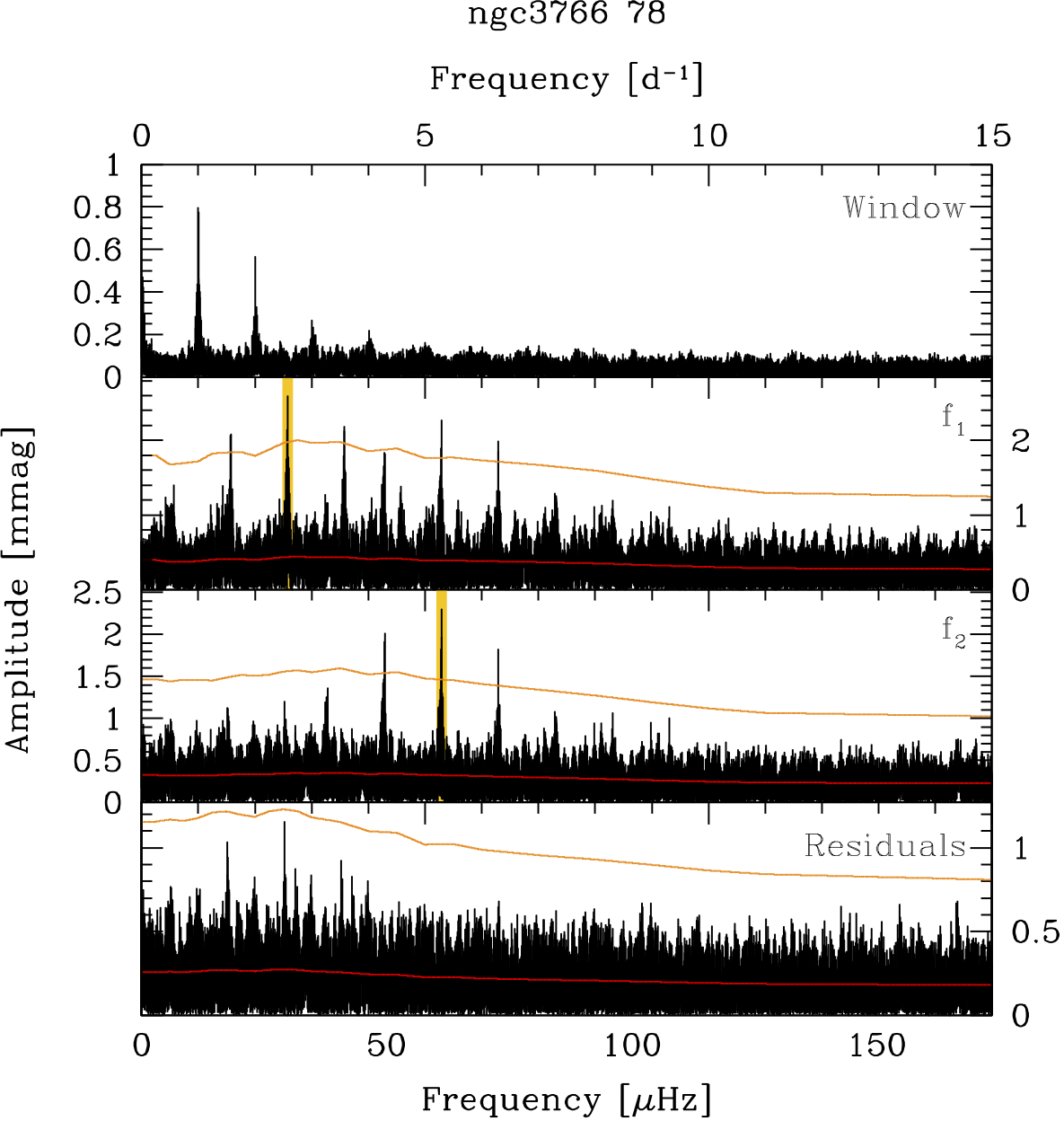}
  \caption{Same as Fig.~\ref{Fig:star36Group1_periodogram}, but for star 78 of group~2.
  }
\label{Fig:star78Group2_periodogram}
\end{figure}

The second group of periodic variables falls in the range $V'=11-14.3$~mag, with periods between $\sim$0.1 and $\sim$1.1~d \toReferee{(but see remark below)}.
Thirty-six stars belong to this group.
Twenty-three of them are monoperiodic, 11 biperiodic and 2 have three frequencies.
Their folded light curves are shown in Figs~\ref{Fig:foldedLcsGroup2_monoperiodic} to \ref{Fig:foldedLcsGroup2_triperiodic} of Appendix~\ref{Appendix:periodicVariables}.
The periodogram of the biperiodic variable 78 (with $P=0.189088$ and 0.38845~d) is shown in Fig.~\ref{Fig:star78Group2_periodogram} as an example.

Two characteristics of this group must be mentioned.
First, the group extends over more than three magnitudes, from $V'=11$ to 14.3~mag, and partly overlaps with the adjacent groups.
The overlap with group~1 was already mentioned in the previous section.
The overlap with group~3 ($\delta$~Sct candidates) occurs at $V'=14.0-14.3$~mag.
The two groups are however clearly distinct in the $(V',P)$ diagram (Fig.~\ref{Fig:VClass}), as was the case between groups~1 and 2.
We find no hybrid case of a multiperiodic variable with periods that would make it belong to groups~2 and 3 at the same time.
Second, about one third of the stars of this group (13 out of 36 members) are detected as being multiperiodic.

\toReferee{
The period distribution of group~2 stars shows a clear concentration between 0.1 and 0.7~d (see Fig.~\ref{Fig:HistoGroup2} in Sect.~\ref{Sect:group2_properties}), with only 4 stars having periods between 0.7 and 1.1~d.
We have therefore distinguished, in all figures, the bulk of group~2 stars from those four `outliers' by plotting the former ones in green and the latter ones in cyan.
}

No classical type of pulsating star is expected on the MS between SPB and $\delta$~Sct stars in the HR diagram \citep[see, e.g., Fig. 22 of][]{Christensen-Dalsgaard04}.
The origin of our group~2 stars thus requires investigation.
Their periods fall between the ones expected for the $p$-mode pulsating $\delta$~Sct and the $g$-mode pulsating SPB stars.
The case is further discussed in Sect.~\ref{Sect:discussion_group2}.

\subsection{Group 3 of periodic variables: $\delta$~Sct candidates}
\label{Sect:group3}

\begin{figure}
  \centering
  \includegraphics[width=0.8\columnwidth]{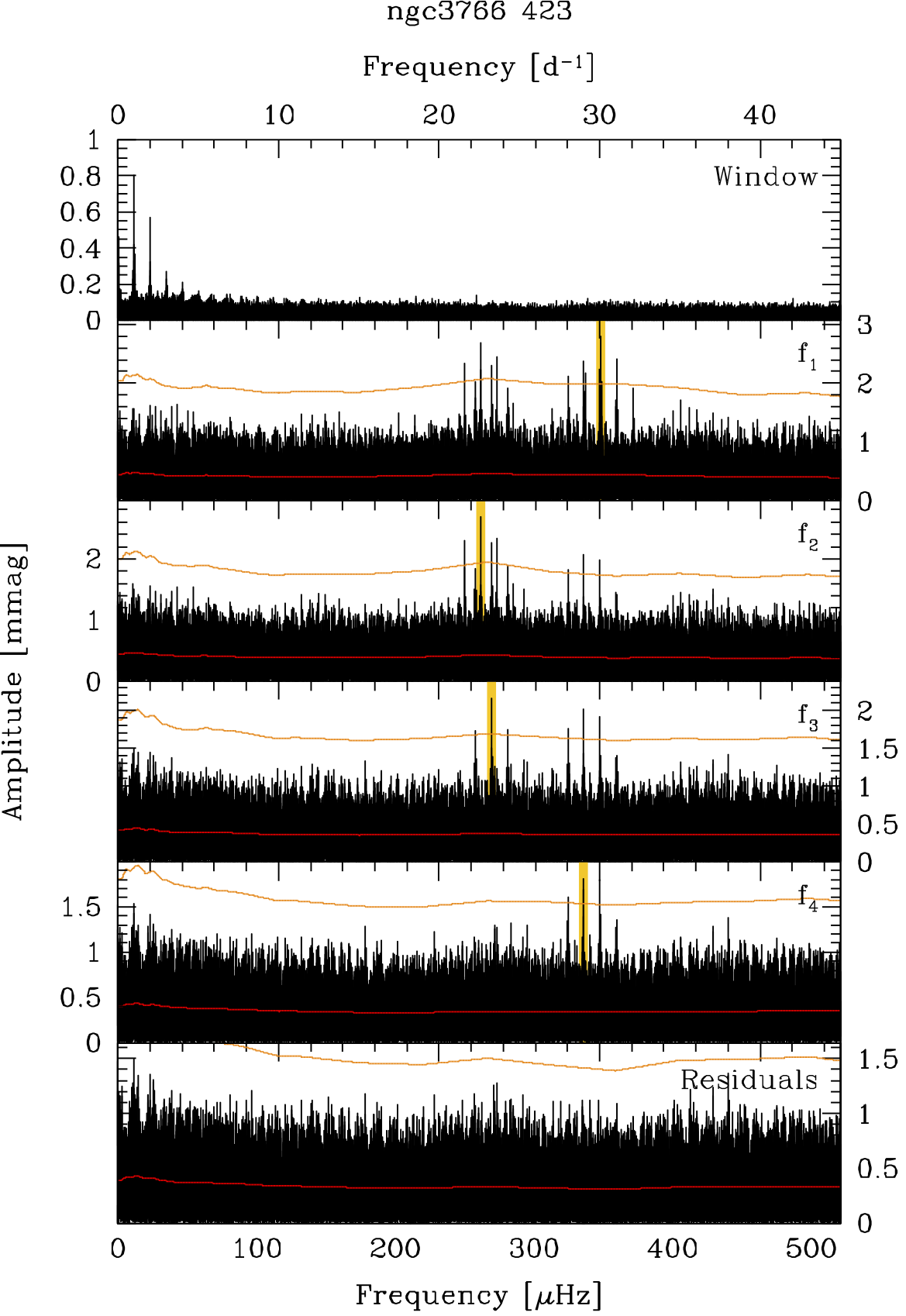}
  \caption{Same as Fig.~\ref{Fig:star36Group1_periodogram}, but for star 423 of group~3.
  }
\label{Fig:star423Group3_periodogram}
\end{figure}

The third group contains 14 stars, 6 of which are monoperiodic, 3 biperiodic, 2 triperiodic, 2 quadriperiodic and 1 has five periods.
Their folded light curves are shown in Figs~\ref{Fig:foldedLcsGroup3_monoperiodic} to \ref{Fig:foldedLcsGroup3_multiperiodic} of Appendix~\ref{Appendix:periodicVariables}.

They are $\delta$~Sct candidates, pulsating in $p$-modes ($\kappa$ mechanism mainly on the second partial ionization zone of He) with periods between 0.02 and 0.25~d.
They are often multiperiodic.
An example of a periodogram typical of those stars is shown in Fig.~\ref{Fig:star423Group3_periodogram} with the quadriperiodic star~423. 
The magnitudes of the $\delta$~Sct candidates in NGC~3766 range between $V'=14$ and 14.9~mag (Fig.~\ref{Fig:VClass}).
All periodic variables in our field of view within this magnitude range belong to this group, except two.
The first exception is star 349 at $V'=14.12$~mag, with a period of 61.51~d.
It is a red giant not belonging to NGC~3766, as shown by its position in the color-magnitude diagram (Fig.~\ref{Fig:cmClass}).
It is put in group~5.
The second exception is star 364 at $V'=14.25$~mag.
With $P=0.91772$~d, it belongs to group~2.

In addition to stars with magnitudes between $V'=14$ and 14.9~mag, two fainter stars are added to group~3.
The first one, star 576 at $V'=15.28$~mag, has three periods, at $P=0.0761747$, 0.0770839 and 0.0773966~d, reminiscent of $\delta$~Sct pulsators.
In the color-color diagram (Fig.~\ref{Fig:ccClass}), it stands at $B'-V'=0.857$ and $U'-B'=0.783$.
It is thus not a member of NGC~3766, but a reddened star located behind the cluster in the Galaxy.
De-reddening would bring it back on the S-shape sequence in that diagram, among the population of $\delta$~Sct candidates.

The second star, star 582, is also reddened according to its position in the color-color diagram.
Both its inferred de-reddened position in that diagram and the values of its four periods, ranging between 0.16 and 0.21~d, make it a strong $\delta$~Sct candidate.

\subsection{Group 4 of periodic variables: $\gamma$~Dor candidates}
\label{Sect:group4}

\begin{figure}
  \centering
  \includegraphics[width=1.0\columnwidth]{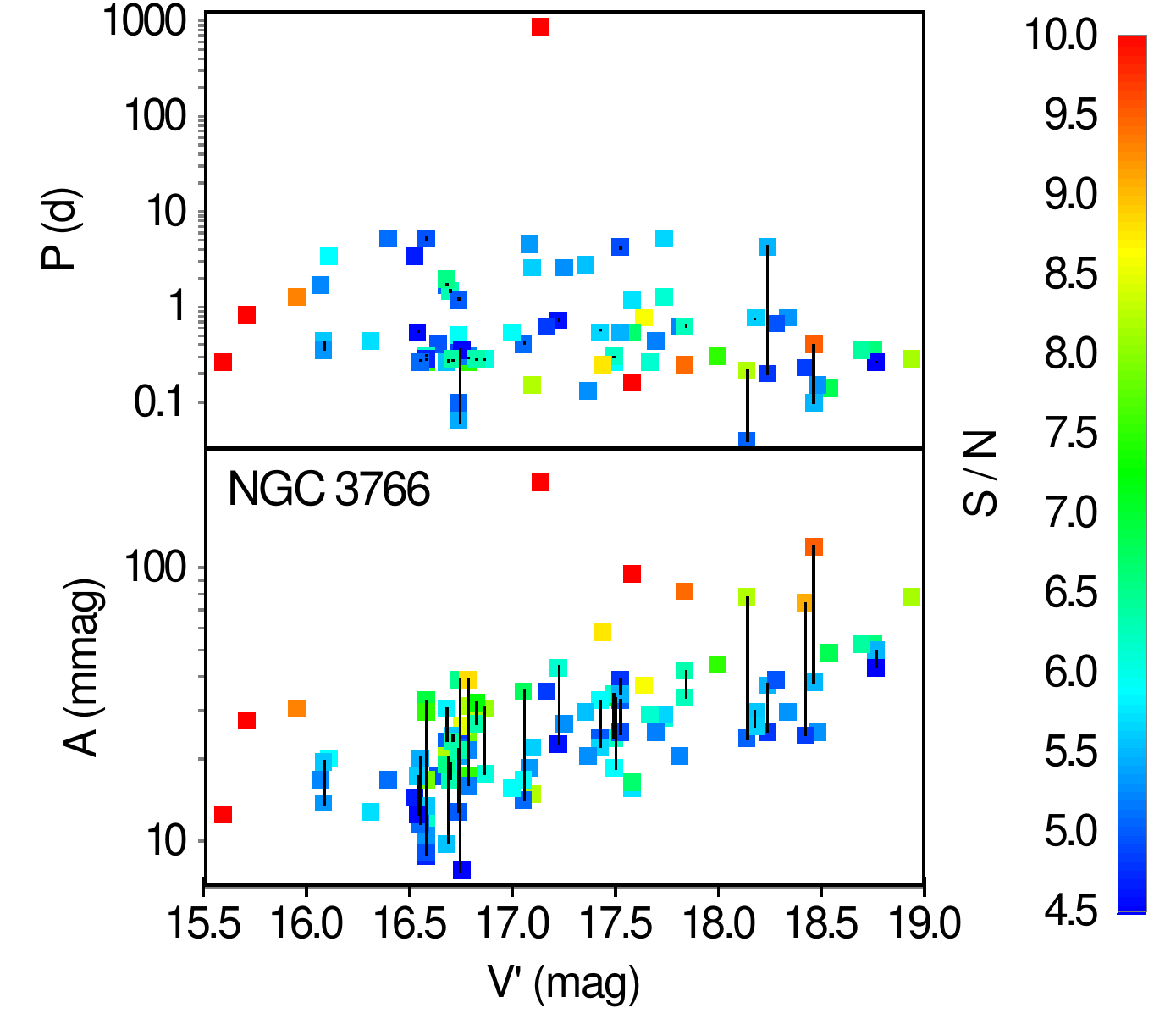}
  \caption{Same as Fig.~\ref{Fig:VPeriodicsBright}, but for the faint periodic variables.
  }
\label{Fig:VPeriodicsFaint}
\end{figure}

The fourth group identified in Fig.~\ref{Fig:VClass} contains 12 stars, 5 of which are monoperiodic, 5  biperiodic and 2 triperiodic.
Their light curves are shown in Figs~\ref{Fig:foldedLcsGroup4_monoperiodic} to \ref{Fig:foldedLcsGroup4_multiperiodic} of Appendix~\ref{Appendix:periodicVariables}.
They are at the faint side of the $\delta$~Sct candidates and cover only half a magnitude from $V'=14.9$ to 15.2~mag.
Their periods range between 0.3 and 2~d, and their amplitudes from few to $\sim$30~mmag.

Among them, star 1584 is an apparent exception at $V'=17$~mag.
It is actually a reddened star not member of the cluster.
De-reddening would bring it in the color-color diagram in the region of group~4 stars (Fig.~\ref{Fig:ccClass}), in agreement with its period of 0.54271~d.

Those stars are $\gamma$~Dor candidates pulsating in $g$-modes.
Their periods range from 0.3 to 1.7~d, though two cases are known at longer periods, up to 2.7~d \citep{HenryFekelHenry07,HenryFekelHenry11}.
Our periods are compatible with this range.

Finally, we remark the very narrow range in magnitudes (of 0.26~magnitudes) covered by the $\gamma$~Dor candidates that are members of the cluster, compared to the magnitude ranges of each of the $\delta$~Sct, group~2 and SPB candidates (of 0.77, 3.2 and $>$1.3 magnitudes, respectively).

\subsection{Group 5 of periodic variables}
\label{Sect:periodicFaint}

Group~5 contains 72 stars, of which 40 are monoperiodic, 25 biperiodic, 6 triperiodic and 1 star has four periods.
It gathers all periodic variables fainter than the $\gamma$~Dor candidates.
Their periods and amplitudes are shown in Fig.~\ref{Fig:VPeriodicsFaint} as a function of $V'$, with the S/N ratio of each detected frequency color coded.
No special feature comes out from the period distribution in the figure.
They all have periods shorter than 10~days, except star 1428, a long period variable (LPV), with a period of 887~d and an amplitude of 206~mmag.
The other LPV, star 349 mentioned in Sect.~\ref{Sect:group3} at $V'=14.12$~mag and with $P=61.51$~d, is also added to this group.
None of those two LPVs belong to the cluster.
Their light curves are shown in Fig.~\ref{Fig:lcsLPVs} of Appendix~\ref{Appendix:periodicVariables}.

The nature of the faint periodic variables is not easy to determine due to, among other reasons, the lack of stellar characterization and cluster membership confirmation.
We therefore omit their further classification.

\section{General properties of the periodic variables}
\label{Sect:discussion_periodics}

We discuss in this section several properties of the periodic variables found in the previous section.
Section~\ref{Sect:discussion_PA} presents the period-amplitude diagram, Sect.~\ref{Sect:discussion_multiperiodics} the frequency separations of the multiperiodic variables, and Sect.~\ref{Sect:discussion_amplitudeRatios} the amplitudes in different photometric bands.

\subsection{Periods versus amplitudes}
\label{Sect:discussion_PA}

\begin{figure}
  \centering
  \includegraphics[width=1.0\columnwidth]{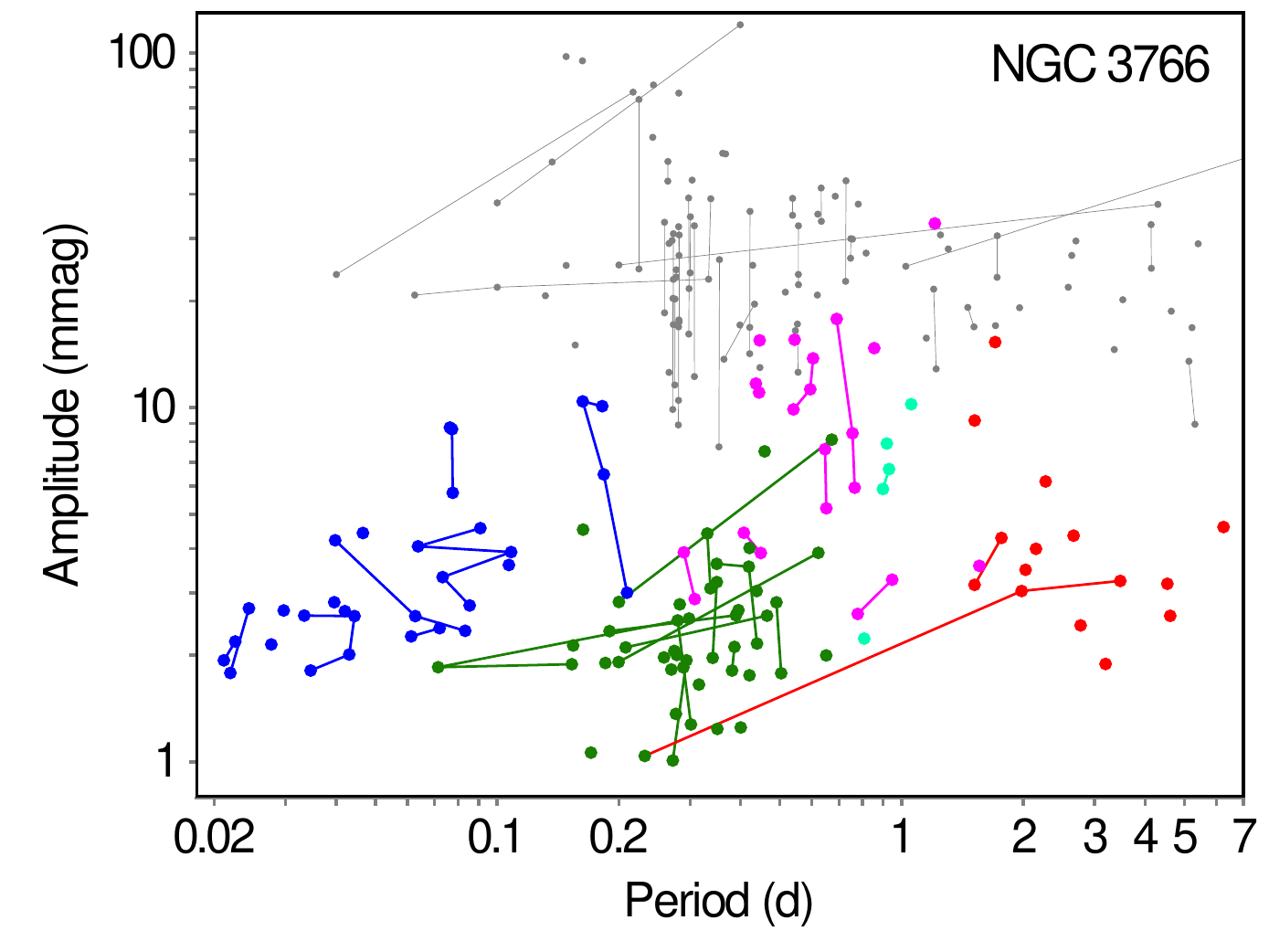}
  \caption{Period -- amplitude diagram, in logarithms, of all periodic variables.
  Multiple periods of a same star are connected with straight lines.
  Colors indicate the group to which the period belongs, coded as in Fig.~\ref{Fig:VClass} except for group~5 which is plotted in small gray points instead of black.
  The X-axis range has been truncated at 7~d for sake of clarity.
  Only two stars fall outside this range, stars 349 and 1428 with $P=61.5$ and 888~d, respectively, both belonging to group~5.
  }
\label{Fig:periodAmplVarTypeAllFreqs}
\end{figure}

A classical visualization diagram of periodic variables is the period versus amplitude diagram.
This is shown in Fig.~\ref{Fig:periodAmplVarTypeAllFreqs} for our periodic variables.
Ignoring group~5 of faint stars which are difficult to classify, we see that groups~1 (red points, SPB candidates), 2 (\toReferee{green and} cyan points) and 3 (blue points; $\delta$~Sct candidates) occupy rather distinct regions in this diagram.
They all have amplitudes below $\sim$10~mmag, down to 2~mmag for $\delta$~Sct and SPB candidates, and down to 1~mmag for group~2 stars, but they have quite distinct periods.
$\delta$~Sct candidates have periods shorter than $\sim$0.1~d, group~2 stars between $\sim$0.1 and $\sim $1.1~d, \toReferee{(with the bulk between $\sim$0.1 and $\sim $0.7~d}), and SPB candidates longer than $\sim$1.5~d.
Group~4 (magenta points; $\gamma$~Dor candidates), on the other hand, has periods similar to group~2.
$\gamma$~Dor candidates have larger amplitudes, on the average, than group~2 stars, but the distribution of the amplitudes of the members of the two groups overlap significantly between 2.5 and 10~mmag.
There is, however, a clear difference in $V'$, of at least one magnitude, between those two groups.

Figure~\ref{Fig:periodAmplVarTypeAllFreqs} also reveals an overall trend of increasing amplitude with period within groups 2, 3 and 4.
In group~2, for example, stars with small amplitudes of $\sim$1~mmag all have periods less than 0.5~d, while those with $A > 5$~mmag all have $P \gtrsim 0.5$~days.

\subsection{Multiperiodic variables}
\label{Sect:discussion_multiperiodics}

\begin{figure}
  \centering
  \includegraphics[width=\columnwidth]{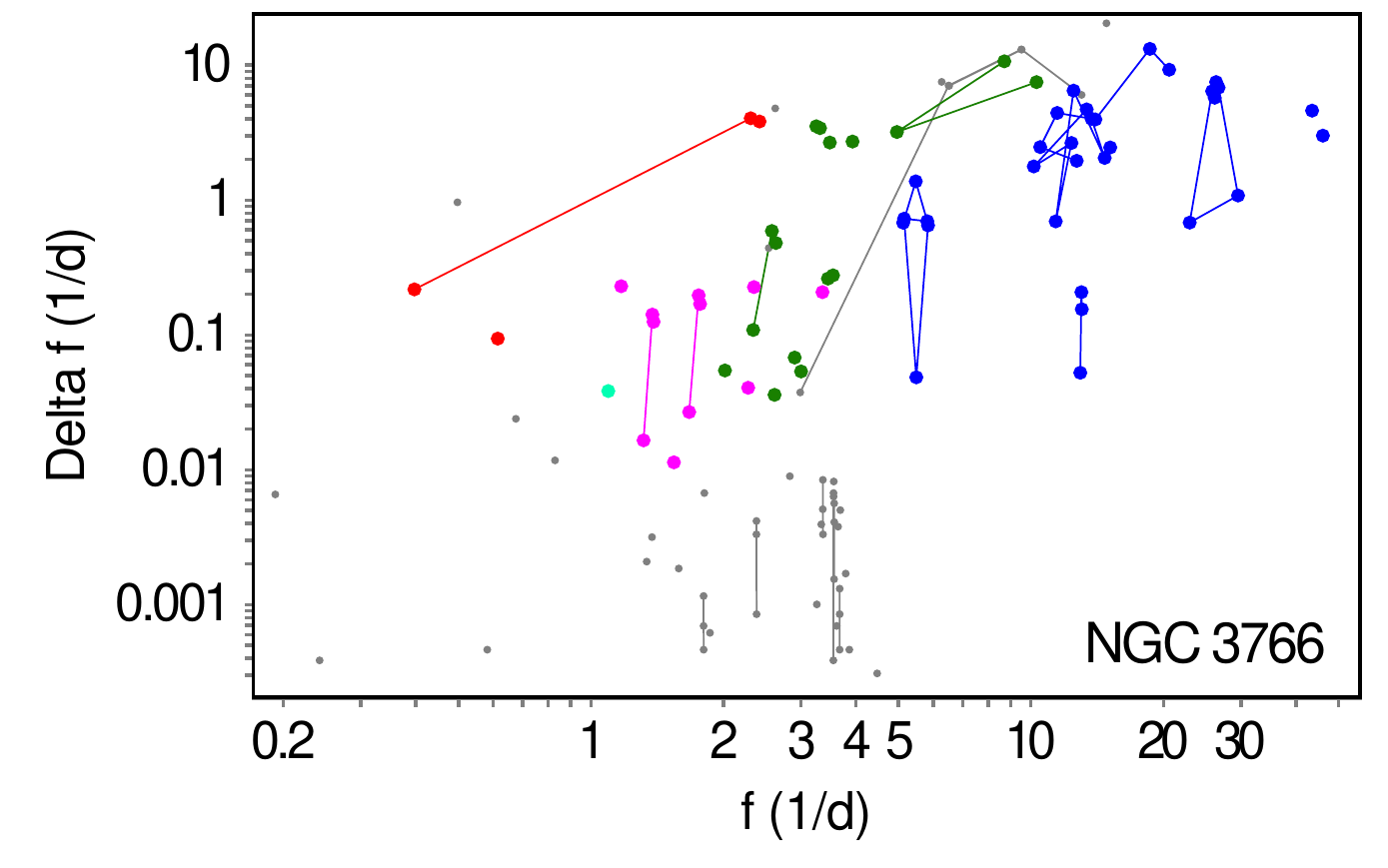}
  \caption{Frequency separations of all frequency pairs belonging to multiperiodic variables, as a function of the mean of the considered pair of frequencies.
  The data for stars with more than two frequencies are connected with a line.
  Colors in all panels indicate the group to which the star belongs, coded as in Fig.~\ref{Fig:VClass}, except for group~5 stars which are colored in gray (small points) instead of black.
  }
\label{Fig:multiperiodicsFreq}
\end{figure}

The frequency separations $|\Delta f|$ between the frequencies of multiperiodic variables are shown in Fig.~\ref{Fig:multiperiodicsFreq}.
Group~2 stars are seen to have frequency separations on line with those of $\gamma$~Dor and $\delta$~Sct candidates.
The relative frequency separations $\Delta f/f$ range from $\sim$0.01 to $\sim$1 for all stars from group~1 to 4, with no distinctive feature characterizing any of those four groups relative to the others.

The multiperiodic variables among the faint stars (group~5), on the other hand, have very small frequency separations for the majority of them, with $\Delta f <0.01$~d$^{-1}$.
This may be attributed to noise in the light curves of those stars.
We note that seven years of observations lead to a typical peak width in the Fourier space of $0.5\times 10^{-3}$~d$^{-1}$.

\subsection{Multi-band variability properties}
\label{Sect:discussion_amplitudeRatios}

The wavelength dependency of the amplitude of a pulsating star has been considered since the late seventies as an observational tool to identify the degree $l$ of pulsation modes \citep[e.g.][]{Dziembowski77,ButaSmith79,StamfordWatson81,Watson88}.
The amplitude ratios at different wavelengths depend on the degree $l$ of the pulsation modes \citep[e.g.][]{HeynderickxWaelkensSmeyers94,DupretDeRidderDeCat_etal03}.
Recent applications of this technique to derive the modes of individual stars include, for example, \cite{AertsKolenberg05}, \cite{TremblayFontaineBrassard06}, \cite{Daszyska-Daszkiewicz08}, \cite{HandlerShobbrookUytterhoeven12}.

\begin{figure}
  \centering
  \includegraphics[width=\columnwidth]{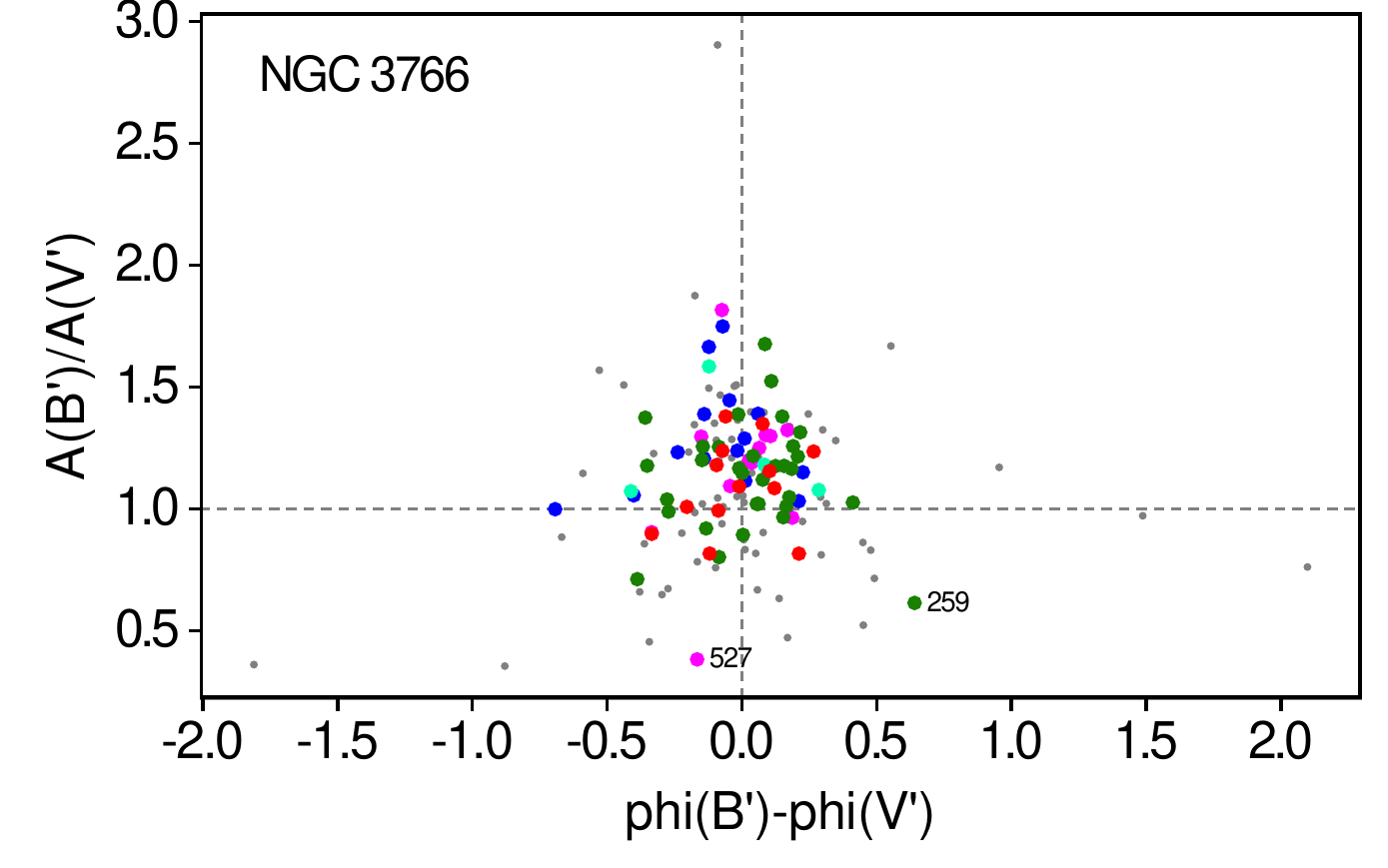}
  \caption{Ratio of the amplitude in $B'$ over that in $V'$ versus phase difference in those two bands.
  The color of each point indicates the group to which the star belongs, coded as in Fig.~\ref{Fig:VClass}, except for group~5 stars which are colored in gray (small points) instead of black.
  The horizontal and vertical dotted lines are added as eye guides to locate identical amplitudes and phases, respectively, in the two photometric bands.
  }
\label{Fig:amplRatios}
\end{figure}

The $A(B')/A(V')$ amplitude ratios as a function of the phase differences $\varphi(B')-\varphi(V')$ are shown in Fig.~\ref{Fig:amplRatios} for the dominant period all our periodic variables, color coded according to the group to which they belong.
The $A(B')$ amplitude in $B'$ is computed by fitting a sine to the $B'$ time series using the dominant period in $V'$.
The majority of the stars have $A(B')/A(V') \gtrsim 1$, as expected.
The existence of a few stars with amplitudes smaller in $B'$ than in $V'$ is understood by the larger uncertainties in the $B'$ amplitudes compared to those in $V'$.
Indeed, the mmag level of the amplitudes requires a sufficient number of observations to be made in order to be detected, but the number of observations in $B'$ is smaller than that in $V'$ by a factor of about six (see Fig.~\ref{Fig:histoNumGoodPoints}).

The more than 1500 observations in $V'$, combined with the stability of the periodic variability over the seven-year duration of the survey, is a key factor in the detection of the variability.
Such a detection could not reliably be achieved in the $B'$ band for the small amplitude variables.
This is especially true for the subset of group~2 stars with amplitudes below 3~mmag.
Stars 259 and 527, for example, which have the smallest $A(B')/A(V')$ ratios among group~1 to 4 stars (0.61 and 0.38, respectively, see Fig.~\ref{Fig:amplRatios}), have amplitudes in $V'$ as small as 1.8 and 3.3~mmag, respectively, and only 1.2 and 1.3~mmag, respectively, in $B'$.
This explains why group~2 stars, which have on average the smallest amplitudes among our periodic variables, also have the largest number of members with $A(B')/A(V')<1$ in Fig.~\ref{Fig:amplRatios}.
It is therefore not possible at this stage to perform a deeper pulsation study based on the multiband variability properties of our $V'$ and $B'$ time series.
The situation is worse with the $U'$ time series, which have both larger uncertainties and fewer observations than the $B'$ time series.
Additional observations at those short wavelengths are necessary to conduct such a study.

The multiband data displayed in Fig.~\ref{Fig:amplRatios} nevertheless reveal an important fact, that there is no significant difference in the distribution of groups~1 to 4 stars in the $A(B')/A(V')$ versus $\varphi(B')-\varphi(V')$ diagram.
This supports the idea that the variability of group~2 stars may be of a similar nature than that of group~1, 3 and 4 stars, i.e. that it may originate from pulsation as well.

\section{The nature of group~2 stars}
\label{Sect:discussion_group2}

The nature of group~2 stars is mysterious.
As mentioned in Sect.~\ref{Sect:group2}, no classical type of pulsating star is expected on the MS of the HR diagram in the region between $\delta$~Sct and SPB stars\toReferee{, which is out of all known instability strips}.
This is exactly where group~2 stars are found.

In this section, we thus further analyze those stars.
We start with a summary of their properties in Sect.~\ref{Sect:group2_properties}, and discuss the possible origin of their variability in Sect.~\ref{Sect:group2_rotation}.
Section~\ref{Sect:literature} briefly explores the literature for potential stars showing similar properties.
\toReferee{Section~\ref{Sect:newClass} finally discusses whether they form a new class of variable stars.}

\subsection{Summary of their variability properties}
\label{Sect:group2_properties}

\begin{figure}
  \centering
  \includegraphics[width=0.9\columnwidth]{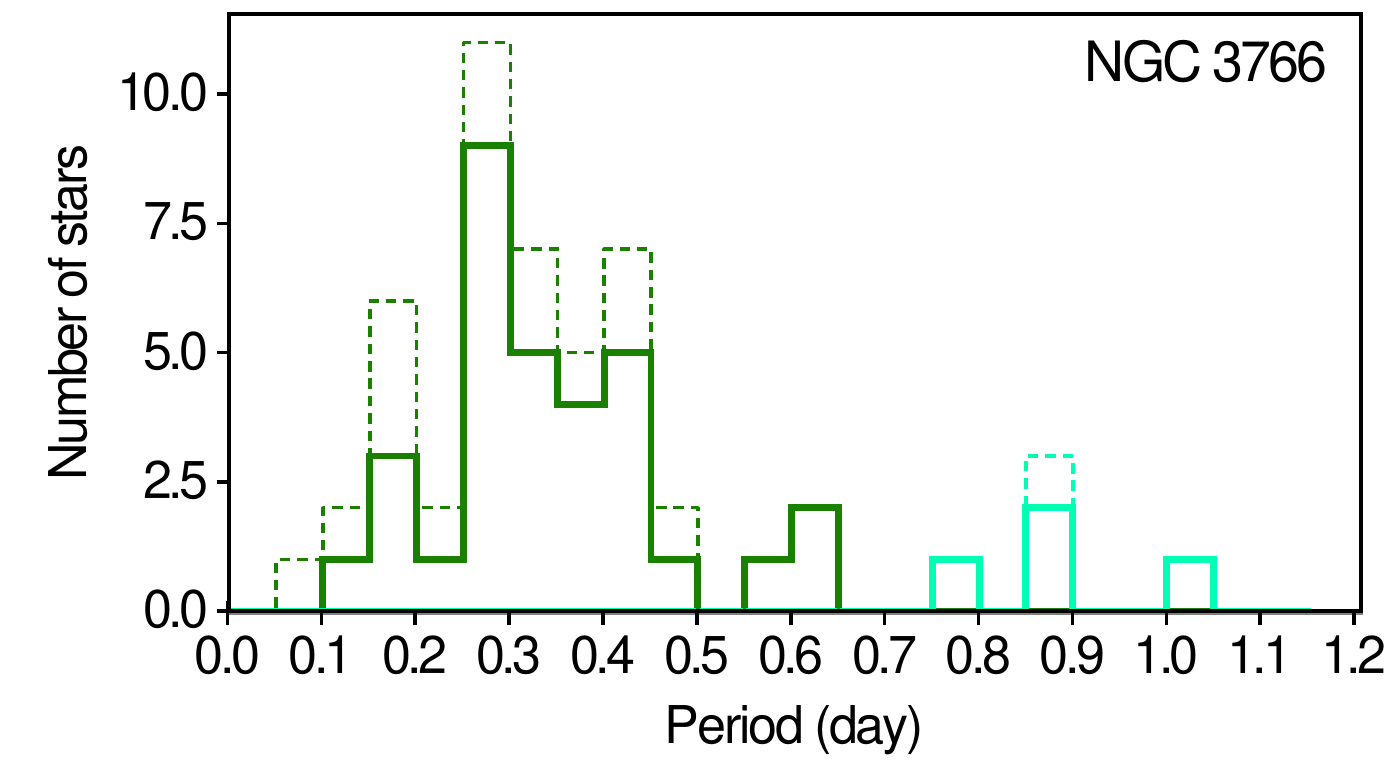}
  \includegraphics[width=0.9\columnwidth]{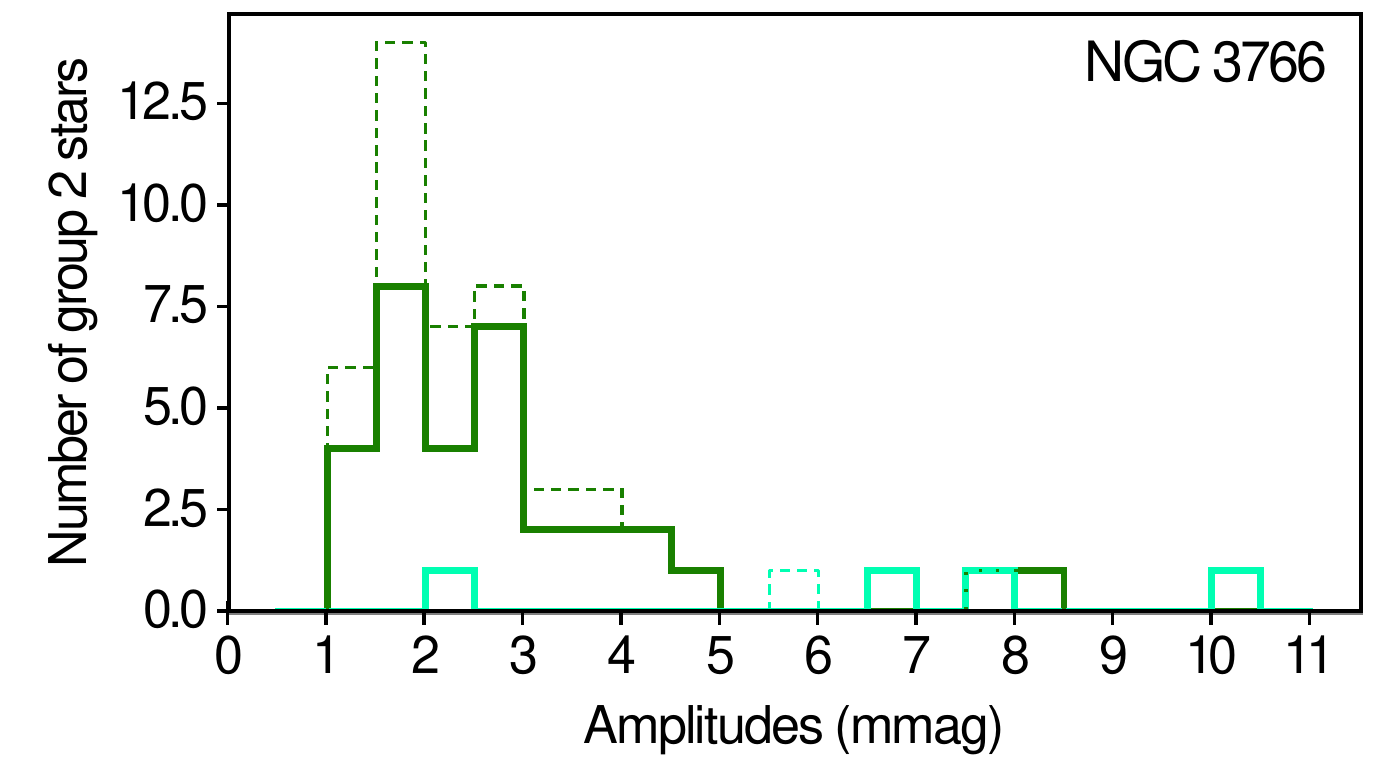}
  \caption{Histograms of the periods (top panel) and amplitudes (bottom panel) of group~2 stars.
  Solid lines give the histograms for the periods and amplitudes of the dominant detected frequency, while dotted lines show the histograms considering all frequencies of the multiperiodic group~2 variables.
  \toReferee{The bulk of stars with periods between 0.1 and 0.7~d are shown in dark green, while the few stars with periods between 0.7 and 1.1~d are displayed in cyan, in agreement with the color codes adopted in Fig.~\ref{Fig:VClass}.}
  }
\label{Fig:HistoGroup2}
\end{figure}

The main properties of group~2 stars from our photometric variability analysis of NGC~3766 can be summarized as follows:
\begin{enumerate}[1)]
\item They are MS stars with magnitudes between those of $\delta$~Sct and SPB candidates (Fig.~\ref{Fig:cmClass}).
\vskip 1mm
\item They are numerous in NGC~3766. There are 36 such stars, which represent about 20\% of all stars in their magnitude range.
\vskip 1mm
\item \toReferee{The bulk of their periods ranges from 0.1 to 0.7~d, with a peak around 0.3~d, as shown in the histogram displayed in Fig.~\ref{Fig:HistoGroup2} (top panel).
      Only four stars have periods between 0.7 and 1.1~d.}
%
\vskip 1mm
\item One third of them are multiperiodic, at our 1~mmag variability detection limit.
Their frequency separations are similar to those of $\delta$~Sct and $\gamma$~Dor candidates (Fig.~\ref{Fig:multiperiodicsFreq}).
\vskip 1mm
\item Their amplitudes are at the mmag level.
The histogram of the amplitudes (Fig.~\ref{Fig:HistoGroup2}, bottom panel) \toReferee{shows that all, except one, of the bulk stars (i.e. with periods less than 0.7~d) have amplitudes below 5~mmag.
It also} suggests that the identification of group~2 stars in our data base is instrument-sensitivity limited.
The number of group~2 stars is thus expected to be higher than what we find, with sub-mmag amplitudes.
\vskip 1mm
\item There is an overall trend of increasing amplitudes with increasing periods (Fig.~\ref{Fig:periodAmplVarTypeAllFreqs}).
\vskip 1mm
\item Their amplitudes of variability are generally larger in the $B'$ band than in the $V'$ (Fig.~\ref{Fig:amplRatios}).
There is no significant phase shift between the two bands.
\label{Item:test}
\vskip 1mm
\item The \toReferee{found periods} are stable over at least the seven years of our monitoring campaign\toReferee{, which enabled their detection despite their small amplitudes}.
\end{enumerate}

\begin{figure}
  \centering
  \includegraphics[width=1.0\columnwidth]{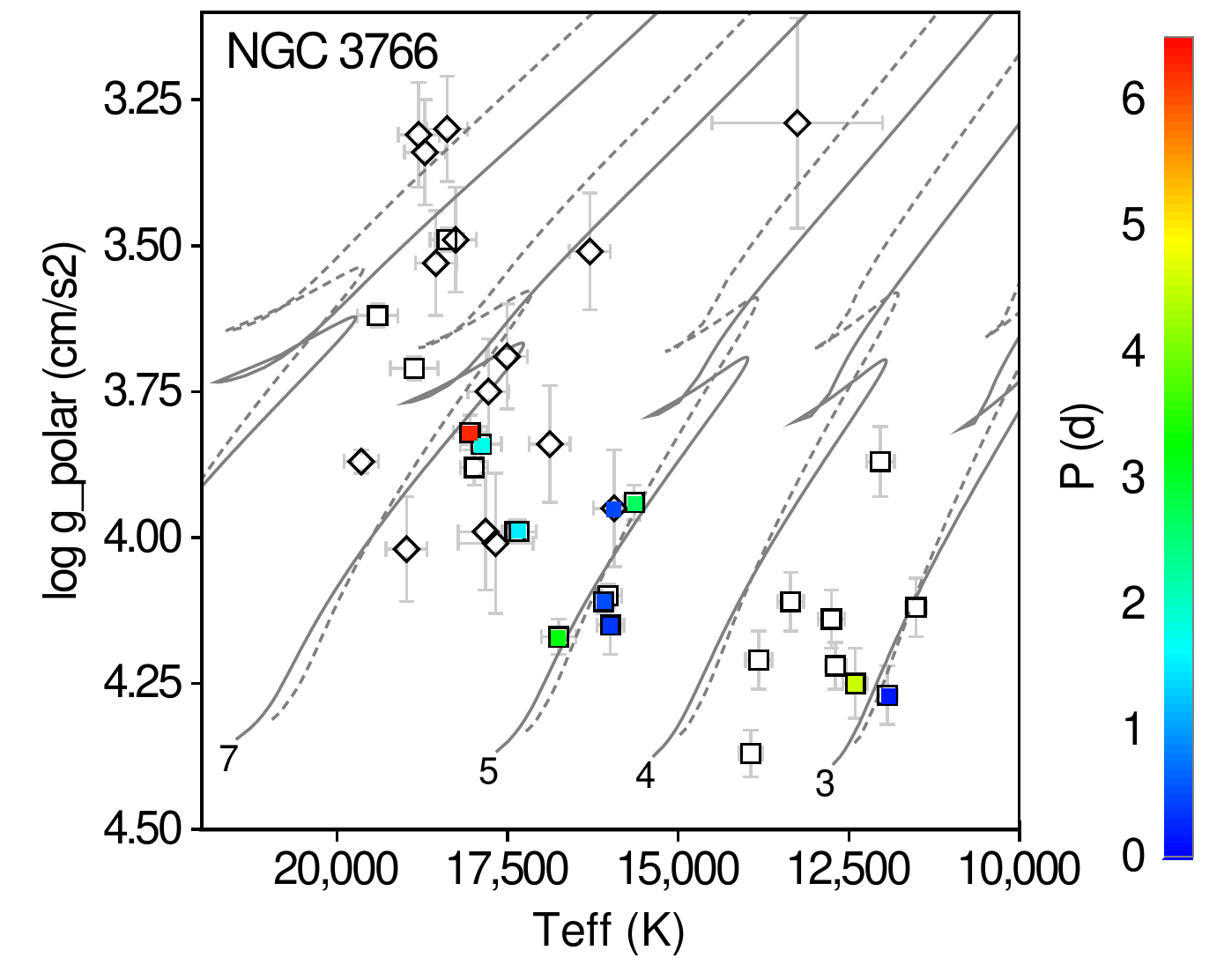}
  \caption{Location in the ($\Teff, \log g_\mathrm{polar}$) HR diagram of the B (squares) and Be (diamonds) stars of \cite{McSwainHuangGies_etal08} that fall in our field of view.
  Open markers identify non-periodic variables, and filled markers periodic variables with the color of the marker related to the period (the dominant one for multiperiodic variables) according to the color scale on the right.
  The four dark blue markers represent group~2 stars (periods less than 1~day), the other six color markers representing SPB candidates.
  Tracks of non-rotating stellar evolution models are shown by continuous lines for various initial stellar masses, as labeled next to the tracks.
  Tracks with initial models rotating at 40\% the critical rotational velocity are shown by dashed lines.
  All tracks are taken from \cite{EkstromGeorgyEggenberger_etal12}.
  }
\label{Fig:HrMcSwain}
\end{figure}

\begin{figure}
  \centering
  \includegraphics[width=1.0\columnwidth]{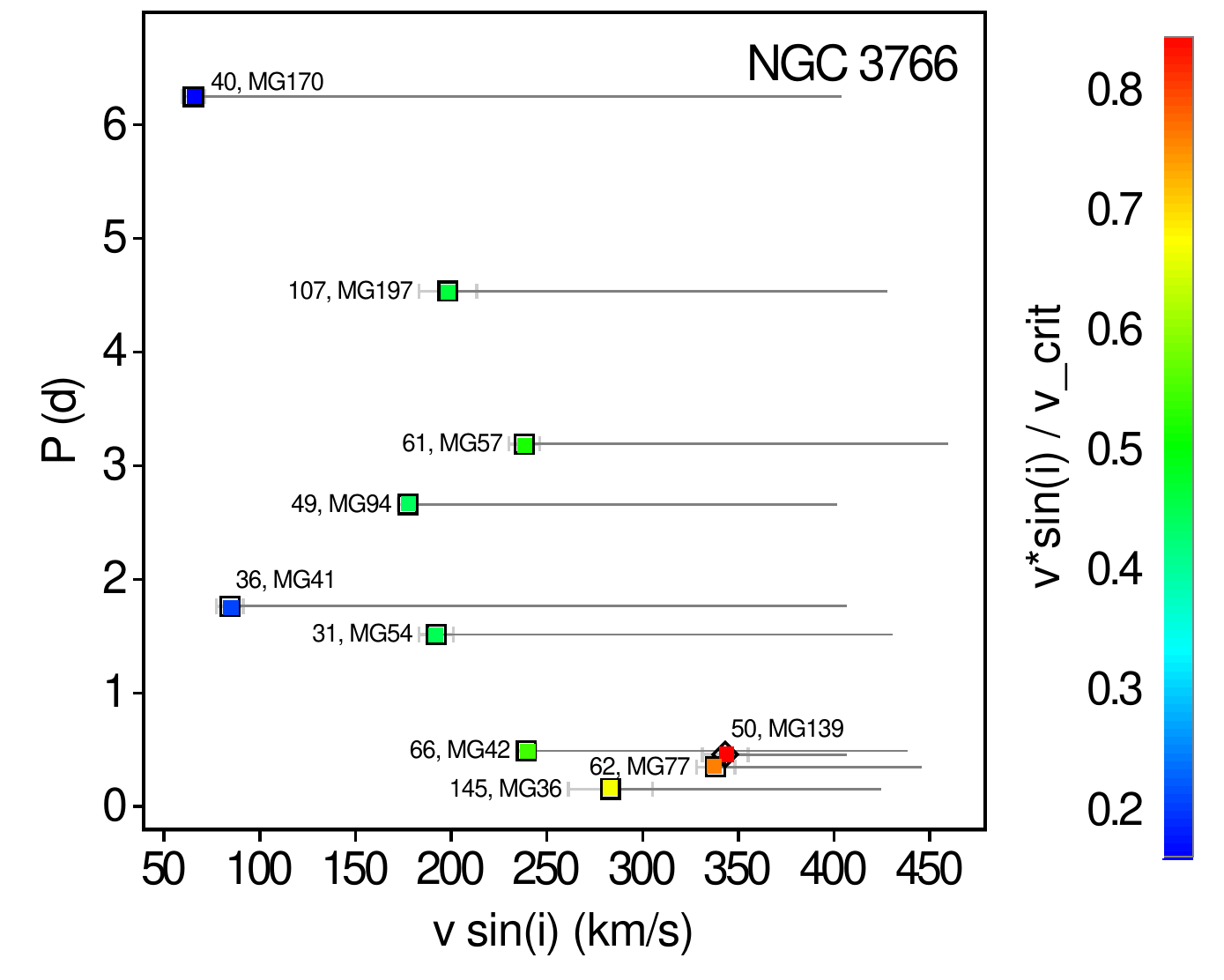}
  \caption{Period as a function of $v \sin(i)$ for the periodic variables in common with the stars spectroscopically observed by \cite{McSwainHuangGies_etal08}.
  The projected rotational velocity $v \sin(i)$ is taken from those authors.
  The markers have the same meaning as those in Fig.~\ref{Fig:HrMcSwain}, with the colors related to the $v \sin(i) / v_\mathrm{crit}$ ratio according to the color scale plotted on the right of the figure, $v_\mathrm{crit}$ being the critical rotational velocity.
  The four group~2 stars are the ones having $P<1$~d, the other ones belonging to group~1.
  Horizontal lines indicate the range of possible rotational velocities $v$ for each star, i.e. from $v \sin(i)$ to $v_\mathrm{crit}$.
  Our star id is indicated next to each marker, followed by the star id from \cite{McSwainGies05b}.
  }
\label{Fig:vsini}
\end{figure}

The overall distribution of group~2 stars in the $(\log P, \log A)$ diagram (Fig.~\ref{Fig:periodAmplVarTypeAllFreqs}) places them at periods \toReferee{longer than those of $\delta$~Sct and shorter than those of SPB stars (their periods are actually similar to those of $\beta$~Cep stars).}
The periods are similar to \toReferee{or slightly below} those of $\gamma$~Dor candidates, but at smaller amplitudes, on average. 
Obviously, the key attribute to distinguish group~2 stars from $\gamma$~Dor candidates \toReferee{(and from $\beta$~Cep stars)} is their magnitude.

\subsection{Origin of their variability}
\label{Sect:group2_rotation}

The properties of group~2 stars, summarized in the previous section, suggest a pulsation origin for their variability.
Among them is the multiperiodic nature of one third of the members of that group, with periods stable over seven years.
The amplitude dependency on wavelength is also compatible with pulsations.
Nevertheless, no pulsation is expected to be sustained in the region of the HR diagram where they are found.

A fraction of the bright stars in NGC~3766 are known to be rapid rotators, a fact supported by the known high number of Be stars present in the cluster.
We can then wonder whether rotation plays a role in sustaining pulsation in group~2 stars.
Studies of the interplay between fast rotation and non-radial pulsations have shown that rotation does impact pulsation in SPB stars.
\cite{UshomirskyBildsten98}, in particular, get to interesting predictions.
First, they show that ``rapid rotation stabilizes some of the g-modes that are excited in a nonrotating star and, conversely, excites g-modes that are damped in the absence of rotation''.
This provides a potential explanation of why pulsating stars can be found outside the instability strips, though detailed models need to confirm the pulsation modes that can be sustained by rotation, if any.
Second, these authors find that ``the fluid velocities and temperature perturbations are strongly concentrated near the equator for most g-modes in rapidly rotating stars, which means that a favorable viewing angle may be required to observe the pulsations''.
This can potentially explain the existence, inside or outside instability strips, of `observationally non-pulsating' stars that, in reality, do pulsate while rotating.


While \cite{UshomirskyBildsten98} illustrate their findings in models of SPB stars, their conclusions are not restricted to those stars.
\cite{Townsend05b} also predicts unstable modes as a result of rotation.
Using a non-adiabatic pulsation code, he finds that ``retrograde mixed modes are unstable in mid- to late-B-type stars, as a result of the same iron-bump opacity mechanism that is usually associated with SPB and $\beta$~Cep stars''.
This even leads him to claim the ``discovery of a wholly new class of pulsational instability in early-type stars''.
The periods he predicts range from 100~days down to a fraction of a day, depending on the azimuthal order $m$ of the mode.
His lowest predicted periods are in agreement with the periods found in our group~2 stars, which would be compatible with $m=3$ and 4.

To further explore the nature of group~2 stars requires the characterization of their stellar properties.
Not many studies helping in this direction are available in the literature.
One of them is the work by \cite{McSwainHuangGies_etal08}, who analyzed the spectra of 42 B stars in NGC~3766, including 16 Be stars.
Thirty-nine of them are present in our list of stars, and ten of them are periodic variables.
Six of them belong to group~1 (stars 31, 36, 40, 49, 61 and 107) and four to group~2 (stars 50, 62, 66 and 145)\footnote{
The correspondence between our star identification numbers and the ones used by \cite{McSwainHuangGies_etal08} is given in Table~\ref{Tab:periodicVariables} of the Appendix.
}.
They are all B stars, except star~50 which is classified as a Be star by \cite{McSwainHuangGies_etal08}.

From the spectra of their stars, \cite{McSwainHuangGies_etal08} derived, among other stellar parameters, the effective temperatures $\Teff$, the surface gravities $g$, and the projected rotational velocities $v \sin(i)$, where $v$ is the equatorial rotational velocity and $i$ is the inclination angle, i.e. the angle between the line of sight and the rotation axis.
They then derived the expected critical rotational velocity by comparison with stellar evolutionary models.
They also estimated the surface gravities $g_\mathrm{polar}$ at the pole of the stars, a quantity that should be less affected by rotation than $g$ is.
The thirty-nine stars in common between their list of stars and ours are shown in the ($\Teff, \log g_\mathrm{polar}$) diagram in Fig.~\ref{Fig:HrMcSwain}.
We also plot in the figure solar metallicity stellar evolution tracks, both with and without rotation, computed by \cite{EkstromGeorgyEggenberger_etal12}.
The ten periodic variables in common between the list of stars of \cite{McSwainHuangGies_etal08} and ours are highlighted in color in the figure.

The periods of our stars are plotted in Fig.~\ref{Fig:vsini} as a function of the projected equatorial velocity obtained by \cite{McSwainHuangGies_etal08}, for the stars in common in the two studies.
We see that the four stars among them that belong to group~2 are spinning at rotation velocities of at least 50\% the critical velocity.
The actual rotation velocities $v$ may be higher than the measured projected velocities, depending on the inclination angle $i$, but they are limited by the critical velocity.
The range of possible velocities for each star between those two limits is shown by the horizontal lines in Fig.~\ref{Fig:vsini}.
Four stars are of course not enough to draw firm conclusions, but that all four of them are fast rotators strongly supports the potential role of rotation in explaining the existence of MS pulsators in the luminosity range between $\delta$~Sct and SPB stars.
Analysis of more stars from group~2 is of course needed to check the validity of this conclusion, but this requires more spectroscopic data.
Star 51, with two periods reminiscent of SPB stars (at 3.47 and 1.98~d) and one of group~2 (at 0.23~d) is an interesting case to be studied in this context.

\subsection{Link with known variables in the literature}
\label{Sect:literature}

In the fifties, \cite{Struve55} hypothesized the presence of a sequence of variable stars in the region of the HR diagram between the locations of the currently known SPB and $\delta$~Sct stars.
The suggestion was based on the detection of radial velocity variations in the B8-type star Maia of the Pleiades, with a period of $\sim$4~h, and in the A-type star $\gamma$~Ursae Minoris, with a period of 2.5~h.
He called those variable stars \textit{Maia stars}.
The variability of the Maia star itself was however disclaimed two years later, by himself \citep{StruveSahadeLynds_etal57}, and the very existence of `Maia variables' became since then a source of debate \citep{McNamara87}.
In the late nineties, \cite{ScholzLehmannHildebrandt_etal98}, despite assessing unambiguous detections of multiperiodic short-term radial velocity variations in two A-type stars, state that ``no proof for the existence of a separate class of variables, designated as Maia variables, are found''.

The high precision of the photometric data gathered by the Hipparcos satellite gave a new impetus in the search for Maia stars, but resulted in inconclusive results.
\cite{PercyWilson00}, for example, conclude after a careful analysis of several hundred Hipparcos variables that ``the Maia variables, if they exist, are very rare and very elusive''.
Seven years later, \cite{DeCatBriquetAerts_etal07} monitored from ground 35 B-type stars during a 3-year high-quality observation campaign.
In their variability analysis, ``all the Maia candidates were reclassified into other variability classes''.

These few examples show the difficulty to confirm the hypothetical sequence of `Maia variables'.
The trouble is not so much to find variable stars in that region of the HR diagram -- individual cases are found --, but that they are rather rare.
The non-prediction of pulsations in this part of the HR diagram, at least based on classical stellar models, did not help to support for a new class of variables either.

The fact that we find in NGC~3766 so many variables at spectral types between those of $\delta$~Sct and SPB stars -- there are 36 stars in our group~2 variables, which represent about 20\% of the stars in that magnitude range --, and with variability properties that distinguish them from their neighboring $\delta$~Sct and SPB candidates, calls for a new attention.
Whether the `Maia' candidates considered in the past and with confirmed short-period variability are of the same nature as our group~2 stars is not clear.
But we believe that we found a new class of variable stars, and that at least some variables discussed in the literature belong to this class.

HD~121190 is an example of a known variable that we believe belongs to our new group~2 stars.
It is of spectral type B9V and has been classified as an SPB star by \cite{WaelkensAertsKestens_etal98}, with a period of 0.38~d derived from Hipparcos epoch photometry.
This makes it an unusually short period SPB star, as recognized by \cite{AertsKolenberg05}, the periods of SPB variables being predicted to be above 0.5~d \citep{Pamyatnykh99}.
\cite{AertsKolenberg05} confirm the short-term periodic variability of this star, with three significant periods detected in multicolor Geneva photometry light curves, at $P=0.3727$, 0.3817 and 0.4046~d and with amplitudes of a few mmag.
They further derive a projected rotation velocity of $v\,\sin i = 118$~km\,s$^{-1}$ from spectroscopic observations, which implies a rotational velocity of at least 26\% of its critical velocity.
This, those authors state, may explain the pulsations found in this star \toReferee{because ``such rotation velocities imply that the effects of the Coriolis force may come into play and that this force introduces significant frequency shifts for the low-frequency gravity modes'', quoting \cite{Townsend03}}.
They also note that HD~121190 is ``the coolest single star of that class (SPB) known to date''.
In view of our findings, we consider HD~121190 to be a member of our new group~2 variables.
It also supports the conclusion of Sect.~\ref{Sect:group2_rotation} that rotation may play a key role in explaining the pulsations of those stars.
\toReferee{The important point to raise here is that group~2 stars lie in the HR diagram out of any classical instability strip, and thus cannot be easily linked to known variability types.}

The small amplitudes of 1 to 3~mmag of the group~2 variables may explain why those stars have been hardly detected until now from ground-based photometry.
Among the published studies of variability in open clusters, the analysis performed by one of us on NGC~884 \citep{SaesenCarrierPigulski_etal10}, based on an 800 days multisite campaign gathering more than 77\,500 CCD images from 15 different instruments, also reached variability detection limits down to the mmag level.
NGC~884, with an age of about 13~My, is a little younger than NGC~3766, but shares with our cluster the property of being rich in B stars.
In NGC~884, the $\delta$~Sct stars are found at $V$ magnitudes between 15 and 16~mag, roughly one magnitude fainter than in NGC~3766.
According to our results, thus, the equivalents of our group~2 stars should be observed in NGC~844, if they exist, at magnitudes between $V=12$ and 15~mag.
Interestingly, \cite{SaesenCarrierPigulski_etal10} also find many periodic variables in that `forbidden' range of magnitudes corresponding to the region between $\delta$~Sct and SPB stars.
Furthermore, many of them have periods between 0.15 and 1~d, and amplitudes of few mmags, like our group~2 stars.
But they also find such variables at brighter magnitudes, down to $V=9$~mag, where SPB stars are expected to be found, as well as variables with periods up to 3 to 10 days at all magnitudes.
It is thus difficult to classify their periodic stars and to analyze their properties in the way we did, without further studies with complementary data.
But one conclusion, we believe, emerges from their database as well, that the range of magnitudes between the $\delta$~Sct and the SPB instability strips in NGC~884 is definitely well populated with periodic variables, contrary to expectations from standard pulsation model predictions.
It must be noted that NGC~884 is also known to hold Be stars.
We refer to \cite{SaesenCarrierPigulski_etal10} for more information on the variability content of this cluster.

Space-based missions such as CoRoT and \textit{Kepler} provided new opportunities in the detection of small amplitude variables.
Very interestingly, \cite{DegrooteAertsOllivier_etal09} report on the ``evidence for a new class of low-amplitude B-type pulsators between the SPB and $\delta$~Sct instability strips''.
The amplitudes of the main peaks in the periodograms of the new B-type pulsators found by those authors are $\sim$0.1\%, with periods between 0.1 and 1~d on average.
The variability properties are thus very similar to those of our group~2 stars.
But a negative conclusion seems to be reached by \cite{BalonaPigulskiCat_etal11} from \textit{Kepler} data, as they state that ``no evidence of pulsation is found in the B-type stars located between the cool edge of the SPB and the hot edge of the $\delta$~Sct instability strips''.
\cite{Balona11}, however, reports on the detection of low-frequency (0.2~d~$<P<$~5~d), low-amplitude ($\sim$40~ppm) variability in most \textit{Kepler} A-type stars.
He tries to explain those periodic variations through known classes of variability, assuming that the variable A-type stars ``are just the hotter counterparts of the spotted F-type stars, since there is no pulsation mechanism which can account for the low frequencies in these stars''.
But he recognizes that, if this is the explanation, ``the idea that A-type atmospheres are quiescent would need to be revised''.
He also detects low frequencies in `hot A-type $\delta$~Sct stars', of about twice the rotational frequency, for which he cannot find any explanation.
We believe that those \textit{Kepler} A-type stars and the CoRoT new B-type pulsators may be of the same nature as our new group~2 variables.

\subsection{\toReferee{A new class of variable stars?}}
\label{Sect:newClass}


\toReferee{
Group~2 stars fall in a region of the HR diagram where no pulsation is predicted to occur.
Their classification in a known class of pulsating stars is therefore not possible.
If we consider a typical member of group~2 stars, for example a pulsating early A-type star with a period of 0.3~d, it cannot be a $\delta$~Sct star both because the period is too long and because it stands out of the classical instability strip.
It cannot be an SPB star either, which are B-type stars and with longer periods.
}

\toReferee{
Of course, a nonpulsating origin for some of these selected stars is not excluded, and this must certainly be the case.
Spots, for example, are known to exist on the surface of A-type stars.
Some group~2 stars may also be ellipsoidal variables.
But many of the group~2 stars are most probably pulsators, as discussed in the previous sections.
The impossibility of classifying them in one of the known classes of variable stars thus calls for a new class of low-amplitude (multi-)periodic stars that contains main sequence A and late B type stars variable at the mmag level (at least in the optical bands) with periods between 0.1 and 0.7~d.
The definition of a new variability class does not imply the need for a new pulsation mechanism, but some physics must be specific to those stars in order to be able to excite pulsation modes outside the borders of known instability strips.
The previous sections have suggested that the interaction between stellar rotation and pulsation (based on standard $\kappa$-mechanism) may fulfill the requirements.
}

\toReferee{
The same conclusion is reached by \cite{DegrooteAertsOllivier_etal09} based on CoRoT data.
Thanks to the high photometric precisions achieved by this space mission, those authors even conclude that this new class may not be uniform.
They suggest the existence of four groups within the class, based on the structure of the frequency spectra of the stars.
In their database, each group contains four to six candidates.
}

\toReferee{
The detection of group~2 stars in an open cluster has the great advantage to enable the identification of stellar types by comparing the location of the stars in the color-magnitude diagram with the locations of identified SPB and $\delta$~Sct members (membership is relatively secured in our case of NGC~3766).
This is not the case for the $\sim$16 CoRoT candidates of this new class, the stellar parameters of which had to be estimated using a procedure that is not straightforward \citep{DegrooteAertsOllivier_etal09}.
They nevertheless concluded those stars to be of type~A.
Our analysis of group~2 stars in NGC~3766 does not suffer from these difficulties and directly supports the claim for a new class of variable stars on the MS between SPB and $\delta$~Sct stars.
}

Eventually, a suitable name for this new group of variables should be found.
We think that the elusiveness of these variables results from their small amplitudes.
Given furthermore that the distribution of their amplitudes may extend to the sub-mmag range, we would propose calling them `low amplitude periodic A and late-B variables', which may actually become `low amplitude \textit{pulsating} A and late-B variables', if confirmed so.

\section{Conclusions}
\label{Sect:conclusions}

This paper presents the periodic variables detected in the field of view of NGC~3766 from a seven-year monitoring campaign performed with the 1.2-m Swiss Euler telescope in La Silla, Chili.
We reach the mmag variability detection level for periodic variables, allowing the discovery of a large number of low-amplitude pulsators.
We find 13 SPB (group~1), 14 $\delta$~Sct (group~3) and 12 $\gamma$~Dor (group~4) candidates, as well as 16 eclipsing binaries and 72 various periodic variables fainter than the $\gamma$~Dor candidates (group~5), not all belonging to the cluster.
All variable stars presented in this paper are new discoveries.
It must be noted that our search does not include any $\beta$~Cep star, even though such stars may be present in NGC~3766, because we excluded from our analysis the brightest, saturated, stars.

Most importantly, we find 36 MS variable (group~2) stars between the red edge of SPB stars and the blue edge of $\delta$~Sct stars, where no pulsation is predicted with standard stellar models.
These form a new group of variables with properties distinct from those of the adjacent SPB and $\delta$~Sct candidates.
About 20\% of the MS stars of NGC~3766 in that magnitude range are found to be group~2 variable candidates, and they represent the most numerous group of variable stars (excluding group~5 of faint stars).
The true number of variable stars in that new group is however expected to be larger, our results being limited by our mmag detection limit.

The properties of this new group of variables, which are summarized in Sect.~\ref{Sect:group2_properties}, support a pulsation origin.
Among those properties are the multiperiodic nature of more than one third of its members and the stability of their variability properties on a time scale of at least seven years.
If pulsation is indeed responsible for their photometric variability, rotation is believed to provide the necessary physical conditions to excite the pulsation modes.
Indeed, four of our candidates in this new group, for which spectra are available in the literature, do rotate at least at half their critical velocity.
Pulsation predictions in models of rotating stars, computed in the last decade, also support this conclusion.

Several observations in the literature support the existence of the new class of variables we find in NGC~3766.
The evidence from ground-based observation is rare, most probably because of the low, mmag level, variability amplitude of those objects, but such evidence is much more convincing from the data gathered by the CoRoT and \textit{Kepler} satellites.
We find it remarkable that our ground-based observations, performed with a one-meter class telescope, is able to provide evidence of that new class of variables at such small amplitudes.
This, we believe, is thanks to the high quality of the observations and of the data reduction, combined with the seven-year baseline of our monitoring campaign.

We encourage searching for this new class of variables in other young open clusters.
That there are so many candidates for this new class in NGC~3766 may be analogous to the large number of Be stars observed in this cluster.
This would be understandable if both phenomena are indeed related to high rotational velocities characterizing the stellar population of the cluster.
If this is true, then the search would be the most efficient in other clusters that are already known to have rich populations of Be stars as well.
Variables of this new class may, for example, already be present in the data of NGC~884 gathered by \cite{SaesenCarrierPigulski_etal10} for NGC~884.

\toReferee{
Finally, we propose (Sect.~\ref{Sect:newClass}) the name of `low amplitude periodic (or pulsating) A and late-B variables' for this new class of variable stars.
}

\begin{acknowledgements}

We would like to thank A. Gautschy for his careful reading and commenting on an early version of the paper.
We also thank G. Burki, F. Carrier and A. Blecha for having set up the Geneva Open Clusters monitoring campaign ten years ago, as well as all the observers that devoted their times to build the resulting high-quality data base of photometric light curves.
The observers include, in alphabetic order, G. Burki, F. Carrier, M. Cherix, C. Greco, M. Spano and M. Varadi in addition to ourselves, as well as other people who occasionally contributed to the monitoring campaign while performing other observation runs at the telescopes.

\end{acknowledgements}

\bibliographystyle{aa}
\bibliography{bibTex}

\newpage
\begin{appendix}
\section{Eclipsing binaries}
\label{Appendix:EBs}

\begin{table*}
\centering
\caption{Eclipsing binaries in our field of view of NGC~3766.
}
\begin{tabular}{r c c c r r c c c}
\hline
 GvaId &  RA & DEC & $V'$  &  $B'$-$V'$  &  $U'$-$B'$  & Period & Type & Potential \\
        & (hour) & (deg) & (mag) &   (mag)   &   (mag)   & (day)  &      &  member   \\
\hline
  50 & 11.611636 & -61.631599 & 11.079 & -0.016 & -0.398 & (100.63957) & EA    & yes \\
 444 & 11.589491 & -61.649919 & 14.734 &  0.513 &  0.240 & 1.194975 & EB    & yes \\
 504 & 11.609959 & -61.689997 & 15.092 &  0.847 &  0.418 & 3.082914 & EA    &  no \\
 542 & 11.610533 & -61.680373 & 15.150 &  0.883 &  0.947 & 3.182190 & EA    &  no \\
 545 & 11.602196 & -61.690321 & 15.209 &  0.505 &  0.249 & 7.716726 & EA    & yes \\
 626 & 11.590774 & -61.582517 & 15.511 &  0.758 &  0.363 & 1.358385 & EA    &  no \\
 685 & 11.609402 & -61.601084 & 15.352 &  0.561 &  0.234 & 1.078387 & EA    & yes \\
 693 & 11.601583 & -61.558911 & 15.646 &  0.653 &  0.689 & 2.990783 & EA    &  no \\
 719 & 11.608576 & -61.616905 & 15.464 &  0.931 &  0.796 & 2.138442 & EB    &  no \\
 934 & 11.591592 & -61.650848 & 16.117 &  1.160 &  1.012 & 2.533318 & EB    &  no \\
1125 & 11.607174 & -61.661810 & 16.007 &  0.777 & -0.016 & 3.218536 & EA    &  no \\
1443 & 11.595281 & -61.552545 & 16.837 &  1.066 &  0.561 & 0.887205 & EW    & yes \\
2000 & 11.600740 & -61.562226 & 17.170 &  1.228 &  0.361 & 0.366388 & EW    &  no \\
2076 & 11.594144 & -61.679395 & 17.553 &  1.238 &  0.110 & 4.175527 & EA    &  no \\
2336 & 11.589583 & -61.580192 & 18.016 &  1.183 & -1.088 & 0.316234 & EW    &  no \\
2663 & 11.613256 & -61.597650 & 17.993 &  1.018 & -0.197 & 0.476018 & EB    &  no \\
\hline
\end{tabular}
\label{Tab:EBs}
\end{table*}

\begin{figure*}
  \centering
  \includegraphics[width=0.66\columnwidth]{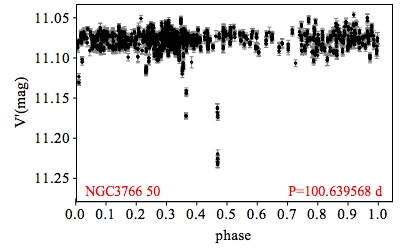}
  \includegraphics[width=0.66\columnwidth]{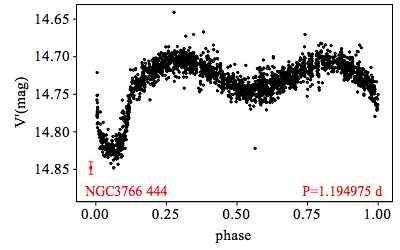}
  \includegraphics[width=0.66\columnwidth]{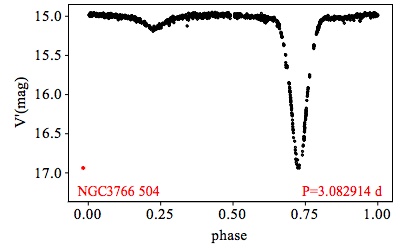}
  \includegraphics[width=0.66\columnwidth]{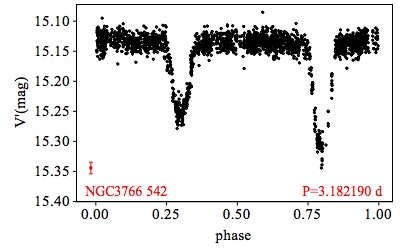}
  \includegraphics[width=0.66\columnwidth]{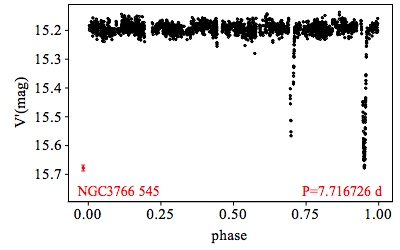}
  \includegraphics[width=0.66\columnwidth]{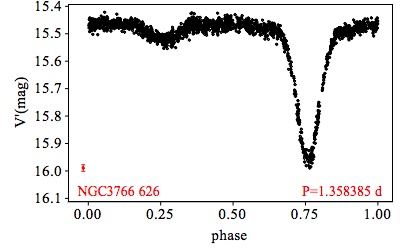}
  \includegraphics[width=0.66\columnwidth]{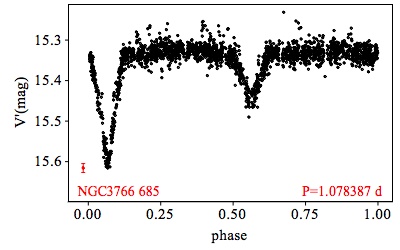}
  \includegraphics[width=0.66\columnwidth]{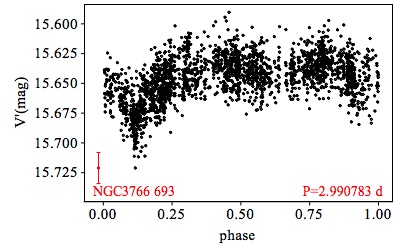}
  \includegraphics[width=0.66\columnwidth]{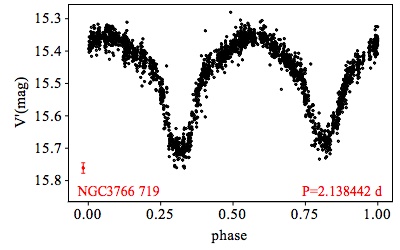}
  \includegraphics[width=0.66\columnwidth]{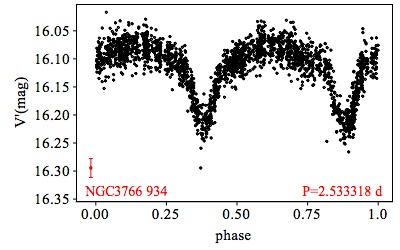}
  \includegraphics[width=0.66\columnwidth]{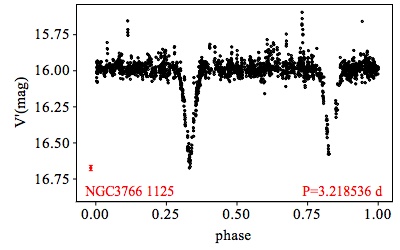}
  \includegraphics[width=0.66\columnwidth]{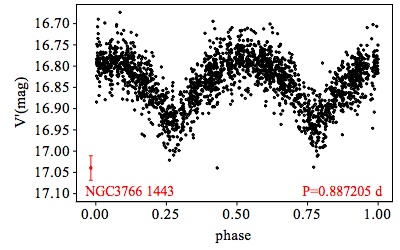}
  \includegraphics[width=0.66\columnwidth]{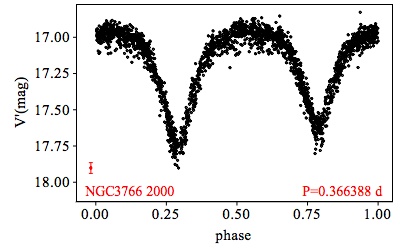}
  \includegraphics[width=0.66\columnwidth]{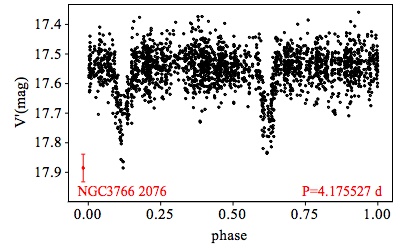}
  \includegraphics[width=0.66\columnwidth]{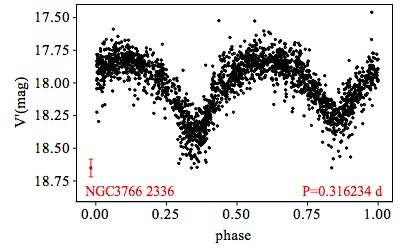}
  \includegraphics[width=0.66\columnwidth]{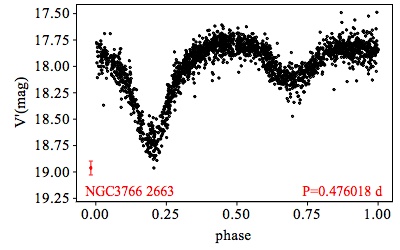}
  \caption{Folded light curves of the eclipsing binaries listed in Table~\ref{Tab:EBs}.
  The error bar displayed in red in the lower-left corner of each plot gives the mean error of the measurements.
  A zoom on the eclipses of candidate 50 (upper left most panel) is shown in Fig.~\ref{Fig:foldedLc50}.}
\label{Fig:foldedLcsEBs}
\end{figure*}

\begin{figure*}
  \centering
  \includegraphics[width=2\columnwidth]{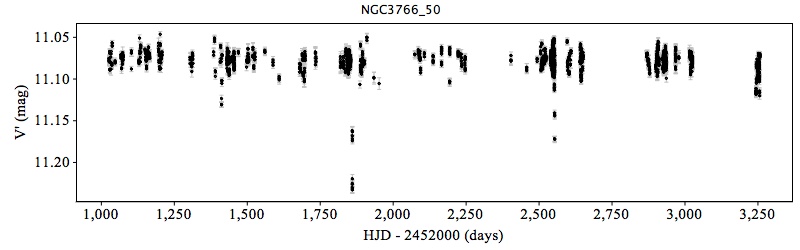}
  \caption{Light curve of the eclipsing binary candidate 50.}
\label{Fig:lc50}
\end{figure*}

\begin{figure*}
  \centering
  \includegraphics[width=\columnwidth]{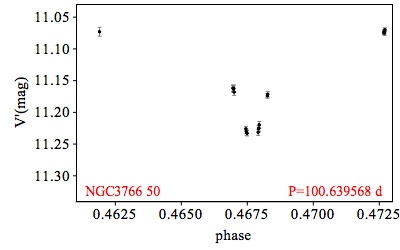}
  \includegraphics[width=\columnwidth]{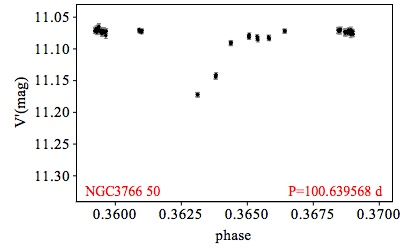}
  \caption{Folded light curve of the eclipsing binary candidate 50, zoomed on the first (left panel) and second (right panel) eclipse.}
\label{Fig:foldedLc50}
\end{figure*}

The list of eclipsing binaries found in our FOV of NGC~3766 is provided in table~\ref{Tab:EBs}.
We give, in the order of the columns presented in the table, their identification number in our numbering scheme, their mean $V'$ magnitude, their colors, their period, their binarity type and their potential cluster membership. 
Membership is evaluated from the position of the star in the color-magnitude and color-color diagrams (Figs~\ref{Fig:cmClass} and \ref{Fig:ccClass}, respectively, in the main body of the paper).

The folded light curves of the eclipsing binaries are shown in Fig.~\ref{Fig:foldedLcsEBs}.

\toReferee{
The eclipsing binary system candidate 50 requires further discussion.
Its light curve, displayed in Fig.~\ref{Fig:lc50}, shows only two eclipsing event candidates, that occurred respectively during the nights 2552.5 and 1858.5~d.
The eclipse durations, if both events are indeed due to an eclipse, are of the order of several hours to half a day.
The third dimming event, around 1410~d, spans two nights and seems less probably due to an eclipsing event.
We therefore assume here that only the first two events are due to an eclipse.
The first of them shows a very nice V-shape light curve, while the second one has only two sets of measurements on the brightening side of event.
}

\toReferee{
It is difficult to estimate a period of an eclipsing binary with only two eclipses.
Furthermore, and unfortunately, no $B'$ or $U'$ measurement is available at the times of the two events.
We therefore proceeded in the following way.
We first determined the most probable period $P_V$ from the $V'$ light curve.
No bright (i.e. at the out-of-eclipse level) measurement should be present at eclipse times, as easily checked in the folded light curves.
This led to $P_V = 6.291223$~d.
We then search the smallest multiple $n \; P_V$, $n$ being an integer, for which the folded light curves in $B'$ and $U'$ are compatible with the two eclipses at their relevant phases.
The final period is thereby estimated to $P=100.63957$~d.
}

\toReferee{
The folded light curve is shown in the upper-left panel of Fig.~\ref{Fig:foldedLcsEBs}, and a zoom on the two eclipses in Fig.~\ref{Fig:foldedLc50}.
The minima of the two eclipses are seen to be distant, in phase, by only 0.1 from each other, pointing to a very eccentric system.
}

\toReferee{
In order to shed some further light on the binary nature of star~50, we took spectra of it with the Euler telescope.
From its location on the MS in the HR diagram, we expected the star to be of spectral type around B3.
Its spectra, however, showed the presence of many iron lines, thus revealing the presence of an A-type companion, possibly of spectral type A3.
We thus conclude on a double system with a B-type primary and an A-type secondary.
More observations are required to better characterize this system.
We further note that this binary candidate 50 is also a member of group~2 stars.
}

\clearpage
\newpage
\section{Group~1 to 5 periodic variables}
\label{Appendix:periodicVariables}

\begin{table*}
\centering
\caption{Periodic stars with secured frequencies in our FOV of NGC~3766.
 Colors are put in parenthesis if the involved time series are not all good.
 The column \texttt{Gr} gives the group (1 to 5), and \texttt{MGid} gives the star id given by \cite{McSwainGies05b}, if available. Eclipsing binaries have their GvaId highlighted in bold.}
\begin{tabular}{r c c c c c c l c c c c}
\hline
 GvaId &  RA & DEC &  $ V' $  & $B'-V'$  &  $U'-B'$   & Gr &  $ P $  &   $ A $  & $\sigma(A)$  & S/N  & MGid\\
        & (hour) & (deg) & (mag) & (mag)  &  (mag)   &       & (day) & \small{(mmag)} & \small{(mmag)} &     &       \\
\hline
  22 & 11.605085 & -61.595067 & 10.023 &  -0.092  &  -0.601  &  1 &     4.610(3) &   2.58 &  0.19 &  4.8 &   78 \\
  25 & 11.601374 & -61.580283 & 10.109 &  -0.070  &  -0.518  &  1 &    2.0228(6) &   3.48 &  0.16 &  6.2 &      \\
  31 & 11.593306 & -61.583515 & 10.398 &  -0.068  &  -0.496  &  1 &   1.51349(2) &   9.19 &  0.13 &  7.3 &   54 \\
  33 & 11.606775 & -61.673301 & 10.496 &  -0.080  &  -0.579  &  1 &    2.7678(0) &   2.43 &  0.15 &  5.0 &  179 \\
  34 & 11.594816 & -61.615589 & 10.560 &  -0.066  &  -0.530  &  1 &    2.1457(3) &   3.99 &  0.13 &  5.2 &  110 \\
  36 & 11.593591 & -61.568012 & 10.650 &  -0.076  &  -0.561  &  1 &    1.7631(6) &   4.29 &  0.13 &  4.7 &   41 \\
     &           &            &        &          &          &    &    1.5127(9) &   3.16 &  0.13 &  4.7 &      \\
  38 & 11.602299 & -61.572019 & 10.718 &  -0.081  &  -0.589  &  1 &    2.2673(1) &   6.18 &  0.13 &  7.9 &   45 \\
  40 & 11.597773 & -61.661004 & 10.720 &  -0.079  &  -0.583  &  1 &     6.246(9) &   4.60 &  0.13 &  6.6 &  170 \\
  49 & 11.613708 & -61.605837 & 11.009 &  -0.024  &  -0.419  &  1 &    2.6587(6) &   4.35 &  0.12 &  5.3 &   94 \\
\textbf{  50} & 11.611636 & -61.631599 & 11.079 &  -0.016  &  -0.398  &  2 &  0.457454(8) &   7.52 &  0.12 &  4.5 &  139 \\
  51 & 11.604386 & -61.602233 & 11.121 &  -0.014  &  -0.323  & 1,2&    3.4692(2) &   3.24 &  0.12 &  6.0 &   89 \\
     &           &            &        &          &          &    &    1.9777(3) &   3.04 &  0.12 &  7.2 &      \\
     &           &            &        &          &          &    &   0.23111(0) &   1.04 &  0.12 &  4.9 &      \\
  58 & 11.602906 & -61.599720 & 11.220 &  -0.030  &  -0.379  &  2 &   0.67052(0) &   8.12 &  0.13 &  8.4 &   86 \\
     &           &            &        &          &          &    &  0.199453(9) &   2.83 &  0.13 &  7.1 &      \\
  59 & 11.599175 & -61.637656 & 11.252 &  -0.015  &  -0.374  &  2 &   0.46378(3) &   2.59 &  0.12 &  4.9 &      \\
     &           &            &        &          &          &    &  0.207271(9) &   2.10 &  0.12 &  5.0 &      \\
  60 & 11.603076 & -61.641252 & 11.280 &  -0.013  &  -0.421  &  2 &   0.26885(4) &   1.82 &  0.12 &  7.9 &  152 \\
  61 & 11.603347 & -61.586258 & 11.325 &  -0.033  &  -0.434  &  1 &     3.190(6) &   1.89 &  0.12 &  4.6 &   57 \\
  62 & 11.611338 & -61.594763 & 11.390 &  -0.015  &  -0.340  &  2 &  0.348156(1) &   3.62 &  0.12 &  6.0 &   77 \\
     &           &            &        &          &          &    &   0.41805(8) &   3.56 &  0.12 &  6.6 &      \\
     &           &            &        &          &          &    &   0.43796(7) &   2.16 &  0.12 &  4.7 &      \\
  66 & 11.610668 & -61.569594 & 11.421 &  -0.030  &  -0.394  &  2 &   0.48931(2) &   2.82 &  0.12 &  5.5 &   42 \\
     &           &            &        &          &          &    &   0.50270(8) &   1.78 &  0.12 &  4.6 &      \\
  68 & 11.601627 & -61.630920 & 11.488 &  -0.010  &  -0.344  &  2 &   0.41983(4) &   4.01 &  0.12 &  8.4 &  138 \\
  69 & 11.611235 & -61.572993 & 11.501 &   0.009  &  -0.285  &  2 &   0.43710(9) &   3.04 &  0.12 &  8.1 &   46 \\
  70 & 11.600999 & -61.597358 & 11.517 &  -0.059  &  -0.450  &  2 &   1.05454(8) &  10.22 &  0.12 &  7.4 &   84 \\
  71 & 11.607325 & -61.593554 & 11.557 &  -0.017  &  -0.319  &  2 &   0.39406(5) &   2.68 &  0.12 &  7.7 &   71 \\
  78 & 11.595582 & -61.601200 & 11.674 &  -0.013  &  -0.287  &  2 &   0.38845(8) &   2.59 &  0.12 &  5.9 &   87 \\
     &           &            &        &          &          &    &  0.189088(1) &   2.34 &  0.12 &  7.1 &      \\
  79 & 11.607461 & -61.602947 & 11.720 &  -0.001  &  -0.270  &  2 &   0.38522(7) &   2.11 &  0.12 &  5.1 &   90 \\
     &           &            &        &          &          &    &   0.37997(0) &   1.81 &  0.12 &  5.1 &      \\
  83 & 11.604403 & -61.513950 & 11.773 &   0.005  &  -0.255  &  2 &  0.184685(8) &   1.90 &  0.13 &  4.8 &   11 \\
  84 & 11.600225 & -61.539624 & 11.781 &  -0.007  &  -0.357  &  2 &   0.92952(6) &   6.70 &  0.12 &  5.3 &   26 \\
     &           &            &        &          &          &    &   0.89754(3) &   5.89 &  0.12 &  6.6 &      \\
  94 & 11.607110 & -61.607262 & 11.925 &   0.037  &  -0.150  &  2 &   0.34949(7) &   1.24 &  0.12 &  5.4 &   97 \\
 105 & 11.604225 & -61.589549 & 12.111 &   0.012  &  -0.174  &  2 &  0.170123(7) &   1.06 &  0.12 &  5.3 &   68 \\
 106 & 11.602525 & -61.609240 & 12.137 &   0.003  &  -0.228  &  2 &   0.39929(4) &   1.25 &  0.12 &  4.8 &   99 \\
 107 & 11.597451 & -61.699041 & 12.146 &   0.032  &  -0.113  &  1 &     4.536(2) &   3.18 &  0.12 &  7.4 &  197 \\
 112 & 11.600021 & -61.574157 & 12.198 &   0.008  &  -0.147  &  2 &  0.162641(8) &   4.52 &  0.12 &  9.0 &   48 \\
 135 & 11.610479 & -61.656502 & 12.509 &   0.040  &  -0.021  &  2 &   0.27596(4) &   1.37 &  0.13 &  5.0 &  166 \\
 136 & 11.592039 & -61.640134 & 12.527 &   0.081  &   0.072  &  2 &   0.31470(2) &   1.65 &  0.13 &  6.2 &  150 \\
 142 & 11.595813 & -61.584695 & 12.649 &   0.034  &   0.031  &  2 &   0.29739(5) &   2.54 &  0.14 &  5.8 &   56 \\
     &           &            &        &          &          &    & 0.0712331(9) &   1.85 &  0.14 &  5.3 &      \\
     &           &            &        &          &          &    &  0.152541(7) &   1.89 &  0.14 &  5.3 &      \\
 144 & 11.609517 & -61.663840 & 12.648 &   0.125  &   0.271  &  2 &   0.27736(7) &   2.00 &  0.14 &  6.3 &  174 \\
 145 & 11.606308 & -61.559413 & 12.675 &   0.050  &   0.037  &  2 &  0.153684(4) &   2.13 &  0.13 &  6.7 &   36 \\
 147 & 11.605854 & -61.694624 & 12.745 &   0.095  &   0.021  &  2 &   0.34838(4) &   3.22 &  0.13 &  6.5 &  195 \\
     &           &            &        &          &          &    &   0.34033(4) &   1.97 &  0.13 &  5.5 &      \\
 149 & 11.607215 & -61.621949 & 12.739 &   0.068  &   0.066  &  2 &   0.29308(2) &   1.93 &  0.13 &  6.7 &  120 \\
     &           &            &        &          &          &    &   0.27112(6) &   1.01 &  0.13 &  5.3 &      \\
 154 & 11.603145 & -61.618639 & 12.800 &   0.073  &  -0.048  &  2 &   0.80672(4) &   2.23 &  0.14 &  6.3 &  113 \\
 161 & 11.608217 & -61.587215 & 12.856 &   0.065  &   0.089  &  2 &  0.278763(2) &   2.51 &  0.14 &  7.7 &   60 \\
     &           &            &        &          &          &    &   0.30067(7) &   1.28 &  0.14 &  5.4 &      \\
 167 & 11.601512 & -61.646867 & 12.895 &   0.095  &   0.136  &  2 &   0.28800(3) &   1.85 &  0.14 &  6.3 &  157 \\
 170 & 11.601938 & -61.603492 & 12.918 &   0.057  &   0.126  &  2 &  0.282046(6) &   2.78 &  0.14 &  7.5 &   91 \\
 175 & 11.603317 & -61.573258 & 12.955 &   0.058  &   0.107  &  2 &  0.330029(6) &   4.41 &  0.14 &  7.9 &      \\
     &           &            &        &          &          &    &   0.33597(0) &   3.09 &  0.14 &  7.6 &      \\
 194 & 11.605677 & -61.602511 & 13.103 &   0.086  &   0.217  &  2 &   0.27351(3) &   2.06 &  0.14 &  6.1 &      \\
 236 & 11.612530 & -61.541098 & 13.385 &   0.119  &   0.313  &  2 &   0.25785(1) &   1.97 &  0.17 &  5.8 &      \\
 259 & 11.597326 & -61.701395 & 13.560 &   0.132  &   0.342  &  2 &   0.41966(4) &   1.76 &  0.17 &  5.4 &      \\
 278 & 11.605333 & -61.513864 & 13.717 &   0.218  &   0.419  &  2 &   0.62115(6) &   3.89 &  0.20 &  6.3 &      \\
     &           &            &        &          &          &    &  0.199088(0) &   1.91 &  0.20 &  4.8 &      \\
 295 & 11.593782 & -61.703412 & 13.856 &   0.432  &  -0.051  &  1 &   1.70109(6) &  15.29 &  0.19 &  9.6 &      \\
 298 & 11.608680 & -61.637095 & 13.823 &   0.267  &   0.398  &  2 &   0.64962(4) &   2.00 &  0.17 &  5.1 &      \\
\hline
\end{tabular}
\label{Tab:periodicVariables}
\end{table*}

\addtocounter{table}{-1}
\begin{table*}
\centering
\caption{Continued.}
\begin{tabular}{r c c c c c c l c c c c}
\hline
 GvaId &  RA & DEC &  $ V' $  & $B'-V'$  &  $U'-B'$   & Gr &  $ P $  &   $ A $  & $\sigma(A)$  & S/N  & MGid\\
        & (hour) & (deg) & (mag) & (mag)  &  (mag)   &       & (day) & \small{(mmag)} & \small{(mmag)} &     &       \\
\hline
 322 & 11.597664 & -61.617539 & 14.036 &   0.404  &   0.370  &  3 & 0.0718470(1) &   2.38 &  0.19 &  6.0 &      \\
     &           &            &        &          &          &    & 0.0610848(4) &   2.26 &  0.19 &  6.1 &      \\
 339 & 11.613240 & -61.654688 & 14.111 &   0.277  &   0.433  &  3 & 0.0295522(7) &   2.67 &  0.19 &  7.1 &      \\
 349 & 11.596891 & -61.698295 & 14.124 &   2.155  &   1.918  &  5 &     61.51(7) & 111.08 &  0.19 &  9.9 &      \\
     &           &            &        &          &          &    &   1.02235(4) &  25.02 &  0.19 &  5.9 &      \\
 364 & 11.597168 & -61.539853 & 14.249 &   0.397  &   0.409  &  2 &   0.91772(4) &   7.91 &  0.20 &  9.5 &      \\
 371 & 11.604864 & -61.641162 & 14.269 &   0.303  &   0.444  &  3 & 0.0394145(8) &   2.82 &  0.20 &  7.0 &      \\
 383 & 11.599914 & -61.595467 & 14.344 &   0.399  &   0.516  &  3 &  0.106571(1) &   3.59 &  0.21 &  7.9 &      \\
 396 & 11.602076 & -61.606627 & 14.444 &   0.438  &   0.346  &  3 & 0.0275385(7) &   2.14 &  0.22 &  5.6 &      \\
 398 & 11.606180 & -61.555198 & 14.456 &   0.334  &   0.417  &  3 & 0.0224275(7) &   2.19 &  0.22 &  5.7 &      \\
     &           &            &        &          &          &    & 0.0210132(8) &   1.93 &  0.22 &  5.2 &      \\
 403 & 11.598990 & -61.683809 & 14.525 &   0.357  &   0.422  &  3 & 0.0242393(7) &   2.71 &  0.22 &  6.0 &      \\
     &           &            &        &          &          &    & 0.0218076(0) &   1.78 &  0.22 &  4.7 &      \\
 404 & 11.592682 & -61.568086 & 14.521 &   0.438  &   0.377  &  3 & 0.0396191(4) &   4.22 &  0.22 &  7.1 &      \\
     &           &            &        &          &          &    & 0.0625392(2) &   2.58 &  0.22 &  5.1 &      \\
     &           &            &        &          &          &    &  0.083101(1) &   2.34 &  0.22 &  5.1 &      \\
 411 & 11.589003 & -61.597937 & 14.562 &   0.351  &   0.328  &  3 & 0.0464042(7) &   4.42 &  0.24 &  9.0 &      \\
 423 & 11.598604 & -61.644744 & 14.646 &   0.455  &   0.345  &  3 & 0.0332194(7) &   2.59 &  0.23 &  6.3 &      \\
     &           &            &        &          &          &    & 0.0442266(4) &   2.58 &  0.23 &  6.2 &      \\
     &           &            &        &          &          &    & 0.0429348(6) &   2.01 &  0.23 &  5.8 &      \\
     &           &            &        &          &          &    & 0.0344554(3) &   1.81 &  0.23 &  5.3 &      \\
 456 & 11.603671 & -61.665864 & 14.788 &   0.370  &   0.471  &  3 & 0.0418736(3) &   2.66 &  0.24 &  5.7 &      \\
 458 & 11.603400 & -61.646823 & 14.795 &   0.474  &   0.492  &  3 & 0.0905980(4) &   4.57 &  0.26 &  5.5 &      \\
     &           &            &        &          &          &    & 0.0635444(7) &   4.06 &  0.26 &  5.5 &      \\
     &           &            &        &          &          &    &  0.107950(2) &   3.91 &  0.26 &  5.9 &      \\
     &           &            &        &          &          &    & 0.0730737(4) &   3.32 &  0.26 &  5.6 &      \\
     &           &            &        &          &          &    &  0.085234(5) &   2.76 &  0.26 &  4.8 &      \\
 482 & 11.589308 & -61.587064 & 14.915 &   0.462  &   0.258  &  4 &   0.40645(7) &   4.43 &  0.25 &  5.1 &      \\
     &           &            &        &          &          &    &   0.44762(0) &   3.88 &  0.25 &  4.9 &      \\
 486 & 11.603423 & -61.574974 & 14.933 &   0.500  &   0.270  &  4 &   0.60284(3) &  13.76 &  0.39 &  6.0 &      \\
     &           &            &        &          &          &    &   0.59326(3) &  11.26 &  0.39 &  6.8 &      \\
     &           &            &        &          &          &    &   0.53903(1) &   9.88 &  0.39 &  7.5 &      \\
 487 & 11.589927 & -61.652501 & 14.918 &   0.532  &   0.252  &  4 &    1.5541(6) &   3.57 &  0.26 &  4.9 &      \\
 503 & 11.611771 & -61.645566 & 15.006 &   0.482  &   0.284  &  4 &  0.435316(7) &  11.68 &  0.26 &  7.7 &      \\
     &           &            &        &          &          &    &   0.44313(4) &  11.02 &  0.26 &  9.6 &      \\
 513 & 11.606852 & -61.562636 & 15.028 &   0.467  &   0.279  &  4 &  0.444622(4) &  15.46 &  0.27 &  9.4 &      \\
 517 & 11.605800 & -61.635845 & 15.053 &   0.502  &   0.315  &  4 &   0.28878(0) &   3.90 &  0.27 &  6.7 &      \\
     &           &            &        &          &          &    &   0.30718(6) &   2.88 &  0.27 &  6.2 &      \\
 521 & 11.605288 & -61.636663 & 15.017 &   0.518  &   0.296  &  4 &   1.20660(7) &  33.06 &  0.27 &  7.5 &      \\
 525 & 11.590774 & -61.694319 & 15.066 &   0.474  &   0.242  &  4 &   0.68957(9) &  17.77 &  0.30 &  7.3 &      \\
     &           &            &        &          &          &    &   0.75461(1) &   8.46 &  0.30 &  5.6 &      \\
     &           &            &        &          &          &    &   0.76413(6) &   5.94 &  0.30 &  4.8 &      \\
 527 & 11.602230 & -61.539634 & 15.097 &   0.494  &   0.274  &  4 &    0.9460(5) &   3.27 &  0.30 &  4.8 &      \\
     &           &            &        &          &          &    &    0.7773(4) &   2.62 &  0.30 &  4.5 &      \\
 532 & 11.590808 & -61.581019 & 15.112 &   0.517  &   0.329  &  4 &   0.85411(4) &  14.70 &  0.35 &  9.6 &      \\
 548 & 11.609013 & -61.594757 & 15.170 &   0.461  &   0.267  &  4 &   0.64536(8) &   7.62 &  0.29 &  8.1 &      \\
     &           &            &        &          &          &    &   0.65012(9) &   5.19 &  0.29 &  5.8 &      \\
 576 & 11.610962 & -61.588967 & 15.279 &   0.857  &   0.783  &  3 & 0.0761747(0) &   8.78 &  0.34 &  7.7 &      \\
     &           &            &        &          &          &    & 0.0770839(3) &   8.68 &  0.34 &  8.7 &      \\
     &           &            &        &          &          &    & 0.0773966(3) &   5.75 &  0.34 &  8.0 &      \\
 582 & 11.609336 & -61.549058 & 15.264 &   0.781  &   0.733  &  3 &  0.181395(8) &  10.09 &  0.31 &  7.5 &      \\
     &           &            &        &          &          &    &  0.162313(8) &  10.40 &  0.31 &  9.0 &      \\
     &           &            &        &          &          &    &  0.183005(0) &   6.48 &  0.31 &  7.9 &      \\
     &           &            &        &          &          &    &  0.208947(2) &   3.00 &  0.31 &  4.6 &      \\
 688 & 11.603246 & -61.651846 & 15.584 &   0.784  &   0.225  &  5 &  0.265345(4) &  12.56 &  0.37 & 10.3 &      \\
 732 & 11.612531 & -61.683247 & 15.699 &   0.587  &   0.138  &  5 &   0.81556(1) &  27.26 &  0.39 & 10.7 &      \\
 829 & 11.597904 & -61.669576 & 15.946 &   0.596  &   0.296  &  5 &   1.24570(7) &  30.69 &  0.48 &  9.3 &      \\
 895 & 11.593339 & -61.636325 & 16.062 &   0.746  &   0.160  &  5 &    1.7040(0) &  17.02 &  0.49 &  5.2 &      \\
 916 & 11.598819 & -61.627093 & 16.078 &   0.855  &   0.170  &  5 &   0.43183(4) &  19.57 &  0.50 &  5.6 &      \\
     &           &            &        &          &          &    &   0.36283(4) &  13.68 &  0.50 &  5.3 &      \\
 954 & 11.591866 & -61.541493 & 16.105 &   0.825  &   0.199  &  5 &    3.5197(7) &  20.14 &  0.56 &  5.9 &      \\
1031 & 11.607701 & -61.577649 & 16.304 &   0.768  &   0.212  &  5 &   0.44537(4) &  12.96 &  0.58 &  5.7 &      \\
1079 & 11.601244 & -61.659534 & 16.393 &   0.763  &   0.119  &  5 &     5.220(4) &  16.80 &  0.61 &  5.1 &      \\
1136 & 11.595086 & -61.555719 & 16.537 &   1.022  &   0.367  &  5 &   0.55137(4) &  17.20 &  0.68 &  5.6 &      \\
     &           &            &        &          &          &    &   0.55342(3) &  12.57 &  0.68 &  4.5 &      \\
1179 & 11.600334 & -61.626781 & 16.547 &   0.878  &   0.029  &  5 &  0.274527(7) &  20.23 &  0.70 &  5.4 &      \\
     &           &            &        &          &          &    &  0.274242(9) &  11.58 &  0.70 &  5.1 &      \\
\hline
\end{tabular}
\end{table*}

\addtocounter{table}{-1}
\begin{table*}
\centering
\caption{Continued.}
\begin{tabular}{r c c c c c c l c c c c}
\hline
 GvaId &  RA & DEC &  $ V' $  & $B'-V'$  &  $U'-B'$   & Gr &  $ P $  &   $ A $  & $\sigma(A)$  & S/N  & MGid\\
        & (hour) & (deg) & (mag) & (mag)  &  (mag)   &       & (day) & \small{(mmag)} & \small{(mmag)} &     &       \\
\hline
1180 & 11.597436 & -61.614476 & 16.577 &   0.966  &   0.349  &  5 &     5.129(4) &  13.50 &  0.71 &  5.6 &      \\
     &           &            &        &          &          &    &     5.308(1) &   8.97 &  0.71 &  4.9 &      \\
1193 & 11.603726 & -61.526970 & 16.521 &   0.964  &   0.221  &  5 &     3.352(2) &  14.57 &  0.66 &  4.6 &      \\
1214 & 11.611543 & -61.646323 & 16.577 &   0.998  &   0.343  &  5 &  0.280210(3) &  16.89 &  0.71 &  7.9 &      \\
     &           &            &        &          &          &    &   0.28008(9) &  10.47 &  0.71 &  5.3 &      \\
     &           &            &        &          &          &    &   0.27976(9) &   8.93 &  0.71 &  4.7 &      \\
1229 & 11.615596 & -61.567463 & 16.576 &   0.953  &   0.256  &  5 &  0.269982(5) &  29.58 &  0.82 &  7.5 &      \\
1243 & 11.605376 & -61.693680 & 16.735 &   0.886  &   0.121  &  5 &    1.1983(5) &  21.57 &  0.79 &  6.3 &      \\
     &           &            &        &          &          &    &    1.2154(4) &  12.85 &  0.79 &  5.0 &      \\
1244 & 11.596807 & -61.654920 & 16.683 &   0.776  &   0.079  &  5 &    1.9549(8) &  19.12 &  0.73 &  6.5 &      \\
1265 & 11.610328 & -61.650383 & 16.578 &   0.842  &   0.193  &  5 &  0.306583(1) &  32.57 &  0.66 &  6.9 &      \\
     &           &            &        &          &          &    &   0.30667(7) &  12.22 &  0.66 &  6.2 &      \\
1277 & 11.599574 & -61.534387 & 16.706 &   0.977  &   0.253  &  5 &  0.276478(7) &  24.46 &  0.75 &  5.6 &      \\
     &           &            &        &          &          &    &  0.276425(6) &  23.34 &  0.75 &  6.4 &      \\
1280 & 11.608598 & -61.593859 & 16.821 &   0.991  &   0.263  &  5 &  0.280277(0) &  32.37 &  0.82 &  7.2 &      \\
     &           &            &        &          &          &    &  0.280921(2) &  26.85 &  0.82 &  6.3 &      \\
1292 & 11.608012 & -61.620849 & 16.683 &   0.890  &   0.238  &  5 &  0.271393(3) &  20.31 &  0.73 &  8.2 &      \\
     &           &            &        &          &          &    &   0.27102(4) &   9.87 &  0.73 &  5.6 &      \\
1293 & 11.610974 & -61.694219 & 16.694 &   0.827  &   0.136  &  5 &    1.4555(0) &  19.16 &  0.71 &  5.8 &      \\
     &           &            &        &          &          &    &    1.5078(5) &  16.89 &  0.71 &  6.3 &      \\
1314 & 11.599004 & -61.658982 & 16.678 &   0.860  &   0.171  &  5 &    1.7212(1) &  30.51 &  0.74 &  5.8 &      \\
     &           &            &        &          &          &    &    1.7198(4) &  23.30 &  0.74 &  5.0 &      \\
1321 & 11.611152 & -61.635980 & 16.742 &   0.956  &   0.282  &  5 &  0.353691(0) &  26.13 &  0.75 &  8.5 &      \\
     &           &            &        &          &          &    &   0.35257(4) &   7.74 &  0.75 &  4.5 &      \\
1355 & 11.598935 & -61.623828 & 16.782 &   0.903  &   0.201  &  5 &  0.296703(6) &  39.00 &  0.82 &  8.8 &      \\
     &           &            &        &          &          &    &  0.297446(2) &  21.65 &  0.82 &  5.2 &      \\
     &           &            &        &          &          &    &  0.296996(1) &  16.12 &  0.82 &  5.2 &      \\
1356 & 11.592101 & -61.530478 & 16.635 &   0.838  &   0.166  &  5 &   0.39744(2) &  17.09 &  0.70 &  5.0 &      \\
1375 & 11.610875 & -61.610815 & 16.740 &   1.037  &   0.324  &  5 &  0.336598(0) &  38.80 &  1.07 &  6.5 &      \\
     &           &            &        &          &          &    &   0.33240(9) &  23.01 &  1.07 &  4.8 &      \\
     &           &            &        &          &          &    &  0.099727(4) &  21.84 &  1.07 &  5.0 &      \\
     &           &            &        &          &          &    & 0.0623187(0) &  20.76 &  1.07 &  5.4 &      \\
1420 & 11.596537 & -61.630730 & 16.732 &   0.867  &   0.129  &  5 &   0.51484(6) &  21.15 &  0.77 &  5.9 &      \\
1426 & 11.605309 & -61.573042 & 16.783 &   0.995  &   0.244  &  5 &  0.272083(0) &  30.92 &  0.79 &  8.2 &      \\
     &           &            &        &          &          &    &  0.272048(7) &  22.99 &  0.79 &  7.8 &      \\
     &           &            &        &          &          &    &  0.271985(9) &  17.13 &  0.79 &  7.5 &      \\
1428 & 11.612314 & -61.637564 & 17.138 & ( 2.692) & (-1.487) &  5 &      887.(9) & 205.92 &  1.02 & 10.3 &      \\
1493 & 11.609096 & -61.633745 & 16.858 &   1.051  &   0.103  &  5 &  0.281220(0) &  30.67 &  0.85 &  8.1 &      \\
     &           &            &        &          &          &    &  0.281189(5) &  17.46 &  0.85 &  5.9 &      \\
     &           &            &        &          &          &    &  0.280689(9) &  17.66 &  0.85 &  6.1 &      \\
1584 & 11.614060 & -61.618346 & 16.993 &   0.972  &   0.787  &  4 &   0.54271(2) &  15.53 &  0.89 &  5.9 &      \\
1619 & 11.594024 & -61.528056 & 17.089 &   1.055  &   0.327  &  5 &    2.5793(2) &  21.84 &  0.98 &  5.6 &      \\
1624 & 11.613816 & -61.616588 & 17.055 &   1.016  &   0.363  &  5 &   0.42090(5) &  35.75 &  0.94 &  6.8 &      \\
     &           &            &        &          &          &    &   0.42031(7) &  16.83 &  0.94 &  5.9 &      \\
     &           &            &        &          &          &    &   0.42016(7) &  14.18 &  0.94 &  5.1 &      \\
1631 & 11.589010 & -61.556211 & 17.165 &   0.983  &   0.019  &  5 &   0.61935(8) &  35.11 &  1.07 &  4.8 &      \\
1632 & 11.614971 & -61.575756 & 17.095 &   1.356  & ( 0.707) &  5 &  0.155513(7) &  15.00 &  1.06 &  8.2 &      \\
1635 & 11.606469 & -61.525441 & 17.085 &   1.000  &   0.256  &  5 &     4.639(0) &  18.69 &  0.97 &  5.3 &      \\
1677 & 11.614275 & -61.650115 & 17.225 &   1.051  &   0.379  &  5 &   0.72732(7) &  43.61 &  1.05 &  6.1 &      \\
     &           &            &        &          &          &    &   0.72565(6) &  22.70 &  1.05 &  4.6 &      \\
1808 & 11.601592 & -61.549937 & 17.252 &   1.040  &   0.135  &  5 &    2.6311(8) &  26.85 &  1.24 &  5.3 &      \\
1840 & 11.603009 & -61.615725 & 17.370 &   1.315  & (-0.005) &  5 &  0.131230(1) &  20.67 &  1.27 &  5.3 &      \\
1851 & 11.601026 & -61.568062 & 17.427 &   1.067  &  -0.067  &  5 &   0.55453(6) &  32.56 &  1.33 &  5.9 &      \\
     &           &            &        &          &          &    &   0.55439(4) &  23.74 &  1.33 &  5.0 &      \\
     &           &            &        &          &          &    &   0.55475(0) &  22.20 &  1.33 &  5.7 &      \\
1943 & 11.607080 & -61.527034 & 17.443 &   1.205  &   0.233  &  5 &  0.241868(2) &  57.85 &  1.30 &  8.8 &      \\
1954 & 11.602310 & -61.608826 & 17.523 &   1.024  & ( 0.179) &  5 &   0.53648(6) &  38.91 &  1.47 &  4.8 &      \\
     &           &            &        &          &          &    &   0.53630(8) &  34.86 &  1.47 &  5.5 &      \\
1957 & 11.611634 & -61.555442 & 17.524 &   1.082  &   0.180  &  5 &     4.134(2) &  32.81 &  1.37 &  5.0 &      \\
     &           &            &        &          &          &    &     4.140(8) &  24.70 &  1.37 &  4.9 &      \\
2022 & 11.602410 & -61.552437 & 17.578 &   1.551  &  -0.348  &  5 &   0.54497(2) &  16.48 &  1.49 &  6.7 &      \\
2046 & 11.598558 & -61.619435 & 17.501 &   1.240  &   0.064  &  5 &  0.258649(0) &  33.32 &  1.42 &  6.9 &      \\
     &           &            &        &          &          &    &   0.25868(0) &  18.49 &  1.42 &  5.9 &      \\
2067 & 11.604188 & -61.646589 & 17.578 &   0.951  &   0.259  &  5 &    1.1490(3) &  15.69 &  1.54 &  5.7 &      \\
2087 & 11.593437 & -61.679804 & 17.633 &   1.405  & ( 0.084) &  5 &   0.77977(4) &  37.48 &  1.47 &  8.6 &      \\
2127 & 11.615326 & -61.541133 & 17.669 &   1.133  &   0.265  &  5 &  0.265405(2) &  29.02 &  1.51 &  6.1 &      \\
\hline
\end{tabular}
\end{table*}

\addtocounter{table}{-1}
\begin{table*}
\centering
\caption{Continued.}
\begin{tabular}{r c c c c c c l c c c c}
\hline
 GvaId &  RA & DEC &  $ V' $  & $B'-V'$  &  $U'-B'$   & Gr &  $ P $  &   $ A $  & $\sigma(A)$  & S/N  & MGid\\
        & (hour) & (deg) & (mag) & (mag)  &  (mag)   &       & (day) & \small{(mmag)} & \small{(mmag)} &     &       \\
\hline
2140 & 11.589755 & -61.552963 & 17.585 &   0.917  &   0.155  &  5 & 0.1620640(5) &  95.09 &  1.41 & 11.8 &      \\
2214 & 11.597199 & -61.630694 & 17.742 &   1.254  & (-0.398) &  5 &     5.405(5) &  28.97 &  1.71 &  5.7 &      \\
2215 & 11.598451 & -61.594279 & 17.698 &   1.168  &  -0.214  &  5 &   0.42733(3) &  25.19 &  1.69 &  5.2 &      \\
2225 & 11.615113 & -61.678138 & 17.736 &   1.070  &   0.101  &  5 &    1.3021(8) &  28.01 &  1.55 &  6.1 &      \\
2259 & 11.609796 & -61.520962 & 17.843 &   1.122  &  -0.021  &  5 &   0.63127(9) &  41.62 &  1.71 &  6.3 &      \\
     &           &            &        &          &          &    &   0.63201(8) &  33.52 &  1.71 &  6.3 &      \\
2269 & 11.612000 & -61.658532 & 17.803 &   0.978  &   0.204  &  5 &   0.61773(3) &  20.76 &  1.65 &  5.2 &      \\
2348 & 11.607309 & -61.585568 & 17.841 &   1.295  & (-0.117) &  5 &  0.242874(9) &  81.28 &  1.83 &  9.4 &      \\
2430 & 11.594418 & -61.641885 & 17.994 &   1.233  &  -0.291  &  5 &  0.302735(8) &  43.81 &  2.06 &  7.5 &      \\
2680 & 11.609580 & -61.658850 & 18.178 &   1.291  & (-0.712) &  5 &   0.74751(3) &  29.94 &  2.47 &  5.6 &      \\
     &           &            &        &          &          &    &    0.7463(5) &  26.38 &  2.47 &  5.6 &      \\
2792 & 11.609594 & -61.581031 & 18.466 &   0.842  & (-0.984) &  5 &  0.398101(2) & 120.19 &  3.16 &  9.5 &      \\
     &           &            &        &          &          &    &  0.099727(4) &  37.79 &  3.16 &  5.5 &      \\
2808 & 11.593834 & -61.554431 & 18.427 &   1.283  & (-0.807) &  5 &  0.223584(6) &  73.96 &  2.93 &  9.1 &      \\
     &           &            &        &          &          &    &   0.22356(9) &  24.60 &  2.93 &  4.8 &      \\
2813 & 11.608429 & -61.615832 & 18.144 &   0.780  & (-1.301) &  5 &  0.216225(5) &  77.57 &  2.47 &  8.2 &      \\
     &           &            &        &          &          &    & 0.0398908(2) &  23.73 &  2.47 &  5.0 &      \\
2856 & 11.602080 & -61.690781 & 18.283 &   1.291  & (-1.396) &  5 &   0.68397(3) &  39.44 &  2.65 &  5.0 &      \\
2859 & 11.603510 & -61.561785 & 18.240 &   1.062  & (-0.389) &  5 &     4.298(8) &  37.42 &  2.60 &  5.5 &      \\
     &           &            &        &          &          &    &  0.199453(9) &  25.24 &  2.60 &  4.8 &      \\
3016 & 11.598079 & -61.690827 & 18.482 & ( 1.217) & (-0.617) &  5 &  0.147699(9) &  25.17 &  3.01 &  5.3 &      \\
3129 & 11.607887 & -61.618213 & 18.344 &   0.868  & (-0.191) &  5 &    0.7546(1) &  29.86 &  2.73 &  5.3 &      \\
3181 & 11.603472 & -61.517611 & 18.544 &   0.929  & (-0.440) &  5 &  0.136340(1) &  49.28 &  3.16 &  6.8 &      \\
3222 & 11.607338 & -61.551968 & 18.705 &   1.044  & (-0.883) &  5 &   0.36628(1) &  51.93 &  3.81 &  6.4 &      \\
3302 & 11.610656 & -61.592919 & 18.772 &   1.074  & (-0.890) &  5 &   0.26367(1) &  49.45 &  4.06 &  5.5 &      \\
     &           &            &        &          &          &    &   0.26355(3) &  43.51 &  4.06 &  4.5 &      \\
3351 & 11.599879 & -61.531930 & 18.759 & ( 1.021) & (-0.992) &  5 &   0.36018(1) &  52.16 &  3.99 &  6.6 &      \\
3485 & 11.591766 & -61.687235 & 18.941 &   0.936  & (-0.837) &  5 &  0.280562(3) &  77.10 &  4.39 &  8.1 &      \\
3525 & 11.595234 & -61.690899 & 19.147 & ( 0.846) & (-0.992) &  5 &  0.147654(5) &  97.74 &  5.49 &  7.2 &      \\
\hline
\end{tabular}
\label{Tab:periodicVariables}
\end{table*}

The list of group~1 to 5 periodic variables with secured frequencies is provided in Table~\ref{Tab:periodicVariables}.
We give, in the order of the columns presented in the table, their identification number in our numbering scheme, their mean $V'$ magnitude and colors, the group to which they belong (as defined in Sect.~\ref{Sect:periodicVariables}), the period, amplitude, uncertainty on the amplitude, and S/N ratio of each of their frequencies determined from their $V'$ time series (one line per period), and the star identification number given by \cite{McSwainGies05b} and used in \cite{McSwainHuangGies_etal08}, if available.

The number of digits printed in Table~\ref{Tab:periodicVariables} for the periods is computed from the uncertainty $\varepsilon_P$ on the period $P$, estimated with the formula, derived by \cite{MontgomeryOdonoghue99},
\begin{equation}
 \varepsilon_P = \frac{\sqrt{6}}{\Pi\sqrt{N_\mathrm{obs}}}
                 \; \frac{1}{T_\mathrm{obs}}
                 \; \frac{\sigma_\mathrm{noise}}{A}
                 \; P^2
                 \;,
\end{equation}
where $N_\mathrm{obs}$ is the number of observations in the time series of duration $T_\mathrm{obs}$, $A$ is the amplitude and $\sigma_\mathrm{noise}$ the mean uncertainty of the measurements.
The uncertain digit is written in parenthesis.

The uncertainty on the amplitude is estimated, also from \cite{MontgomeryOdonoghue99}, with the formula
\begin{equation}
 \sigma(A) = \frac{\sqrt{2}}{\sqrt{N_\mathrm{obs}}}
                 \; \sigma_\mathrm{noise}
                 \;.
\end{equation}

\begin{figure*}
  \centering
  \includegraphics[width=0.66\columnwidth]{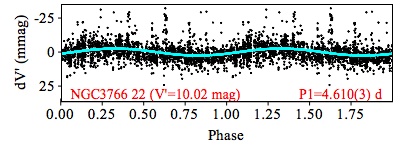}
  \includegraphics[width=0.66\columnwidth]{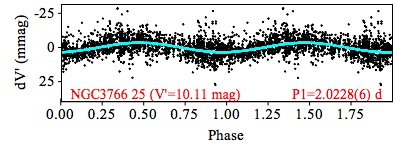}
  \includegraphics[width=0.66\columnwidth]{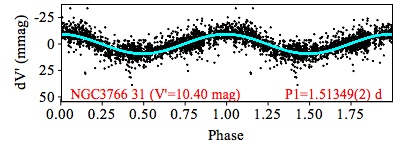}
  \includegraphics[width=0.66\columnwidth]{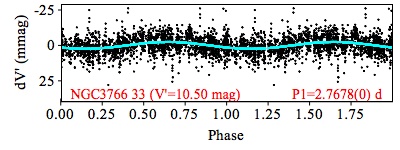}
  \includegraphics[width=0.66\columnwidth]{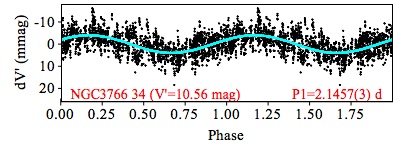}
  \includegraphics[width=0.66\columnwidth]{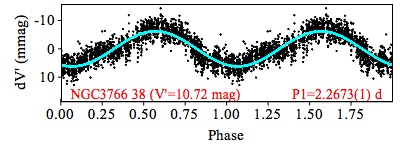}
  \includegraphics[width=0.66\columnwidth]{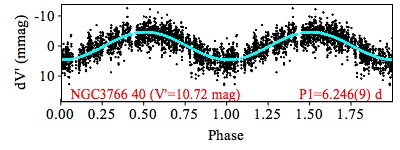}
  \includegraphics[width=0.66\columnwidth]{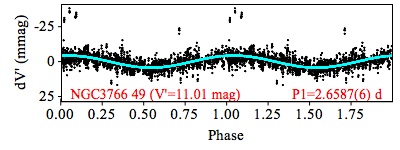}
  \includegraphics[width=0.66\columnwidth]{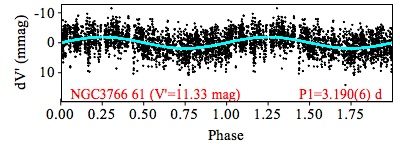}
  \includegraphics[width=0.66\columnwidth]{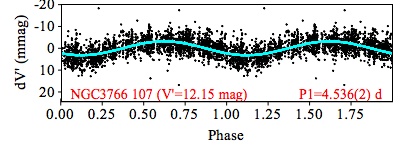}
  \includegraphics[width=0.66\columnwidth]{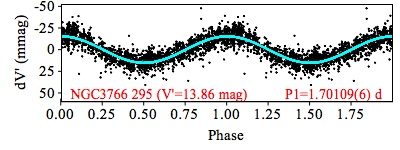}
  \caption{Folded light curves in $V'$ of the monoperiodic variables of group~1 (SPB candidates).
    The mean $V'$ magnitude has been subtracted from the light curves.
    Cyan lines represent a sine fit to the data using the frequency found by the period search algorithm.
    The star identification is written in the lower left corner of each figure, with the mean $V'$ magnitude in parenthesis.
    The period used to fold the data is written in the lower right corner of each figure.}
\label{Fig:foldedLcsGroup1_monoperiodic}
\end{figure*}

\begin{figure*}
  \centering
  \includegraphics[width=0.66\columnwidth]{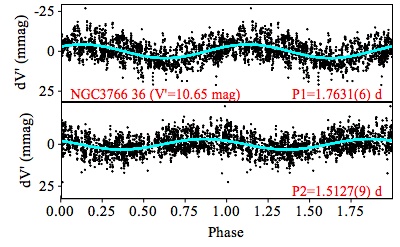}
  \includegraphics[width=0.66\columnwidth]{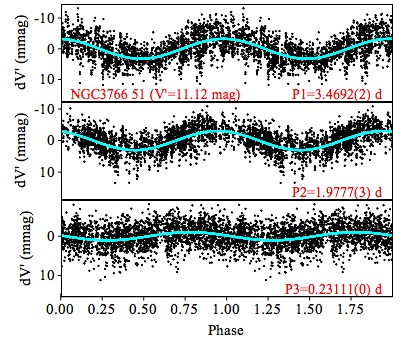}
  \caption{Same as Fig.~\ref{Fig:foldedLcsGroup1_monoperiodic}, but for the multiperiodic variables of group~1 (SPB candidates).
    All significant frequencies are shown, one panel per frequency.
    The data shown for each frequency is the residual obtained by subtracting from the original data all earlier frequencies, from top to bottom panels.
    The period used to fold the data in each panel is written in the lower right corner of each panel.}
\label{Fig:foldedLcsGroup1_multiperiodic}
\end{figure*}

\begin{figure*}
  \centering
  \includegraphics[width=0.66\columnwidth]{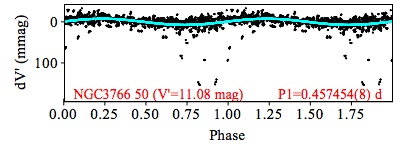}
  \includegraphics[width=0.66\columnwidth]{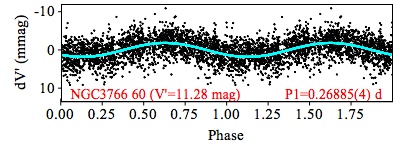}
  \includegraphics[width=0.66\columnwidth]{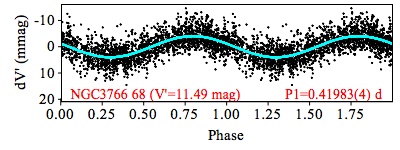}
  \includegraphics[width=0.66\columnwidth]{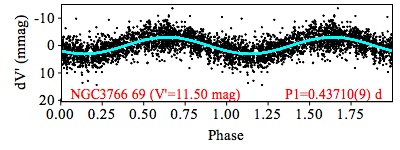}
  \includegraphics[width=0.66\columnwidth]{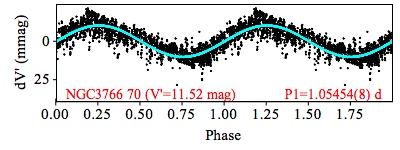}
  \includegraphics[width=0.66\columnwidth]{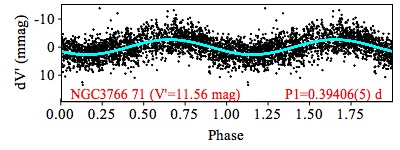}
  \includegraphics[width=0.66\columnwidth]{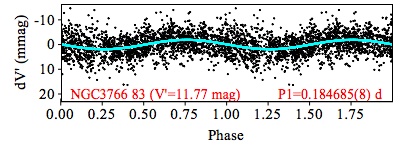}
  \includegraphics[width=0.66\columnwidth]{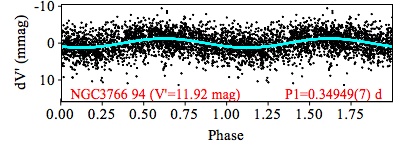}
  \includegraphics[width=0.66\columnwidth]{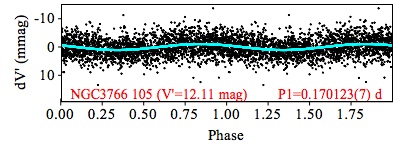}
  \includegraphics[width=0.66\columnwidth]{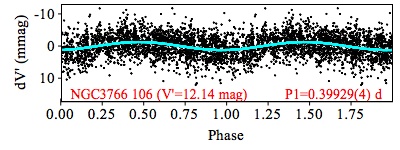}
  \includegraphics[width=0.66\columnwidth]{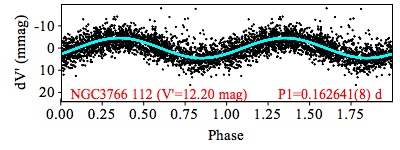}
  \includegraphics[width=0.66\columnwidth]{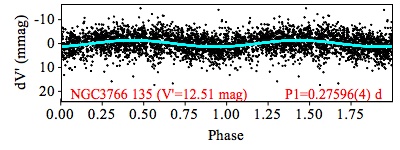}
  \includegraphics[width=0.66\columnwidth]{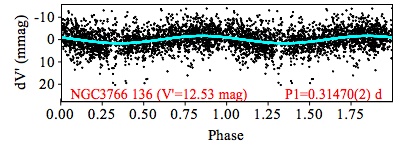}
  \includegraphics[width=0.66\columnwidth]{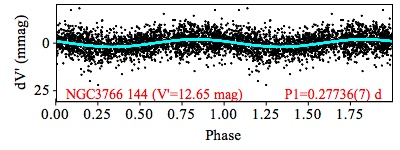}
  \includegraphics[width=0.66\columnwidth]{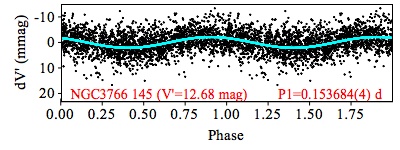}
  \includegraphics[width=0.66\columnwidth]{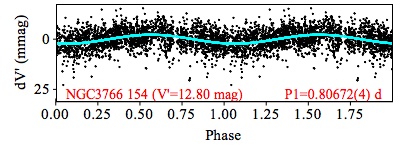}
  \includegraphics[width=0.66\columnwidth]{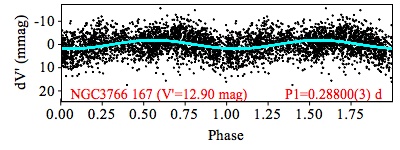}
  \includegraphics[width=0.66\columnwidth]{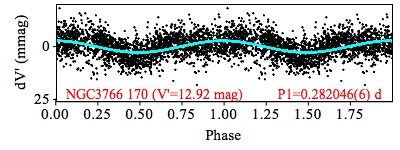}
  \includegraphics[width=0.66\columnwidth]{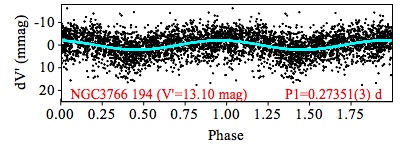}
  \includegraphics[width=0.66\columnwidth]{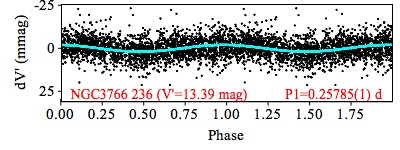}
  \includegraphics[width=0.66\columnwidth]{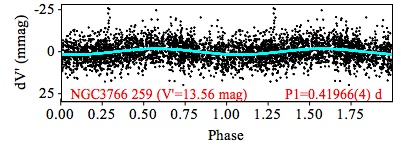}
  \includegraphics[width=0.66\columnwidth]{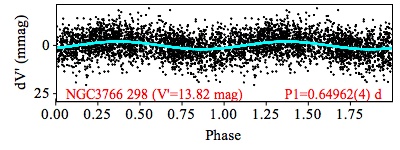}
  \includegraphics[width=0.66\columnwidth]{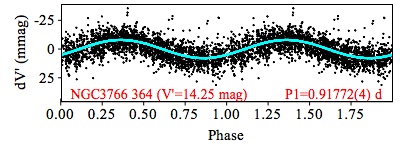}
  \caption{Same as Fig.~\ref{Fig:foldedLcsGroup1_monoperiodic}, but for the monoperiodic variables of group~2.}
\label{Fig:foldedLcsGroup2_monoperiodic}
\end{figure*}

\begin{figure*}
  \centering
  \includegraphics[width=0.66\columnwidth]{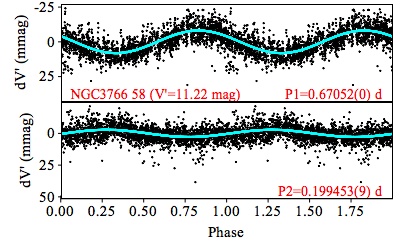}
  \includegraphics[width=0.66\columnwidth]{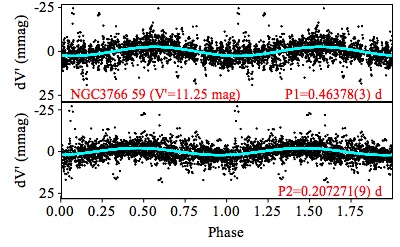}
  \includegraphics[width=0.66\columnwidth]{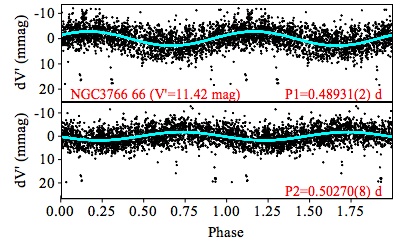}
  \includegraphics[width=0.66\columnwidth]{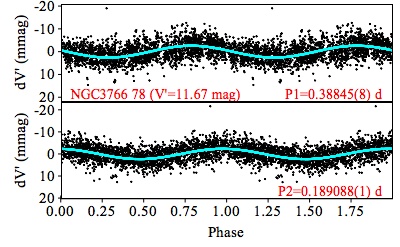}
  \includegraphics[width=0.66\columnwidth]{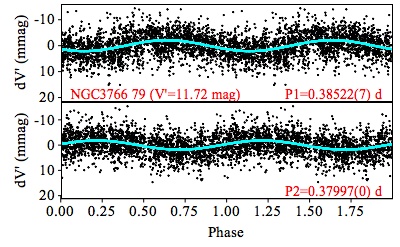}
  \includegraphics[width=0.66\columnwidth]{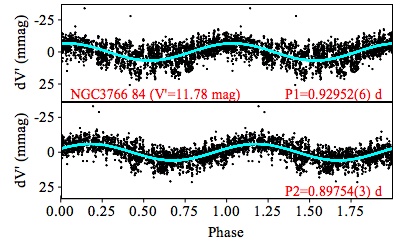}
  \includegraphics[width=0.66\columnwidth]{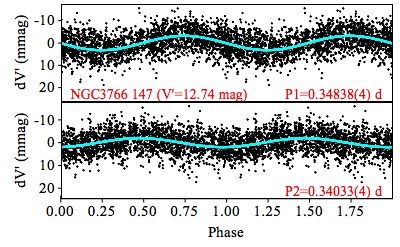}
  \includegraphics[width=0.66\columnwidth]{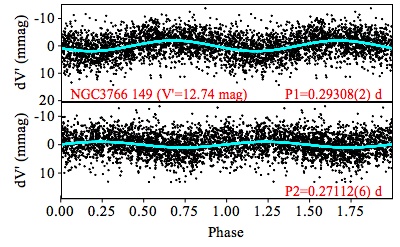}
  \includegraphics[width=0.66\columnwidth]{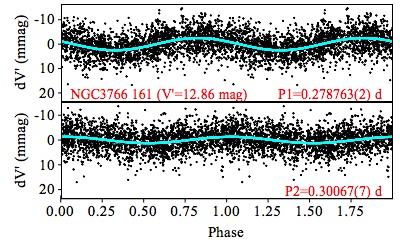}
  \includegraphics[width=0.66\columnwidth]{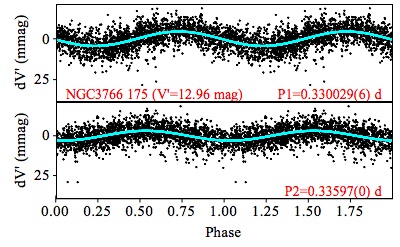}
  \includegraphics[width=0.66\columnwidth]{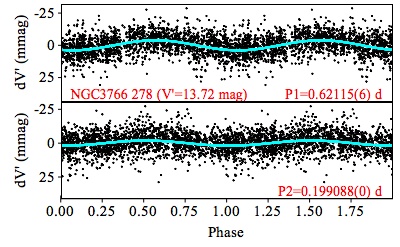}
  \caption{Same as Fig.~\ref{Fig:foldedLcsGroup1_multiperiodic}, but for the biperiodic variables of group~2.}
\label{Fig:foldedLcsGroup2_biperiodic}
\end{figure*}

\begin{figure*}
  \centering
  \includegraphics[width=0.66\columnwidth]{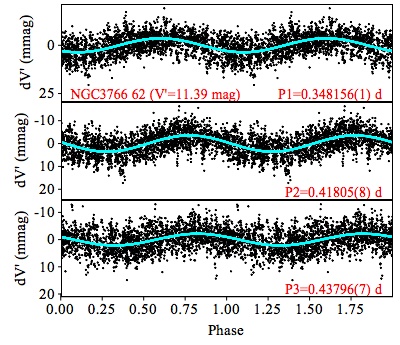}
  \includegraphics[width=0.66\columnwidth]{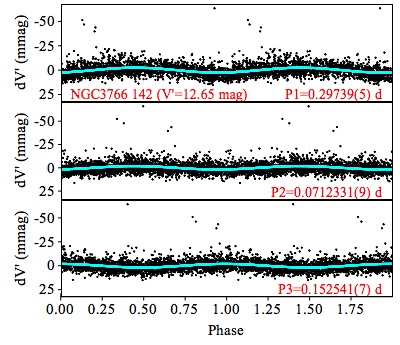}
  \caption{Same as Fig.~\ref{Fig:foldedLcsGroup1_multiperiodic}, but for the triperiodic variables of group~2.}
\label{Fig:foldedLcsGroup2_triperiodic}
\end{figure*}

\begin{figure*}
  \centering
  \includegraphics[width=0.66\columnwidth]{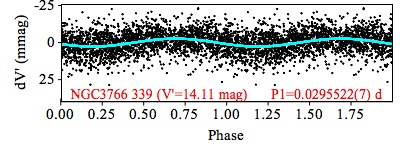}
  \includegraphics[width=0.66\columnwidth]{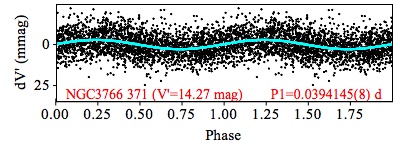}
  \includegraphics[width=0.66\columnwidth]{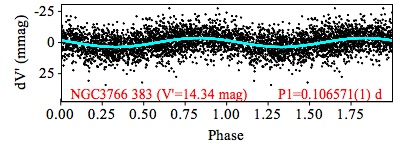}
  \includegraphics[width=0.66\columnwidth]{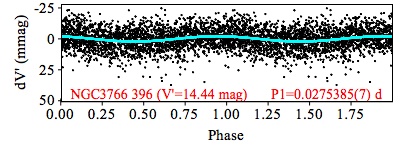}
  \includegraphics[width=0.66\columnwidth]{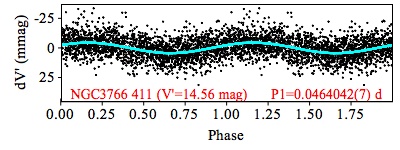}
  \includegraphics[width=0.66\columnwidth]{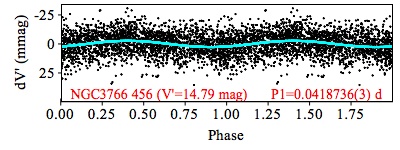}
  \caption{Same as Fig.~\ref{Fig:foldedLcsGroup1_monoperiodic}, but for the monoperiodic variables of group~3 ($\delta$~Sct candidates).}
\label{Fig:foldedLcsGroup3_monoperiodic}
\end{figure*}

\begin{figure*}
  \centering
  \includegraphics[width=0.66\columnwidth]{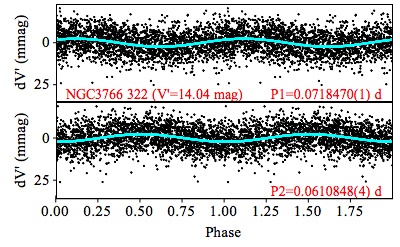}
  \includegraphics[width=0.66\columnwidth]{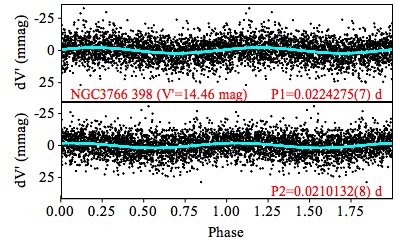}
  \includegraphics[width=0.66\columnwidth]{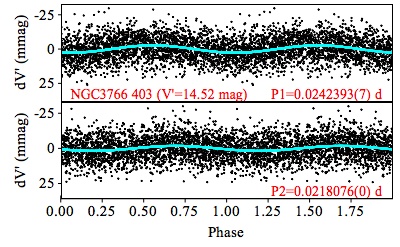}
  \caption{Same as Fig.~\ref{Fig:foldedLcsGroup1_multiperiodic}, but for the biperiodic variables of group~3 ($\delta$~Sct candidates).}
\label{Fig:foldedLcsGroup3_biperiodic}
\end{figure*}

\begin{figure*}
  \centering
  \includegraphics[width=0.66\columnwidth]{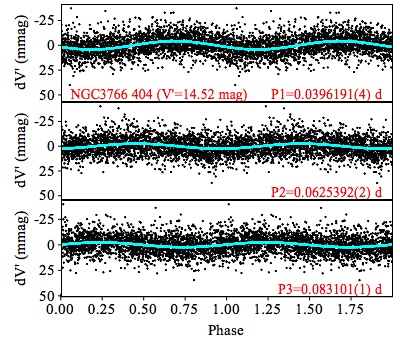}
  \includegraphics[width=0.66\columnwidth]{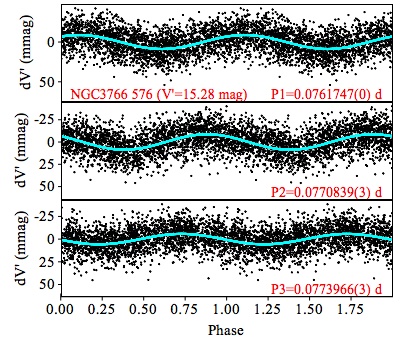}
  \caption{Same as Fig.~\ref{Fig:foldedLcsGroup1_multiperiodic}, but for the triperiodic variables of group~3 ($\delta$~Sct candidates).}
\label{Fig:foldedLcsGroup3_triperiodic}
\end{figure*}

\begin{figure*}
  \centering
  \includegraphics[width=0.66\columnwidth]{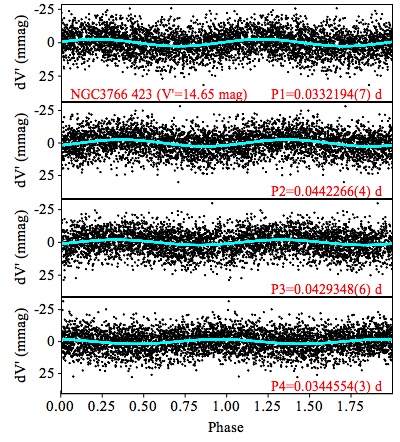}
  \includegraphics[width=0.66\columnwidth]{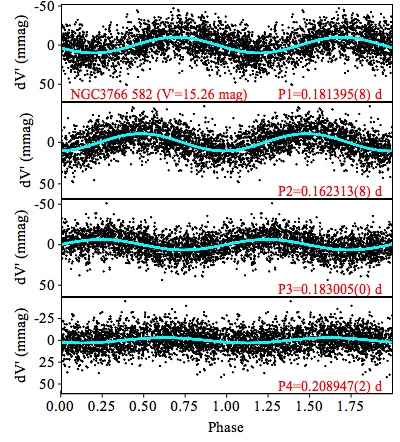}
  \includegraphics[width=0.64\columnwidth]{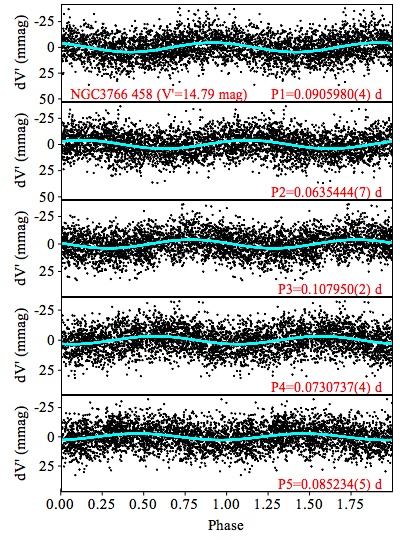}
  \caption{Same as Fig.~\ref{Fig:foldedLcsGroup1_multiperiodic}, but for the multiperiodic variables of group~3 ($\delta$~Sct candidates) with more than three periods.}
\label{Fig:foldedLcsGroup3_multiperiodic}
\end{figure*}

\begin{figure*}
  \centering
  \includegraphics[width=0.66\columnwidth]{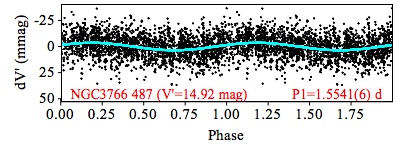}
  \includegraphics[width=0.66\columnwidth]{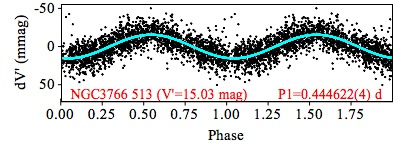}
  \includegraphics[width=0.66\columnwidth]{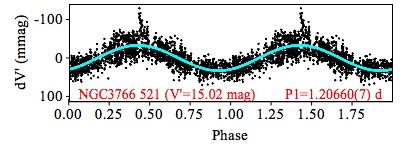}
  \includegraphics[width=0.66\columnwidth]{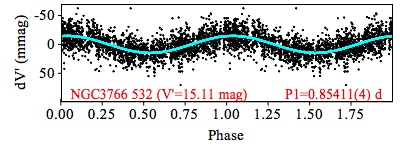}
  \includegraphics[width=0.66\columnwidth]{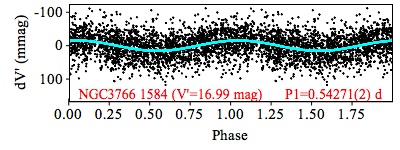}
  \caption{Same as Fig.~\ref{Fig:foldedLcsGroup1_monoperiodic}, but for the monoperiodic variables of group~4 ($\gamma$~Dor candidates).}
\label{Fig:foldedLcsGroup4_monoperiodic}
\end{figure*}

\begin{figure*}
  \centering
  \includegraphics[width=0.66\columnwidth]{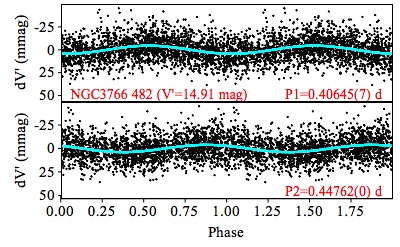}
  \includegraphics[width=0.66\columnwidth]{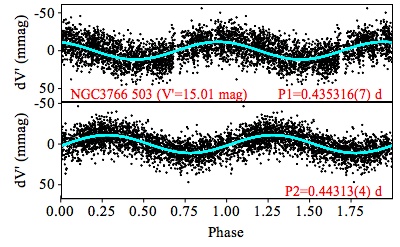}
  \includegraphics[width=0.66\columnwidth]{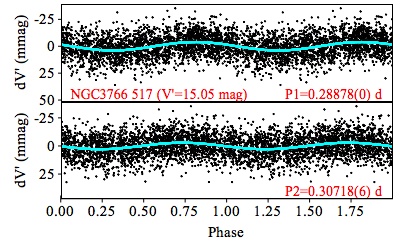}
  \includegraphics[width=0.66\columnwidth]{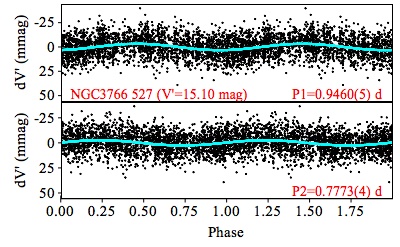}
  \includegraphics[width=0.66\columnwidth]{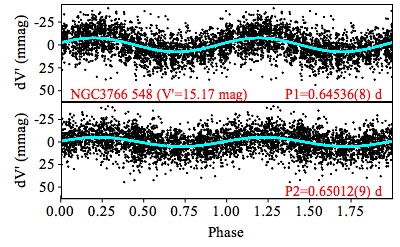}
  \caption{Same as Fig.~\ref{Fig:foldedLcsGroup1_multiperiodic}, but for the biperiodic variables of group~4 ($\gamma$~Dor candidates).}
\label{Fig:foldedLcsGroup4_biperiodic}
\end{figure*}

\begin{figure*}
  \centering
  \includegraphics[width=0.66\columnwidth]{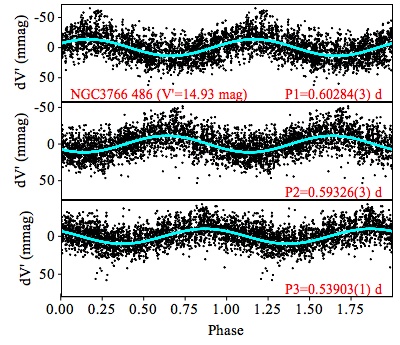}
  \includegraphics[width=0.66\columnwidth]{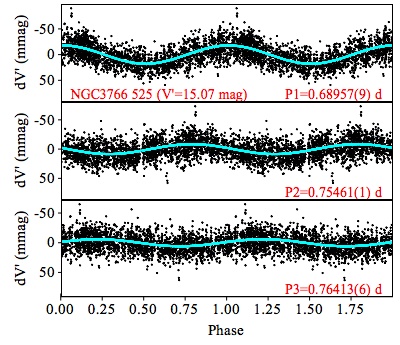}
  \caption{Same as Fig.~\ref{Fig:foldedLcsGroup1_multiperiodic}, but for the multiperiodic variables of group~4 ($\gamma$~Dor candidates) with more than two periods.}
\label{Fig:foldedLcsGroup4_multiperiodic}
\end{figure*}

The folded light curves of group~1 stars are shown in Figs~\ref{Fig:foldedLcsGroup1_monoperiodic} and \ref{Fig:foldedLcsGroup1_multiperiodic}, of group~2 stars in Figs~\ref{Fig:foldedLcsGroup2_monoperiodic} to \ref{Fig:foldedLcsGroup2_triperiodic}, of group~3 stars in Figs~\ref{Fig:foldedLcsGroup3_monoperiodic} to \ref{Fig:foldedLcsGroup3_multiperiodic}, and of group~4 stars in Figs~\ref{Fig:foldedLcsGroup4_monoperiodic} to \ref{Fig:foldedLcsGroup4_multiperiodic}.
\toReferee{The outliers seen in the folded light curve of star~49 (Fig.~\ref{Fig:foldedLcsGroup1_monoperiodic}) are due to an outburst occurring at the end of our observation campaign.
It will be studied in more details a forthcoming paper (Mowlavi et al., in preparation).
}

\begin{figure*}
  \centering
  \includegraphics[width=1.8\columnwidth]{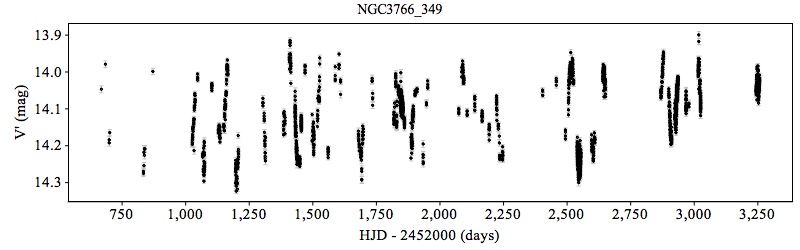}
  \includegraphics[width=1.8\columnwidth]{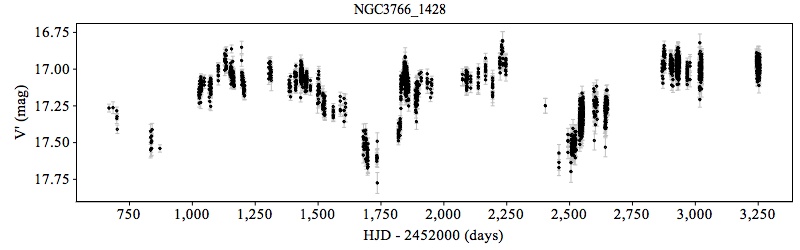}
  \caption{$V'$ light curves of the LPV stars 349 (top figure) and 1428 (bottom figure).
  }
\label{Fig:lcsLPVs}
\end{figure*}

The light curves of the two LPV stars in our sample are shown in Fig.~\ref{Fig:lcsLPVs}.

\end{appendix}

\end{document}